\documentclass[a4paper,iop,twocolumn,numberedappendix]{emulateapj} 
\usepackage[colorlinks=true,linkcolor=blue,anchorcolor=blue,citecolor=blue,urlcolor=blue,draft=false]{hyperref}\usepackage{color}
\usepackage{graphicx}
\usepackage{amsmath} 
\usepackage{amssymb}

\usepackage{placeins}
\usepackage{times}

\usepackage{xcolor}

\usepackage{natbib}
 
\newcommand{\simgt}{\lower.5ex\hbox{$\; \buildrel > \over \sim \;$}}
\newcommand{\simlt}{\lower.5ex\hbox{$\; \buildrel < \over \sim \;$}}
\newcommand{\llangle}{\langle\!\langle}
\newcommand{\rrangle}{\rangle\!\rangle}

\newcommand{\Nbin}{N_\mathrm{GL}}
\newcommand{\Nwl}{N_\mathrm{WL}}
\newcommand{\Nsl}{N_\mathrm{SL}}

\def\bc{\mbox{\boldmath $c$}}
\def\btheta{\mbox{\boldmath $\theta$}}

\def\bSigma{\mbox{\boldmath $\Sigma$}}

\def\by{\mbox{\boldmath $y$}}
\def\bx{\mbox{\boldmath $x$}}

\def\bs{\mbox{\boldmath $s$}}
\def\bp{\mbox{\boldmath $p$}}

\lefthead{Umetsu et al.}
\righthead{CLASH: Strong-lensing, Weak-lensing Shear and Magnification Analysis} 

\usepackage{datetime}
\usepackage{version}    

\begin{document}

\title{CLASH: Joint Analysis of Strong-Lensing, Weak-Lensing
Shear and Magnification Data for 20 Galaxy Clusters\altaffilmark{*}} 

\author{Keiichi Umetsu\altaffilmark{1}}  
\author{Adi Zitrin\altaffilmark{2,3}}       
\author{Daniel Gruen\altaffilmark{4,5,6,7}} 
\author{Julian Merten\altaffilmark{8}}      
\author{Megan Donahue\altaffilmark{9}}      
\author{Marc Postman\altaffilmark{10}}      

\altaffiltext{*}
 {Based in part on data collected at the Subaru Telescope,
  which is operated by the National Astronomical Society of Japan.}
\email{keiichi@asiaa.sinica.edu.tw}
\altaffiltext{1}
 {Institute of Astronomy and Astrophysics, Academia Sinica,
  P.~O. Box 23-141, Taipei 10617, Taiwan}
\altaffiltext{2}{Cahill Center for Astronomy and Astrophysics,
California Institute of Technology, MS 249-17, Pasadena, CA 91125, USA}
\altaffiltext{3}{Hubble Fellow}
\altaffiltext{4}{Universit\"ats-Sternwarte, M\"unchen, Scheinerstr. 1,
D-81679 M\"unchen. Germany}
\altaffiltext{5}{Max-Planck-Institut f\"ur extraterrestrische Physik,
Giessenbachstr. 1, D-85748 Garching} 
\altaffiltext{6}{SLAC National Accelerator Laboratory, Menlo Park, CA
94025, USA; Einstein Fellow} 
\altaffiltext{7}{Kavli Institute for Particle Astrophysics and
Cosmology, Stanford University, Palo Alto, CA 94305, USA}
\altaffiltext{6}{Department of Physics, University of Oxford, Keble
Road, Oxford OX1 3RH, UK}
\altaffiltext{7}{Department of Physics and Astronomy, Michigan State
University, East Lansing, MI 48824, USA} 
\altaffiltext{8}{Space Telescope Science Institute, 3700 San Martin
Drive, Baltimore, MD 21208, USA} 


\begin{abstract}
We present a comprehensive analysis of strong-lensing,
 weak-lensing shear and magnification data for  
a sample of 16 X-ray-regular and 4 high-magnification galaxy clusters
at $0.19\simlt z\simlt 0.69$
 selected from the Cluster Lensing And Supernova survey with Hubble
 (CLASH).
Our analysis combines constraints from 16-band {\em Hubble Space
 Telescope} observations 
 and wide-field multi-color imaging taken primarily with Suprime-Cam on
 the Subaru Telescope, spanning a wide range of cluster radii
 ($10\arcsec$--$16\arcmin$).
We reconstruct surface mass density profiles of individual clusters from
 a joint analysis of the full lensing constraints, and determine masses
 and concentrations for all clusters.
We find internal consistency of the ensemble mass calibration to be
$\le 5\%\pm 6\%$ 
in the one-halo regime ($200$--$2000$\,kpc\,$h^{-1}$) 
by comparison with the CLASH weak-lensing-only measurements of Umetsu et 
 al.
For the X-ray-selected subsample of 16 clusters,
we examine the concentration--mass ($c$--$M$) relation and its
 intrinsic scatter using a Bayesian regression approach.
Our model yields a mean concentration of $c|_{z=0.34}=3.95\pm 0.35$
 at $M_\mathrm{200c} \simeq 14\times 10^{14}M_\odot$ and an intrinsic
 scatter of $\sigma(\ln{c_{\mathrm{200c}}})=0.13\pm 0.06$,
in excellent agreement with $\Lambda$ cold dark matter predictions when
 the CLASH selection function based on X-ray morphological regularity
 and the projection effects are taken into account. 
We also derive an ensemble-averaged surface mass density profile for the
 X-ray-selected subsample  
by stacking their individual profiles.
The stacked lensing signal is detected at $33\sigma$ significance over the
 entire radial range $\le 4000$\,kpc\,$h^{-1}$,
accounting for the effects of intrinsic
 profile variations and uncorrelated large-scale structure along the
 line of sight.
The stacked mass profile is well described by a family of 
 density profiles predicted for cuspy dark-matter-dominated halos in
gravitational equilibrium, namely, the Navarro--Frenk--White (NFW),
 Einasto, and DARKexp models, whereas the single power-law, cored
 isothermal and Burkert density profiles are disfavored by the data.
We show that cuspy halo models that include the large-scale two-halo term
provide improved agreement with the data.
For the NFW halo model, we measure a mean concentration of
 $c_\mathrm{200c}=3.79^{+0.30}_{-0.28}$
at $M_\mathrm{200c}= 14.1^{+1.0}_{-1.0}\times 10^{14}M_\odot$, 
demonstrating 
consistency between complementary analysis methods.
\end{abstract}  
 


\section{Introduction}
\label{sec:intro}

Clusters of galaxies represent the largest and rarest class of
self-gravitating systems formed in the universe.
In the context of hierarchical models of structure formation,
the evolution of the cluster abundance with cosmic time is a sensitive
probe of the amplitude and growth rate of the primordial fluctuation
spectrum because cluster-sized systems populate the exponential tail of
the cosmic mass function of dark-matter (DM, hereafter) halos
\citep{2001ApJ...553..545H}. 
Therefore, large cluster samples with well-defined selection functions   
can, in principle, provide an independent test of 
any viable cosmological model, including the
current concordance $\Lambda$ cold dark matter ($\Lambda$CDM)
model defined in the framework of general relativity, 
complementing cosmic microwave background (CMB) anisotropy, large-scale
galaxy clustering, supernova, CMB lensing and cosmic shear experiments.
Currently, cluster samples are often defined by optical, X-ray, or
Sunyaev--Zel'dovich effect (SZE) observables 
\citep[e.g.,][]{Vikhlinin+2009CCC3,Planck2014XX}, so that the 
masses are indirectly inferred from scaling relations.
In the last few years, a systematic effort has been conducted to enable
a self-consistent calibration of mass--observable relations 
using robust cluster lensing measurements
\citep{vonderLinden2014calib,Gruen2014,Umetsu2014clash,Ford2014cfhtlens,Melchior2015DES,Jimeno2015,Hoekstra2015CCCP,Merten2015clash} 
and well-defined selection functions \citep{JPAS2014,Miyazaki2015}.

Considerable progress has been made in understanding the formation and
evolution of DM halos in an expanding universe, governed by nonlinear
gravitational growth of cosmic density perturbations.     
$N$-body simulations of collisionless CDM established a nearly
self-similar form for the spherically-averaged density profile
$\rho_\mathrm{h}(r)$ of DM halos \citep[][hereafter
Navarro--Frenk--White, NFW]{1997ApJ...490..493N}
over a wide range of halo masses and radii, with some intrinsic variance
associated with the mass accretion histories of individual halos
\citep{Jing+Suto2000,Tasitsiomi+2004,Graham+2006,Navarro+2010,Ludlow+2013,Diemer+Kravtsov2014,Adhikari2014}.
The degree of mass concentration,
$c_\mathrm{200c}=r_\mathrm{200c}/r_{-2}$,\footnote{The quantity $r_\mathrm{200c}$
is defined as the radius within which the mean interior density is
200 times the critical density $\rho_\mathrm{c}(z)$ of the universe at
the cluster redshift $z$, and $r_{-2}$ is a scale radius at which
$d\ln{\rho_\mathrm{h}}/dln{r}=-2$.}
is predicted to correlate with halo mass because the scale radius
$r_{-2}$ stays approximately constant when the halo leaves the fast
accretion regime, whereas $r_\mathrm{200c}$ does still grow, thus
increasing concentration. 
Cluster-sized halos are thus predicted to be less concentrated
than less massive halos and to have typical concentrations
of $c_\mathrm{200c}=3$--$4$
\citep{Bhatt+2013,Dutton+Maccio2014,Meneghetti2014clash}.

Galaxy clusters act as powerful gravitational lenses
\citep[e.g.,][]{Kneib+Natarajan2011},
offering a direct probe for testing these well defined predictions of
DM halo structure.
The critical advantage of cluster gravitational lensing 
is its ability to map the mass distribution of individual systems
independent of assumptions about their physical and dynamical state.
Clusters produce a variety of detectable lensing effects,
including deflection, shearing, and magnifying of the images of
background sources \citep{2001PhR...340..291B}.
In the weak regime where the lensing signal is approximately linearly
related to the potential,
lensing can be used to probe the mass distribution of clusters
in a model-independent manner 
\citep[e.g.,][]{1993ApJ...404..441K,1994ApJ...437...56F,Kaiser1995,1999PThPS.133...53U,2000ApJ...539..540C}.
In the strong regime, several sets of
multiple images with known redshifts  allow us to tightly constrain the
central mass distribution \citep[e.g.,][]{Zitrin+2012M1206,Jauzac2014}. 
For a massive cluster, the two
complementary regimes contribute similar logarithmic radial coverage
\citep[][see their Figure 6]{Umetsu+2011}.
Hence, combining strong and weak lensing can
significantly improve constraints on the cluster mass distribution for
full radial coverage
\citep[][]{Bradac2006,2007ApJ...668..643L,Merten+2009,Diego2015a1689}.

The Cluster Lensing And Supernova survey with Hubble
\citep[CLASH,][]{Postman+2012CLASH}\footnote{\href{http://www.stsci.edu/~postman/CLASH/}{http://www.stsci.edu/~postman/CLASH/}}
is a 524-orbit Multi-Cycle Treasury program that has been designed to
probe the mass distribution of 25 galaxy clusters using their
gravitational lensing properties, providing a sizable sample of mass
calibrators for accurate cluster cosmology. 
All CLASH clusters were observed in 16 filters
with the {\em Hubble Space Telescope} ({\em HST}).
This 16-band {\em HST} photometry has enabled us to uncover many sets of
multiple images \citep{Zitrin+2012M1206,Zitrin2015clash}, whose
spectroscopic redshifts have been determined from a dedicated
spectroscopic survey conducted with the VIMOS spectrograph on
the Very Large Telescope \citep[VLT;][]{Biviano+2013,Balestra2013,Rosati2014VLT,Girardi2015}.
CLASH has produced combined strong- and weak-lensing analyses of 
{\em HST} and Subaru Telescope observations, allowing for detailed mapping of the
cluster mass distributions 
\citep{Umetsu+2012,Coe+2012A2261,Medezinski+2013,Zitrin2015clash,Merten2015clash}. 
\citet{Donahue2014clash} derived radial profiles of temperature, gas
mass, and hydrostatic mass for the full CLASH sample 
using {\em XMM-Newton} and {\em Chandra} observations.

A major goal of the CLASH survey is to test models of structure
formation by using the halo concentration-mass ($c$--$M$) relation
determined from cluster gravitational lensing.
For this aim, twenty CLASH clusters were selected to have X-ray
temperatures greater than 5\,keV and to show a smooth X-ray
morphology, with no lensing information used a priori 
\citep{Postman+2012CLASH}. 
A further sample of five clusters were selected by their high lens
magnification properties to study magnified high-redshift galaxies
behind the clusters \citep{Zheng+2012,Coe2013,Monna2014}.

Recently, we have carried out a systematic study of the
CLASH sample 
to obtain measurements of 
mass and concentration from cluster lensing  
\citep{Meneghetti2014clash,Umetsu2014clash,Merten2015clash}.  
\citet[][]{Meneghetti2014clash} presented a detailed
characterization of the CLASH X-ray-selected clusters with
numerical simulations to make predictions about their 
intrinsic and observational properties.
\citet[][]{Umetsu2014clash} conducted a joint
shear-and-magnification weak-lensing analysis of a subsample of the
CLASH clusters, using wide-field multi-color imaging taken primarily
with Subaru/Suprime-Cam.
\citet[][]{Merten2015clash} presented a two-dimensional
strong- and weak-lensing (hereafter {\sc SaWLens}) analysis of 19
CLASH X-ray-selected clusters,  
by combining strong and weak-shear lensing data from 16-band {\em HST}  
imaging with wide-field weak-shear data analyzed by \citet{Umetsu2014clash}.
In both analyses, we find excellent agreement between the data and
$\Lambda$CDM predictions 
when the projection effects and the selection function based on X-ray
morphology are taken into account. 
More recently, 
\citet{Xu2015clash} carried out an observational and
theoretical study of the abundance of gravitationally lensed arcs in the
CLASH survey, finding full agreement between the observations and
simulations  using an automated, objective arcfinding algorithm. 

In this paper we present a comprehensive joint analysis of
strong-lensing, weak-lensing shear and magnification data for a sample
of 16 X-ray-regular and 4 high-magnification 
clusters selected from the CLASH survey. 
Our extended analysis combines the constraints from the CLASH 
Subaru and {\em HST} lensing data sets of \citet{Umetsu2014clash} and
\citet{Zitrin2015clash}. 
We aim at combining these
complementary lensing constraints to construct individual cluster
surface mass density profiles, from which to determine the $c$--$M$
relation. This improved mass-profile data set also allows us to obtain an
ensemble calibration of cluster masses and to probe the
ensemble-averaged cluster mass distribution.

The paper is organized as follows. 
In Section \ref{sec:basics} we briefly describe the basic theory of
cluster gravitational lensing. After summarizing the properties of the CLASH
sample, we outline in Section \ref{sec:method} the formalism and
procedure for constructing surface mass density profiles from a joint
analysis of strong-lensing, weak-lensing shear and magnification
constraints. 
In Section \ref{sec:profile} we revisit the mass profile analysis of
individual CLASH clusters by combining {\em HST} and ground-based
lensing measurements. In Section \ref{sec:stack}, we conduct stacked
lensing analysis of the X-ray-selected subsample to study their 
ensemble-averaged mass distribution. In Section \ref{sec:cM} we examine
the concentration--mass relation for the X-ray-selected subsample using
Bayesian regression methods.
Section \ref{sec:discussion} is devoted to the discussion of the
results. Finally, a summary is given in Section \ref{sec:summary}.
  
Throughout this paper, we adopt a concordance $\Lambda$CDM cosmology
with $\Omega_\mathrm{m}=0.27$, $\Omega_{\Lambda}=0.73$, and
$h=0.7h_{70}=0.7$ \citep{Komatsu+2011WMAP7},
where $H_0 = h\times 100$\,km\,s$^{-1}$\,Mpc$^{-1}=h_{70}\times 70$\,km\,s$^{-1}$\,Mpc$^{-1}$.
We use the standard notation $M_{\Delta_\mathrm{c}}$ ($M_{\Delta_\mathrm{m}}$)
to denote the mass enclosed within a sphere of radius
$r_{\Delta_\mathrm{c}}$ ($r_{\Delta_\mathrm{m}}$), within 
which the mean overdensity is $\Delta_\mathrm{c}$ ($\Delta_\mathrm{m}$)
times the critical density $\rho_\mathrm{c}(z)$
(mean background density $\overline{\rho}_\mathrm{m}(z)$) at the cluster
redshift $z$. 
All quoted errors are 68.3\% ($1\sigma$)
confidence limits (CL) unless otherwise stated.

\section{Cluster Lensing Basics}
\label{sec:basics}

In the cluster lensing regime \citep{Umetsu2010Fermi}, the convergence
$\kappa=\Sigma/\Sigma_\mathrm{c}$ is the projected surface mass density
$\Sigma(\btheta)$ in units of the critical surface mass density for
lensing, 
$\Sigma_\mathrm{c}=(c^2 D_\mathrm{s})/(4\pi G D_\mathrm{l}D_\mathrm{ls})\equiv c^2/(4\pi G D_\mathrm{l}\beta)$,
where $D_\mathrm{l}$, $D_\mathrm{s}$, and $D_\mathrm{ls}$ 
are the lens, source, and lens-source proper angular diameter distances,
respectively, and
$\beta(z, z_\mathrm{l}) = D_\mathrm{ls}/D_\mathrm{s}$ is the geometric lensing
strength as a function of source redshift $z$ and lens redshift $z_\mathrm{l}$.

The gravitational shear $\gamma$ can be directly observed from
ellipticities of background galaxies in the regime where $\kappa\ll 1$.    
The tangential shear component $\gamma_{+}$
averaged around a circle of radius $\theta$
satisfies the following {\it identity}
\citep{Kaiser1995}:
\begin{equation} 
\label{eq:loop}
\gamma_{+}(\theta) =
\kappa(<\theta)-\kappa(\theta) \equiv 
\Delta\Sigma(\theta)/\Sigma_\mathrm{c}, 
\end{equation}
with $\kappa(\theta)=\Sigma(\theta)/\Sigma_\mathrm{c}$ 
the azimuthally averaged convergence at radius $\theta$,
$\kappa(<\theta)=\Sigma(<\theta)/\Sigma_\mathrm{c}$
the average convergence interior to $\theta$, and
$\Delta\Sigma(\theta)=\Sigma(<\theta)-\Sigma(\theta)$
the differential surface mass density.

The observable quantity for quadrupole weak lensing
in general 
is not the shear but the {\it reduced} gravitational shear,
\begin{equation}
\label{eq:redshear}
g(\btheta)=\frac{\gamma(\btheta)}{1-\kappa(\btheta)},
\end{equation}
which is invariant under 
$\kappa(\btheta) \to \lambda \kappa(\btheta) + 1-\lambda$ 
and
$\gamma(\btheta) \to \lambda \gamma(\btheta)$
with an arbitrary constant $\lambda\ne 0$,
known as the mass-sheet degeneracy
\citep[][]{Falco1985,Gorenstein1988,Schneider+Seitz1995}.
This degeneracy can be broken, for example,\footnote{Alternatively, one
may constrain the constant $\lambda$ such that the enclosed mass within a
certain aperture is consistent with mass estimates from
independent observations \citep[e.g.,][]{2000ApJ...539L...5U}. See also
Section \ref{subsec:rec}.
}
by measuring the magnification factor,  
\begin{equation}
\label{eq:mu2d}
\mu(\btheta) = \frac{1}{[1-\kappa(\btheta)]^2-|\gamma(\btheta)|^2},
\end{equation}
which transforms as $\mu(\btheta)\to \lambda^2\mu(\btheta)$.


We consider a population of source galaxies described by
their redshift distribution function, 
$\overline{N}(z)$, for statistical weak-lensing measurements.
The mean lensing depth for a given population ($X=g,\mu$) is given by
\begin{equation}
\label{eq:depth}
\langle\beta\rangle_X =\left[
\int_0^\infty\!dz\, \overline{N}_X(z) \beta(z)\right]
\left[
\int_0^\infty\!dz\,\overline{N}_X(z)
\right]^{-1}.
\end{equation}
In general, we apply different size, magnitude, and color cuts in source selection
for measuring the shear and magnification effects,
leading to different $\overline{N}_X(z)$. In contrast to the former effect, the
latter does not require source galaxies to be spatially  resolved, but
it does require a stringent flux limit against incompleteness effects.

We introduce the relative lensing depth of a source population 
with respect to a fiducial source in the far background as
$\langle W\rangle_X = \langle\beta\rangle_X  / \beta_\infty$
with $\beta_\infty\equiv \beta(z\to \infty, z_\mathrm{l})$
\citep{2001PhR...340..291B}.
The associated critical surface mass density is
$\Sigma_{\mathrm{c},\infty}=c^2/(4\pi G D_\mathrm{l}\beta_{\infty})$.
The source-averaged convergence and shear fields are then expressed as
$\langle\kappa \rangle_X =\langle W\rangle_X \kappa_\infty$ and 
$\langle\gamma \rangle_X =\langle W\rangle_X \gamma_\infty$, 
using those in the far-background limit.
Hereafter, we use the far-background lensing fields,
$\kappa_\infty(\btheta)$ and $\gamma_\infty(\btheta)$, 
to describe the projected mass distribution of clusters.

\section{Cluster Sample, Data, and Methodology}
\label{sec:method}

In this section, we outline the analysis procedure used to combine
strong-lensing, weak-lensing shear and magnification data for direct
reconstruction of cluster surface mass density profiles.
In Section \ref{subsec:sample}, we present a summary of the properties
of our cluster sample.
In Section \ref{subsec:back}, we describe the background galaxy
selection for the weak-lensing shear and magnification analysis. 
In Sections \ref{subsec:gt}, \ref{subsec:magbias}, and \ref{subsec:sl}, 
we describe our methods for measuring cluster lensing profiles as a function of
clustercentric radius.
In Section \ref{subsec:bayesian} we outline the joint likelihood approach of
\citet{Umetsu2013} to perform a mass profile reconstruction from
multi-probe lensing data.

\subsection{Cluster Sample}
\label{subsec:sample}

\begin{deluxetable*}{lrrrrccccc}
\centering
\tabletypesize{\scriptsize}
\tablecolumns{10}
\tablecaption{
\label{tab:sample}
Properties of the cluster sample
}
\tablewidth{0pt}
\tablehead{
 \multicolumn{1}{c}{Cluster} &
 \multicolumn{1}{c}{$z_{\rm l}$} &
 \multicolumn{1}{c}{R.A. \tablenotemark{a}} &
 \multicolumn{1}{c}{decl.\tablenotemark{a}} &
 \multicolumn{1}{c}{$k_{\rm B}T_X$\tablenotemark{b}} &
 \multicolumn{1}{c}{$\theta_\mathrm{Ein}$\tablenotemark{c}} &
 \multicolumn{4}{c}{$M_\mathrm{2D}$ ($10^{13}M_\odot\, h_{70}^{-1}$)\tablenotemark{d}}
 \\ 
\colhead{} &
\colhead{} &
\multicolumn{1}{c}{(J2000.0)} &
\multicolumn{1}{c}{(J2000.0)} &
\multicolumn{1}{c}{(keV)} &
\multicolumn{1}{c}{($\arcsec$)} &
\multicolumn{1}{c}{$\theta=10\arcsec$} &
\multicolumn{1}{c}{$\theta=20\arcsec$} &
\multicolumn{1}{c}{$\theta=30\arcsec$} &
\multicolumn{1}{c}{$\theta=40\arcsec$}
}
\startdata
X-ray Selected:\\
         ~~Abell 383 & $0.187$ & 02:48:03.40 & -03:31:44.9 & $6.5\pm0.24$ & $15.1$ & $ 1.15 \pm  0.17$ & $ 3.04 \pm  0.51$ & $ 4.98 \pm  0.96$ & $ 6.77 \pm  1.35$\\
         ~~Abell 209 & $0.206$ & 01:31:52.54 & -13:36:40.4 & $7.3\pm0.54$ & $ 8.9$ & $ 0.93 \pm  0.14$ & $ 2.36 \pm  0.45$ & $ 3.96 \pm  0.89$ & $ 5.60 \pm  1.40$\\
        ~~Abell 2261 & $0.224$ & 17:22:27.18 & +32:07:57.3 & $7.6\pm0.30$ & $23.1$ & $ 1.91 \pm  0.31$ & $ 4.79 \pm  0.69$ & $ 7.67 \pm  1.26$ & $10.42 \pm  1.84$\\
    ~~RXJ2129.7+0005 & $0.234$ & 21:29:39.96 & +00:05:21.2 & $5.8\pm0.40$ & $12.9$ & $ 1.13 \pm  0.18$ & $ 3.36 \pm  0.44$ & $ 5.95 \pm  0.74$ & $ 8.67 \pm  1.10$\\
         ~~Abell 611 & $0.288$ & 08:00:56.82 & +36:03:23.6 & $7.9\pm0.35$ & $18.1$ & $ 1.84 \pm  0.25$ & $ 5.25 \pm  0.83$ & $ 9.37 \pm  1.82$ & $14.26 \pm  3.06$\\
       ~~MS2137-2353 & $0.313$ & 21:40:15.17 & -23:39:40.2 & $5.9\pm0.30$ & $17.1$ & $ 2.25 \pm  0.33$ & $ 5.23 \pm  0.81$ & $ 7.99 \pm  1.38$ & $10.76 \pm  2.00$\\
    ~~RXJ2248.7-4431 & $0.348$ & 22:48:43.96 & -44:31:51.3 & $12.4\pm0.60$ & $31.1$ & $ 2.47 \pm  0.45$ & $ 7.36 \pm  1.02$ & $13.14 \pm  1.96$ & $19.41 \pm  3.07$\\
  ~~MACSJ1115.9+0129 & $0.352$ & 11:15:51.90 & +01:29:55.1 & $8.0\pm0.40$ & $18.1$ & $ 1.85 \pm  0.37$ & $ 5.79 \pm  0.84$ & $11.09 \pm  1.50$ & $16.98 \pm  2.37$\\
  ~~MACSJ1931.8-2635 & $0.352$ & 19:31:49.62 & -26:34:32.9 & $6.7\pm0.40$ & $22.2$ & $ 2.91 \pm  0.60$ & $ 7.37 \pm  1.12$ & $12.15 \pm  1.86$ & $17.21 \pm  2.90$\\
    ~~RXJ1532.9+3021 & $0.363$ & 15:32:53.78 & +30:20:59.4 & $5.5\pm0.40$ &        --- &        --- &        --- &        --- &        ---\\
  ~~MACSJ1720.3+3536 & $0.391$ & 17:20:16.78 & +35:36:26.5 & $6.6\pm0.40$ & $20.1$ & $ 2.65 \pm  0.35$ & $ 7.20 \pm  1.08$ & $12.39 \pm  2.17$ & $17.97 \pm  3.36$\\
  ~~MACSJ0429.6-0253 & $0.399$ & 04:29:36.05 & -02:53:06.1 & $6.0\pm0.44$ & $15.7$ & $ 2.22 \pm  0.39$ & $ 6.80 \pm  0.96$ & $12.55 \pm  1.86$ & $18.87 \pm  3.02$\\
  ~~MACSJ1206.2-0847 & $0.440$ & 12:06:12.15 & -08:48:03.4 & $10.8\pm0.60$ & $26.8$ & $ 3.37 \pm  0.50$ & $ 9.51 \pm  1.39$ & $16.37 \pm  2.50$ & $23.18 \pm  3.88$\\
  ~~MACSJ0329.7-0211 & $0.450$ & 03:29:41.56 & -02:11:46.1 & $8.0\pm0.50$ & $24.1$ & $ 3.40 \pm  0.62$ & $ 8.66 \pm  1.26$ & $14.36 \pm  2.16$ & $21.12 \pm  3.17$\\
    ~~RXJ1347.5-1145 & $0.451$ & 13:47:31.05 & -11:45:12.6 & $15.5\pm0.60$ & $33.0$ & $ 3.56 \pm  0.76$ & $11.55 \pm  2.14$ & $20.40 \pm  3.18$ & $29.25 \pm  4.12$\\
  ~~MACSJ0744.9+3927 & $0.686$ & 07:44:52.82 & +39:27:26.9 & $8.9\pm0.80$ & $24.3$ & $ 4.70 \pm  0.92$ & $13.55 \pm  2.13$ & $23.64 \pm  3.35$ & $34.79 \pm  4.57$\\
\hline High Magnification:\\
  ~~MACSJ0416.1-2403 & $0.396$ & 04:16:08.38 & -24:04:20.8 & $7.5\pm0.80$ & $25.9$ & $ 1.33 \pm  0.18$ & $ 5.35 \pm  0.79$ & $11.32 \pm  1.68$ & $17.43 \pm  2.51$\\
  ~~MACSJ1149.5+2223 & $0.544$ & 11:49:35.69 & +22:23:54.6 & $8.7\pm0.90$ & $20.4$ & $ 2.88 \pm  0.51$ & $ 9.29 \pm  1.39$ & $17.80 \pm  2.87$ & $28.09 \pm  5.07$\\
  ~~MACSJ0717.5+3745 & $0.548$ & 07:17:32.63 & +37:44:59.7 & $12.5\pm0.70$ & $55.0$ & $ 2.01 \pm  0.19$ & $ 8.04 \pm  0.74$ & $18.79 \pm  2.01$ & $35.80 \pm  5.83$\\
  ~~MACSJ0647.7+7015 & $0.584$ & 06:47:50.27 & +70:14:55.0 & $13.3\pm1.80$ & $26.4$ & $ 3.62 \pm  0.77$ & $11.75 \pm  1.84$ & $21.59 \pm  3.32$ & $31.95 \pm  5.45$
\enddata
\tablenotetext{a}{The cluster center is taken to be the location of the BCG when a single dominant central galaxy is found. Otherwise, in the case of MACSJ0717.5$+$3745 and MACSJ0416.1$-$2403, it is defined as the center of the brightest red-sequence-selected cluster galaxies \citep{Umetsu2014clash}. 
}
\tablenotetext{b}{X-ray temperature from \citet{Postman+2012CLASH}.}
\tablenotetext{c}{Effective Einstein radius for a fiducial source at $z_\mathrm{s}=2$ determined from the {\em HST} strong and weak-shear lensing analysis by \citet{Zitrin2015clash}. The reported values are the average of two different models (where available) of \citet{Zitrin2015clash}. The typical model uncertainty in $\theta_\mathrm{Ein}$ is $10\%$.}
\tablenotetext{d}{Lensing estimates of the projected cluster mass $M_\mathrm{2D}(<\theta)$ enclosed within a cylinder of radius $\theta$. The data here are constructed for each cluster by combining two different lens models (where available) of \citet{Zitrin2015clash}. For details, see Section \ref{subsec:sl}.}
\end{deluxetable*}

Our cluster sample stems from the CLASH shear-and-magnification
weak-lensing analysis of \citet{Umetsu2014clash} based primarily on
Subaru multi-color imaging. 
This sample comprises two subsamples, one with 16 X-ray regular clusters 
and another with four high-magnification clusters, both   
taken from the CLASH sample of \citet{Postman+2012CLASH}. 

Table \ref{tab:sample} gives a summary of the properties of 20 clusters
in our sample.  
Following \citet{Umetsu2014clash}, we adopt the brightest cluster galaxy
(BCG) position as the cluster center for our mass profile analysis.   
As discussed by \citet{Umetsu2014clash}, our sample exhibits, on
average, a small positional offset between the BCG and X-ray peak, 
characterized by an rms offset of 
$\sigma_\mathrm{off}\simeq 30$\,kpc\,$h^{-1}$. 
For the X-ray-selected subsample, 
$\sigma_\mathrm{off}\simeq 11$\,kpc\,$h^{-1}$.
This level of offset is negligible compared to the range of
overdensity radii of interest for mass measurements
($\Delta_\mathrm{c}\simlt 2500$).
Hence, smoothing from the miscentering effects will not significantly
affect our cluster mass profile measurements 
\citep{Johnston+2007b,Umetsu+2011stack}.

\subsection{Background Galaxy Selection for Weak Lensing}
\label{subsec:back}

A careful selection of background galaxies is critical for a cluster
weak-lensing analysis, so that unlensed cluster members and
foreground galaxies do not dilute the background lensing signal.
\citet{Umetsu2014clash} used the color-color (CC) selection
method of \citet{Medezinski+2010},
typically using the Subaru/Suprim-Cam $B_\mathrm{J}R_\mathrm{C}z'$
photometry where available \citep[][Tables 1 and 2]{Umetsu2014clash},
which spans the full optical wavelength range. 
The photometric zero points were precisely calibrated to an accuracy of
$\sim 0.01$\,mag, 
using the {\em HST} photometry of cluster elliptical galaxies and
with the help of galaxies with measured spectroscopic redshifts.

For shape measurements,
\citet{Umetsu2014clash} combined two distinct populations that encompass
the red and blue branches of background galaxies in CC-magnitude space,
having typical redshift distributions peaked around $z\sim 1$ and $\sim 2$, 
respectively \citep[see Figures 1, 5, and 6 of][]{Medezinski+2011}. 
For validation purposes, we have compared our blue+red background
samples with spectroscopic samples obtained from the CLASH-VLT large
spectroscopic program with VIMOS \citep{Rosati2014VLT} 
providing thousands of spectroscopic redshifts for cluster members and
intervening galaxies along the line of sight, including lensed background
galaxies \citep[e.g.,][]{Biviano+2013,Balestra2013}. 
Combining CLASH-VLT spectroscopic redshifts with the Subaru
multi-band photometry available for 10 southern CLASH clusters,
we find a mean contamination fraction of $(2.4\pm 0.7)\%$ in our
blue+red background regions in CC-magnitude space, where the error  
accounts for Poisson statistics. 

Our magnification-bias measurements are based on flux-limited samples of
red background galaxies  
\citep[$R_\mathrm{C}-z'\simgt 0.5$,][]{Medezinski+2011}.
Apparent faint magnitude cuts ($m_\mathrm{lim}$) were applied for each
cluster in the reddest CC-selection band (typically the Subaru $z'$
band) to avoid incompleteness near the detection limit.
The threshold $m_\mathrm{lim}$ was chosen at the magnitude where the
source number counts turn over, and it corresponds on average to the 
$6\sigma$ limiting magnitude within $2\arcsec$ diameter aperture.  
Our CC selection does not cause incompleteness at the faint end in the
bluer filters \citep[for a general discussion, see][]{Hildebrandt2015}
because we have deeper photometry in the bluer bands 
\citep[this is by design, so as to detect the red galaxies; see][]{Broadhurst1995}
and our ``CC-red'' galaxies are relatively blue in
$B_\mathrm{J}-R_\mathrm{C}$ \citep[Figure 1 of][]{Medezinski+2011}.


The mean depths $\langle \beta\rangle$ and $\langle\beta^2\rangle$
of the background samples were estimated using
photometric redshifts of individual galaxies determined with the BPZ
code \citep{Benitez2000,Benitez+2004} from our point-spread-function
(PSF) corrected multi-band photometry \citep[typically with 5 Subaru
filters; Table 1 of][]{Umetsu2014clash}.   
An excellent statistical agreement was found between the depth estimates
$\langle\beta\rangle$ from our BPZ measurements in the cluster fields
and those from the COSMOS photometric-redshift catalog
\citep{Ilbert+2009COSMOS}, 
with a median relative offset of 0.27\% and an rms field-to-field
scatter of 5.0\% \citep{Umetsu2014clash}.

\subsection{Reduced Tangential Shear}
\label{subsec:gt}

We use the azimuthally averaged radial profile of the 
reduced tangential shear $g_+=\gamma_+/(1-\kappa)$ as the primary
constraint from wide-field weak-lensing observations.
We adopt the following approximation for the nonlinear corrections to
the source-averaged reduced tangential shear
$\langle g_+\rangle =
 \left[\int_0^\infty\!dz\,\overline{N}_{g}(z) g_+(z) \right] 
 \left[\int_0^\infty\!dz\,\overline{N}_{g}(z) \right]^{-1}$
\citep{Seitz1997}:
\begin{eqnarray}
\label{eq:nlcor}
\langle g_+ \rangle
\approx
\frac{\langle W\rangle_{g} 
\left[
\kappa_\infty(<\theta) -\kappa_\infty(\theta)\right]}
{1- \kappa_\infty(\theta) 
\langle W^2\rangle_{g}/\langle W\rangle_{g}}
=
\frac{\langle\gamma_+\rangle}{1-f_{W,g}\langle \kappa\rangle},
\end{eqnarray}
where $\langle W\rangle_{g}$ is the relative lensing strength 
(Section \ref{sec:basics}) averaged over the population $N_{g}(z)$ of
source galaxies, 
$f_{W,g}\equiv \langle W^2\rangle_{g}/\langle W\rangle_{g}^2$ 
is a dimensionless quantity of the order unity,  
$\langle \kappa\rangle = \langle W\rangle_{g} \kappa_\infty$,
and
$\langle \gamma_+\rangle = \langle W\rangle_{g} \gamma_{+,\infty}$.

In the present study, we use the weak lensing shear data obtained by
\citet{Umetsu2014clash}. 
The shear analysis pipeline of \citet{Umetsu2014clash} was implemented
based on the procedures described in \citet{Umetsu+2010CL0024} and on
verification tests with mock ground-based observations
\citep{2007MNRAS.376...13M,Oguri+2012SGAS}. 
The key feature of the shear calibration method of
\citet{Umetsu+2010CL0024} is that we use galaxies detected with very
high significance, $\nu>30$, to model the PSF isotropic correction
as function of object size and magnitude.
Here $\nu$ is the peak significance given by the IMCAT peak-finding
algorithm {\em hfindpeaks}.
Recently, a very similar procedure was used 
by the Local Cluster Substructure Survey (LoCuSS) collaboration in their
Subaru weak-lensing study of 50 clusters \citep{Okabe+Smith2015}. 
Another important feature is that we select isolated galaxies for the
 shape measurement to minimize the impact of crowding and blending
 \citep{Umetsu2014clash}. To do this, we first identify objects having
 any detectable neighbor within $3r_g$, with $r_g$ the Gaussian scale
 length given by {\em hfindpeaks}. All such close pairs of objects
 are rejected. After this close-pair
 rejection, objects with low detection significance $\nu<10$ are
 excluded from our analysis. All galaxies with usable shape measurements
 are then matched with sources in our 
CC-magnitude-selected
 background galaxy samples (Section \ref{subsec:back}), ensuring that
 each galaxy is detected in both the reddest CC-selection band and the 
 shape-measurement band.  

Using simulated Subaru/Suprime-Cam images
\citep[][]{2007MNRAS.376...13M,Oguri+2012SGAS}, 
\citet{Umetsu+2010CL0024} find that the lensing signal can be recovered
with $|m|\sim 0.05$ of the multiplicative shear calibration bias and
$c\sim 10^{-3}$ of the residual shear offset 
\citep[as defined by][]{2006MNRAS.368.1323H,2007MNRAS.376...13M}.
Accordingly, \citet{Umetsu2014clash} included for each galaxy a shear
calibration factor of $1/(1+m)$ ($g\to g/0.95$) to account for residual
calibration. 
As noted by \citet[][]{Umetsu+2012}, 
the degree of multiplicative bias $m$ depends modestly on the seeing
conditions and the PSF quality 
\citep[][see Section 3.2 for their image simulations using Gaussian and Moffat PSF
profiles with $0.5\arcsec$--$1.1\arcsec$\,FWHM]{Oguri+2012SGAS}.
This variation with the PSF properties   
limits the shear calibration accuracy to $\delta m\sim 0.05$
\citep[][Section 3.3]{Umetsu+2012}.

From shape measurements of background galaxies,
the averaged reduced tangential shear was measured
in a set of concentric annuli ($i=1,2,...,N_\mathrm{WL}$) centered on
each cluster as  
\begin{equation}
 \langle g_{+,i}\rangle =
\left[
\displaystyle\sum_{k\in i}
w_{(k)}\,g_{+(k)}
\right]
\left[
\displaystyle\sum_{k\in i} 
w_{(k)}
\right]^{-1},
\end{equation}
where the index $k$ runs over all objects located within the $i$th
annulus, $g_{+(k)}$ is an estimate of $g_+$ for the $k$th object, and
$w_{(k)}$ is its statistical weight given by  
$w_{(k)}=1/(\sigma_{g(k)}^2+\alpha_g^2)$, 
with $\sigma_{g(k)}$ the uncertainty in the estimate of reduced shear
$g_{(k)}$ and $\alpha_g$ the softening constant taken to be $\alpha_g=0.4$
\citep{Umetsu2014clash}, 
a typical value of the mean dispersion
$(\overline{\sigma_g^2})^{1/2}$ in Subaru observations
\citep[e.g.,][]{Umetsu+2009,Oguri+2009Subaru,Okabe+2010WL}.
Here $\alpha$ includes both intrinsic shape and measurement noise
contributions. 
The statistical uncertainty $\sigma_{+,i}$ in $\langle g_{+,i}\rangle$
was estimated from bootstrap resampling of the background source catalog
for each cluster.

The reduced tangential shear profiles analyzed in this study are
presented in Figure 2 of \citet{Umetsu2014clash}.
For all clusters in our sample, the estimated values for 
$\langle \beta\rangle_g$, and $f_{W,g}$ are summarized in 
Table 3 of \citet{Umetsu2014clash}. 
We marginalize over the calibration uncertainty of 
$\langle \beta\rangle_g$ in our joint likelihood analysis of multiple
lensing probes (Section \ref{subsubsec:likelihood}).

\subsection{Magnification Bias}
\label{subsec:magbias}

A fundamental limitation of measuring shear only is the mass-sheet
degeneracy (Section \ref{sec:basics}; see also Section \ref{subsec:mass}).
We can break this degeneracy by using the complementary combination of
shear and magnification 
\citep{Schneider+2000,UB2008,Rozo+Schmidt2010,Umetsu2013}.  

Deep multi-color photometry enables us to explore the faint end of the
luminosity function of red quiescent galaxies at $z\sim 1$  
\citep[][]{Ilbert2010}.
For such a population,
the effect of magnification bias is dominated by the geometric
area distortion because few fainter galaxies can be magnified up into
the flux-limited sample; this results in a net depletion of source counts \citep{1998ApJ...501..539T,BTU+05,UB2008,Umetsu+2010CL0024,Umetsu+2011,Umetsu+2012,Umetsu2014clash,Umetsu2015A1689,Ford2012,Coe+2012A2261,Medezinski+2013,Radovich2015planck}. 
In the regime of negative magnification bias, a practical advantage is
that the effect is not sensitive to the exact form of the source
luminosity function \citep{Umetsu2014clash}.

If the magnitude shift $\delta m=2.5\log_{10}\mu$ of an object due to
magnification is small compared to that on which the logarithmic slope
of the luminosity function varies,
the number counts can be locally approximated by a power law at
the limiting magnitude $m_\mathrm{lim}$.
The magnification bias at redshift $z$ is
then given by \citep{1995ApJ...438...49B}
\begin{equation} 
\label{eq:magbias}
N_\mu(\btheta,z; <m_\mathrm{lim}) = \overline{N}_\mu(z) \,\mu^{2.5s-1}(\btheta,z)
\equiv\overline{N}_\mu(z) b_\mu(\btheta,z), 
\end{equation}  
where $\overline{N}_\mu(z)=\overline{N}_\mu(z;<m_\mathrm{lim})$ is the
unlensed mean source counts and
$s$ is the logarithmic count slope evaluated at $m_\mathrm{lim}$,
$s=[d\log_{10} \overline{N}_\mu(z; <m)/dm]_{m=m_\mathrm{lim}}$.
In the regime where $2.5s\ll 1$, a net count depletion results.
Accounting for the spread of $\overline{N}_\mu(z)$, we express the
population-averaged magnification bias as 
$\langle b_\mu\rangle =
\left[\int_0^\infty\!dz\,\overline{N}_{\mu}(z)b_\mu(z) \right] 
\left[\int_0^\infty\!dz\,\overline{N}_{\mu}(z) \right]^{-1}$.
Following \citet{Umetsu2014clash}, 
we interpret the observed number counts on a grid of equal-area cells
($n=1,2,...$) as (see Appendix A.2 of \citet{Umetsu2013}) 
\begin{equation}
\label{eq:bmu_approx}
 \langle b_\mu(\btheta_n)\rangle = N_\mu(\btheta_n;
  <m_\mathrm{lim})/\overline{N}_\mu(<m_\mathrm{lim})
\approx \langle \mu^{-1}(\btheta_n)\rangle^{1-2.5s_\mathrm{eff}}
\end{equation}
with $s_\mathrm{eff}=[d\log_{10}\overline{N}_{\mu}(<m)/dm]_{m=m_\mathrm{lim}}$ the effective
count slope defined in analogy to Equation (\ref{eq:magbias}).
Equation (\ref{eq:bmu_approx}) is exact for $s_\mathrm{eff}=0$ and gives a
good approximation for depleted populations with $s_\mathrm{eff}\ll 0.4$.

The covariance matrix 
$\mathrm{Cov}[N(\btheta_m),N(\btheta_n)] \equiv (C_N)_{mn}$
of the source counts includes the clustering and
Poisson contributions \citep{Hu+Kravtsov2003} as
$(C_N)_{mn}=(\overline{N}_\mu)^2\omega_{mn} +\delta_{mn}N_\mu(\btheta_m)$,
with $\omega_{mn}$ the cell-averaged angular correlation function of
source galaxies \citep[see Equation (14) of][]{Umetsu2015A1689}. 
The angular correlation length of background galaxies can be small
\citep{Connolly+1998,2000MNRAS.313..524V}
compared to the typical resolution $\sim 1\arcmin$ of cluster weak
lensing, so that the correlation between different cells can be
generally ignored, whereas the unresolved and nonvanishing correlation
on small angular scales accounts for increase of the variance of
counts. We thus approximate $C_N$ by \citep{Umetsu2015A1689}
\begin{equation}
\label{eq:covN}
\left(C_N\right)_{mn} \approx
\left[
\langle \delta N_\mu^2(\btheta_m)\rangle +  
N_\mu(\btheta_m)
\right]\delta_{mn},
\end{equation}
with $\langle \delta N_\mu^2(\btheta_m)\rangle$ the total variance of
the $m$th counts.
To enhance the signal-to-noise ratio (S/N),
we calculate the surface number density 
$n_\mu(\theta)=dN_\mu(\theta)/d\Omega$
of source galaxies as a function of clustercentric radius, by averaging
the counts in concentric annuli centered on the cluster. 
The source-averaged magnification bias is then expressed as 
$\langle n_\mu(\theta)\rangle=\overline{n}_\mu \langle\mu^{-1}(\theta)\rangle^{1-2.5s_\mathrm{eff}}$
with $\overline{n}_\mu$
the unlensed mean surface number density.

We measure the magnification bias signal in each annulus
($i=1,2,...,N_\mathrm{WL}$) as \citep{Umetsu2015A1689}
\begin{equation}
\label{eq:nb}
 \langle n_{\mu,i}\rangle =
  \frac{1}{(1-f_{\mathrm{mask},i})\Omega_\mathrm{cell}}
  \sum_{m} {\cal P}_{im}N_\mu(\btheta_m)
\end{equation}
with 
$\Omega_\mathrm{cell}$ the solid angle of the cell
and
${\cal P}_{im}=(\sum_m A_{mi})^{-1}A_{mi}$ the projection matrix
normalized in each annulus as $\sum_m {\cal P}_{im}=1$;
$A_{mi}$ is the fraction of the area of
the $m$th cell lying within the $i$th annular bin  ($0\le A_{mi} \le 1$), 
and $f_{\mathrm{mask},i}$ is the mask correction factor for
the $i$th annular bin, 
$(1-f_{\mathrm{mask},i})^{-1} \equiv \left[ \sum_{m} (1-f_m)A_{mi}\right]^{-1} \sum_{m} A_{mi}$,
with $f_m$ the fraction of the mask area
in the $m$th cell, due to bad pixels, saturated objects, foreground and
cluster galaxies. 
The intrinsic clustering plus statistical Poisson contributions to the
uncertainty in $\langle n_{\mu,i}\rangle$ are given as
\begin{equation}
\label{eq:Covn}
(\sigma_{\mu,i}^\mathrm{int})^2 + (\sigma_{\mu,i}^\mathrm{stat})^2
=
\frac{1}{(1-f_{\mathrm{mask},i})^2\Omega_\mathrm{cell}^2}
\sum_{m}
{\cal P}_{im}^2
\left(C_N\right)_{mm},
\end{equation}
where $(\sigma_\mu^\mathrm{int})^2$ and $(\sigma_\mu^\mathrm{stat})^2$
account for the contributions from the first and second terms of
Equation (\ref{eq:covN}), respectively.

In the present work, we use magnification measurements from flux-limited  
samples of red background galaxies obtained by \citet{Umetsu2014clash}.
The analysis procedure used in \citet{Umetsu2014clash} is summarized
as follows: 
The magnification bias analysis was limited to the 
$24\arcmin\times 24\arcmin$ region centered on the cluster.
The clustering error term $\sigma_{\mu,i}^\mathrm{int}$ was
estimated empirically from the variance in each annulus due to
variations of the counts along the azimuthal direction.
For the estimation of $\langle n_{\mu,i}\rangle$,
a positive tail of $>\nu\sigma$ cells with $\nu=2.5$ was
excluded in each annulus by iterative $\sigma$ clipping to reduce the
bias due to intrinsic angular clustering of source galaxies. 
The Poisson error term $\sigma_{\mu,i}^\mathrm{stat}$ was estimated from
the clipped mean counts. 
The systematic change between the mean counts estimated with and without
$\sigma$ clipping was then taken as a systematic error, 
$\sigma_{\mu,i}^\mathrm{sys}=|n_{\mu,i}^{(\nu)}-n_{\mu,i}^{(\infty)}|/\nu$,
where $n_{\nu,i}^{(\nu)}$ and $n_{\mu,i}^{(\infty)}$ represent the
clipped and unclipped mean counts in the $i$th annulus, respectively. 
As noted by \citet{Umetsu2014clash}, the $\sigma_\mu^\mathrm{sys}$
term is sensitive to large-scale variations of source counts and can in
principle account for projection effects from background clusters along
the line of sight and spurious excess counts due perhaps to spatial
variation of the photometric zero point and/or to residual flat-field
errors.   
These errors were combined in quadrature as 
\begin{equation}
 \sigma_{\mu,i}^2 = (\sigma_{\mu,i}^\mathrm{int})^2 +
  (\sigma_{\mu,i}^\mathrm{stat})^2 +(\sigma_{\mu,i}^\mathrm{sys})^2.
\end{equation}
Since we include the $\sigma_\mu^\mathrm{sys}$ term in the error
analysis, our magnification bias measurements are stable and insensitive
to the particular choice of $\nu$. The smaller the $\nu$ value, the
larger the resulting total errors as dominated by the Poisson and
systematic terms.  

The count normalization and slope parameters 
$(\overline{n}_\mu, s_\mathrm{eff})$ were estimated from the source
counts in the outskirts at 
$10\arcmin \le \theta \le \theta_\mathrm{max}$,
with $\theta_\mathrm{max}=16\arcmin$
except $\theta_\mathrm{max}=14\arcmin$
for RX J2248.7$-$4431
observed with ESO/WFI
\citep{Umetsu2014clash}.
The mask-corrected magnification bias profile 
$\langle n_{\mu,i}\rangle/\overline{n}_\mu$ is thus proportional to 
$(1-f_\mathrm{mask,back})/(1-f_{\mathrm{mask},i})\equiv 1+\Delta f_{\mathrm{mask},i}$
with 
$f_{\mathrm{mask,back}}=f_{\mathrm{mask}}(10\arcmin\le
\theta\le\theta_\mathrm{max})$ 
the masked area fraction in the reference background region. 
Masking of observed sky was accounted for using the method outlined in
\citet[][Method B of Appendix A]{Umetsu+2011}. 
This method can be fully automated to achieve similar performance to
conservative approaches \citep[e.g., Method A of][]{Umetsu+2011} in terms of
the masked area fraction once the configuration parameters of
SExtractor \citep{SExtractor} are optimally tuned.
We tuned the SExtractor configuration parameters by setting 
${\tt DETECT\_THRESH=5}$ and 
${\tt DETECT\_MINAREA = 300}$
as found by \citet{Umetsu+2011}
to detect foreground objects, cluster galaxies, and defects (e.g.,
saturated stars and stellar trails) in the coadded images 
($0.2\arcsec$\,pixel$^{-1}$ sampling). We marked the pixels that belong
to these objects in the ${\tt CHECKIMAGE\_TYPE = OBJECT}$ mode. 
Recently, \citet{Chiu2015magbias} adopted this method to calculate the
masked area fraction for their magnification bias measurements, finding
that the SExtractor configuration of \citet{Umetsu+2011} is optimal for
their observations with Megacam on the Magellan Clay telescope.  

We find that the unmasked area fraction 
$1-f_\mathrm{mask}$ is on average $93\%$ of the sky at  $\theta\ge 10\arcmin$,
decreasing toward the cluster center down to $88\%$ at 
$1\arcmin \simlt \theta \simlt 2\arcmin$
($210$\,kpc\,$h^{-1}\simlt R\simlt 420$\,kpc\,$h^{-1}$ at $z=0.35$).
The typical variation of the mask correction factor across the radial
range is thus 
$\approx \mathrm{max}(f_\mathrm{mask})-f_\mathrm{mask,back}\sim 5\%$,
which is much smaller than the typical depletion signal
$\delta n_\mu/\overline{n}_\mu\sim -0.3$
in the innermost radial bin $[0.9\arcmin, 1.2\arcmin]$.
Hence, the uncertainty in the mask correction is of the second order.
Furthermore, the net effect of the mask correction depends on the
difference of the $f_\mathrm{mask}$ values, 
$\Delta f_{\mathrm{mask},i}\approx f_{\mathrm{mask},i}-f_\mathrm{mask,back}$
and is insensitive to the particular choice of the SExtractor
configuration parameters. Accordingly, the systematic uncertainty on the
mask correction is negligible.

The magnification-bias profiles used in this study are presented in Figure
2 of \citet{Umetsu2014clash}.
For all clusters, the estimated values and errors for 
$\langle \beta\rangle_\mu$, $\overline{n}_\mu$, and $s_\mathrm{eff}$ are
summarized in 
Table 4
of \citet{Umetsu2014clash}.
The observed values of $s_\mathrm{eff}$ range from 0.11 to 0.20, with a
mean of $\langle s_\mathrm{eff}\rangle = 0.153$ and 
a typical uncertainty of $33\%$ per cluster field.
We marginalize over the calibration parameters ($\langle
\beta\rangle_\mu, \overline{n}_\mu, s_\mathrm{eff}$) for each cluster in
our joint likelihood analysis of multiple lensing probes (Section
\ref{subsubsec:likelihood}).

\subsection{Strong Lensing}
\label{subsec:sl}

Detailed strong-lens modeling using many sets of multiple images with
known redshifts allows us to determine the critical curves with great
accuracy, which then provides  
accurate
estimates of the
projected mass enclosed within the critical area $A_\mathrm{c}$ of an
effective Einstein radius $\theta_\mathrm{Ein}=\sqrt{A_\mathrm{c}/\pi}$
\citep[][]{Zitrin2015clash}.\footnote{For an axisymmetric lens, the average mass density within
this critical area is equal to $\Sigma_\mathrm{c}$, thus enabling us to directly
estimate the enclosed projected mass by
$M_\mathrm{2D}(<\theta_\mathrm{Ein})=\pi(D_\mathrm{l}\theta_\mathrm{Ein})^2\Sigma_\mathrm{c}$. 
}
The enclosed projected mass profile
\begin{equation}
\label{eq:M2d}
 \begin{aligned}
 M_\mathrm{2D}(<\theta) &= \Sigma_\mathrm{c} D_\mathrm{l}^2
  \int_{|\btheta'|\le
  \theta}\kappa(\btheta')d^2\theta'\\
&=\pi
  (D_\mathrm{l}\theta)^2\Sigma_\mathrm{c,\infty}\kappa_\infty(<\theta).
  \end{aligned}
\end{equation}
at the location $\theta$ around $\theta_\mathrm{Ein}$
is less sensitive to modeling assumptions and approaches 
\citep[see][]{Coe+2010,Umetsu+2012,Oguri+2012SGAS},
serving as a fundamental
observable quantity in the strong-lensing regime \citep[][]{Coe+2010}. 

In this work, we use the recent {\em HST} results from a joint analysis
of CLASH strong- and weak-lensing data presented by
\citet{Zitrin2015clash}, who have obtained detailed mass models for each  
cluster core using two distinct parameterizations (when applicable):
One assumes light-traces-mass for both galaxies and DM,
while the other adopts an analytical elliptical NFW form for the DM-halo
components.  
\citet{Zitrin2015clash} performed a detailed comparison of the two
models to obtain a realistic, empirical assessment of the true
underlying errors,
finding that the projected mass enclosed within the critical curves
($z_\mathrm{s}=2$) agrees typically within $\sim 15\%$. 
\citet{Zitrin2015clash} recommended replacing the statistical errors of
their models with the actual (and much larger) uncertainties that
account for model-dependent systematics.

We combine for each cluster the constraints on $M_\mathrm{2D}(<\theta)$
from two distinct mass models (where available)
of \citet[][]{Zitrin2015clash}.
To do this, we take a conservative approach that accounts for
model-dependent systematic uncertainties.
First, 
at each projected clustercentric distance $\theta$, we improve our
estimation of $M_\mathrm{2D}(<\theta)$ using the average of two
different methods as $M_\mathrm{2D}=(M_\mathrm{Z1} + M_\mathrm{Z2})/2$.
Next, we combine the $1\sigma$ confidence intervals of the two models,
by taking the maximum range spanned by their respective 
confidence limits as an overall total uncertainty, 
$\sigma_M=[\mathrm{max}(M_\mathrm{Z1}+\sigma_\mathrm{Z1},M_\mathrm{Z2}+\sigma_\mathrm{Z2})-\mathrm{min}(M_\mathrm{Z1}-\sigma_\mathrm{Z1},M_\mathrm{Z2}-\sigma_\mathrm{Z2})]/2$.
%
Finally, we rescale the error profile $\sigma_M(\theta)$ such that the 
fractional uncertainty on $M_\mathrm{2D}(<\theta_\mathrm{Ein})$ is $15\%$, as
motivated by the findings of \citet{Zitrin2015clash}. For all clusters,
this has resulted in increased errors.

Now, the question is how to determine an effective sampling radius
$\Delta\theta$
of projected mass constraints $M_\mathrm{2D}(<\theta)$, avoiding
oversampling and reducing correlations between adjacent bins. 
Following \citet{Coe+2010}, we estimate here the effective resolution
$\Delta\theta$ based on the surface density of observed multiple images
as $N_\mathrm{im} (\Delta\theta)^2 = \pi \theta_\mathrm{Ein}^2$
with $N_\mathrm{im}$ the number of multiple images.
For our cluster sample,
we find a median of $\overline{N}_\mathrm{im}=17$ multiple images per cluster
and a median effective Einstein radius of 
$\overline{\theta}_\mathrm{Ein}=22.3\arcsec$ for a fiducial source
redshift of $z_\mathrm{s}=2$ (Table \ref{tab:sample}),
yielding $\Delta\theta\approx 10\arcsec$. This is consistent with the
typical map resolutions in the strong-lensing regime adopted by the {\sc
SaWLens} analysis of \citet[][see their Table 5]{Merten2015clash}.
For the X-ray-selected subsample, we find a median value of 
$\overline{\theta}_\mathrm{Ein}=20.1\arcsec$ at $z_\mathrm{s}=2$.
The maximum integration radius is taken to be $40\arcsec$, which is
equal to approximately twice the
median Einstein radius ($z_\mathrm{s}=2$), the region where multiple
images form \citep[see Section 3 of][]{Zitrin+2012CCL}.

To summarize, for each cluster except RXJ1532.9$+$3021 
for which no secure identification of multiple images has been made
\citep[][]{Zitrin2015clash},
we obtain enclosed projected mass
constraints  
$\{M_{\mathrm{2D},i}\}_{i=1}^{N_\mathrm{SL}}\equiv\{M_\mathrm{2D}(<\theta_i)\}_{i=1}^{N_\mathrm{SL}}$ 
for a set of $N_\mathrm{SL}=4$ fixed integration radii
$\theta_i=10\arcsec, 20\arcsec, 30\arcsec$, and $40\arcsec$ 
from the {\em HST} lensing analysis of
\citet{Zitrin2015clash}.
A summary of the {\em HST} lensing mass estimates is given in Table 
\ref{tab:sample}.

\subsection{Cluster Mass Profile Reconstruction}
\label{subsec:bayesian}

\subsubsection{Lensing Constraints}
\label{subsubsec:gldata}

We consider multiple complementary lensing information available
in the cluster regime,
namely enclosed projected mass estimates from strong lensing, 
source-averaged 
tangential distortion 
and magnification bias 
measurements:
\begin{equation}
\label{eq:observable}
\{M_{\mathrm{2D},i}\}_{i=1}^{N_\mathrm{SL}}, 
\{\langle g_{+,i}\rangle\}_{i=1}^{\Nwl},
\{\langle n_{\mu,i}\rangle\}_{i=1}^{\Nwl}.
\end{equation}
Hence, there are a total of $\Nbin \equiv \Nsl + 2\Nwl$
independent lensing constraints for each cluster.

\citet[][]{Umetsu2014clash} measured the shear and magnification
effects   
in $N_\mathrm{WL}=10$ log-spaced clustercentric radial bins over the
range $[0.9\arcmin,16\arcmin]$ for all clusters, 
except $[0.9\arcmin,14\arcmin]$ for RX J2248.7$-$4431 observed with
ESO/WFI (Section \ref{subsec:magbias}).
We have $\Nsl=4$ projected mass estimates (Table \ref{tab:sample})
from the {\em HST} lensing analysis of \citet{Zitrin2015clash} for all
clusters except RXJ1532.9$+$3021 without strong-lensing constraints.

\subsubsection{Joint Likelihood Function}
\label{subsubsec:likelihood}

In the Bayesian framework of \citet[][see also \citet{Umetsu+2011}]{Umetsu2013},
the lensing signal is described by a vector $\bs$ of parameters 
containing the binned convergence profile
$\{\kappa_{\infty,i}\}_{i=1}^{N}$ 
with $N\equiv \Nsl + \Nwl$,
given by  
$N$ binned $\kappa$ values
and the average convergence enclosed by the innermost aperture radius
$\theta_\mathrm{min}$ for strong-lensing mass estimates,
$\kappa_{\infty,\mathrm{min}}\equiv\kappa_{\infty}(<\theta_\mathrm{min})$,\footnote{
If no strong-lensing constraint is available ($N_\mathrm{SL}=0$),
$\kappa_{\infty,\mathrm{min}}$ represents the average convergence within
the inner radial boundary of weak-lensing observations,
$\theta_\mathrm{min}=0.9\arcmin$ \citep{Umetsu2014clash}.
}
so that 
\begin{equation}
\bs=\{\kappa_{\infty, \mathrm{min}}, \kappa_{\infty,i}\}_{i=1}^{N}
\equiv \Sigma_{\mathrm{c},\infty}^{-1}\bSigma
\end{equation}
specified by $(N+1)$ parameters. 
We have $N=14$ for all clusters except
$N=N_\mathrm{WL}=10$ for RXJ1532.9$+$3021 (Section \ref{subsubsec:gldata}).
The number of degrees of freedom (dof) is $\Nbin-(N+1)=\Nwl-1=9$ for all
clusters.

The joint likelihood function ${\cal L}_\mathrm{GL}(\bs)$ for
multi-probe lensing observations is given as a product of 
their separate likelihoods, 
\begin{equation}
\label{eq:Ltot}
 {\cal L}_\mathrm{GL} = {\cal L}_\mathrm{SL} {\cal L}_\mathrm{WL} = 
 {\cal L}_\mathrm{SL} {\cal L}_g {\cal   L}_\mu,
\end{equation}
with ${\cal L}_\mathrm{SL}$, ${\cal L}_{g}$, and ${\cal L}_{\mu}$
the likelihood functions for 
$\{M_{\mathrm{2D},i}\}_{i=1}^{\Nsl}$,
$\{\langle g_{+,i}\rangle\}_{i=1}^{\Nwl}$,
and
$\{\langle n_{\mu,i}\rangle\}_{i=1}^{\Nwl}$,  
respectively,
defined as 
\begin{equation}
\label{eq:chisq}
 \begin{aligned}
   \ln{\cal L}_\mathrm{SL} &= -\frac{1}{2}\sum_{i=1}^{N_\mathrm{SL}} \frac{[M_{\mathrm{2D},i}-\hat{M}_{\mathrm{2D},i}(\bs)]^2}{\sigma_{M,i}^2},\\
  \ln{\cal L}_{g} &= -\frac{1}{2}\sum_{i=1}^{N_\mathrm{WL}}\frac{[\langle g_{+,i}\rangle-\hat{g}_{+,i}(\bs,\bc)]^2}{\sigma^2_{+,i}},\\
\ln{\cal L}_\mu &= -\frac{1}{2}\sum_{i=1}^{N_\mathrm{WL}}\frac{[\langle
  n_{\mu,i}\rangle-\hat{n}_{\mu,i}(\bs,\bc)]^2}{\sigma^2_{\mu,i}},
  \end{aligned}
\end{equation}
with ($\hat{M}_\mathrm{2D}, \hat{g}_+,\hat{n}_\mu$) the
theoretical predictions for the corresponding observations (Appendix
\ref{appendix:estimators}) and $\bc$ the calibration nuisance parameters
to marginalize over, 
\begin{equation} 
\label{eq:cparam}
\bc=\{\langle W\rangle_{g}, f_{W,g}, \langle W\rangle_\mu, \overline{n}_\mu, s_\mathrm{eff}\}.
\end{equation}

For each parameter of the model $\bs$, we consider a flat uninformative
prior with a lower bound of $\bs=0$.
Additionally, we account for the calibration uncertainty in the
observational parameters $\bc$.

\subsubsection{Estimators and Covariance Matrix}
\label{subsubsec:cmat}

We implement our method using a Markov Chain Monte Carlo
algorithm with Metropolis--Hastings sampling
following the prescription outlined in \citet{Umetsu+2011}. 
The method has been tested \citep{Umetsu2013} with synthetic
weak-lensing catalogs from simulations of analytical 
NFW lenses performed using the public package {\sc glafic}
\citep{Oguri2010glafic}.  
The results suggest that, when the mass-sheet degeneracy is
broken, both maximum-likelihood (ML) and marginal maximum a posteriori
probability (MMAP) solutions provide reliable reconstructions with
unbiased profile measurements. 
Hence, this method is not sensitive to the choice and form of priors.
In the presence of a systematic bias in the background-density constraint
($\overline{n}_\mu$), the global ML estimator is less sensitive
to systematic effects than MMAP, and provides 
more accurate reconstructions \citep{Umetsu2014clash}.

On the basis of our simulations, we thus use the global ML estimator for 
determination of the mass profile.  In our error analysis we take into
account statistical, systematic, cosmic-noise, and intrinsic-variance
contributions to the total covariance matrix 
$C_{ij}\equiv\mathrm{Cov}(s_i,s_j)$ for 
$\kappa_\infty=\Sigma/\Sigma_\mathrm{c,\infty}$ as 
\begin{equation}
\label{eq:Ckappa}
C= C^\mathrm{stat} + C^\mathrm{sys} + C^\mathrm{lss} + C^\mathrm{int},
\end{equation}
where $C^\mathrm{stat}$ is the posterior covariance matrix that is
derived from the data (Equations (\ref{eq:Ltot}) and (\ref{eq:chisq})) 
by calculating the sample covariance matrix
$C_{ij}^\mathrm{stat}=\langle (s_i-\langle s_i\rangle) (s_j-\langle s_j\rangle)\rangle$
using MCMC-sampled posterior distributions;
$C^\mathrm{sys}$ accounts for systematic errors due primarily to the
mass-sheet uncertainty \citep{Umetsu2014clash},
\begin{equation} 
\label{eq:Csys}
(C^\mathrm{sys})_{ij}=(\bs_\mathrm{ML}-\bs_\mathrm{MMAP})_i^2\delta_{ij},
\end{equation}
with $\bs_\mathrm{ML}$ and $\bs_\mathrm{MMAP}$ the ML and MMAP solutions,
respectively;
$C^\mathrm{lss}$ is due to uncorrelated large-scale structure (LSS)
projected along the line of sight \citep{2003MNRAS.339.1155H,Umetsu+2011stack};
$C^\mathrm{int}$ accounts for the intrinsic variations of the
projected cluster mass profile \citep{Gruen2015}. 

The cosmic-noise covariance due to projected uncorrelated LSS is given
as \citep{Umetsu+2011stack}
\begin{equation}
 (C^\mathrm{lss})_{ij}=\int\!\frac{ldl}{2\pi}\,C^{\kappa\kappa}_{(ij)}
\hat{J}_0(l\theta_i)\hat{J}_0(l\theta_j),
\end{equation}
where $C^{\kappa\kappa}_{(ij)}$ is the weak-lensing cross power
spectrum as a function of angular multipole $l$ evaluated for a given
pair of source populations in the $i$th and $j$th radial bins;
$\hat{J}_0(\l\theta_i)$ is the Bessel function of the first
kind and order zero ($J_0$) averaged over the $i$th annulus between
$\theta_{i,1}$ and $\theta_{i,2} (>\theta_{i,1})$, given as \citep{Umetsu+2011stack}
\begin{equation}
 \hat{J}_0(l\theta_i) = \frac{2}{(l\theta_{i,2})^2-(l\theta_{i,1})^2}
\left[ l \theta_{i,2} J_1(l\theta_{i,2}) - l\theta_{i,1} J_1(l\theta_{i,1})
\right].
\end{equation}
We compute the elements of the $C^\mathrm{lss}$ matrix for a given pair of source
populations,
using the nonlinear matter power spectrum of 
\citet{Smith+2003halofit} 
for the {\em Wilkinson Microwave Anisotropy Probe} ({\em WMAP})
seven-year cosmology \citep{Komatsu+2011WMAP7}, and then scale to the
reference far-background source plane.
For the inner radial bins constrained by {\em HST} strong lensing, we
assume a source plane at $z_\mathrm{s}=2$, which is a typical
redshift of strongly lensed background galaxies \citep{Zitrin2015clash}.
For the outer weak-lensing bins, we use for each cluster the effective mean
source redshift ($\overline{z}_\mathrm{eff}$) estimated with 
multi-band photometric redshifts \citep[see Table 3 of][]{Umetsu2014clash}. 
The cross covariance terms between the strong- and weak-lensing bins are
computed using the lensing window functions of the respective populations
\citep{Takada+White2004}.
For a given depth of observations, the impact of cosmic
noise is most important where the cluster signal itself is small
\citep{2003MNRAS.339.1155H}.

The intrinsic covariance matrix $C^\mathrm{int}$ accounts for the 
intrinsic variations of the projected cluster density profile 
(i.e., cluster signal itself) due to the $c$--$M$ variance, 
halo asphericity, and the presence of correlated halos 
\citep{Gruen2015}.\footnote{\citet{Gruen2015} formally include the
contribution from the projected uncorrelated LSS in $C^\mathrm{int}$. In
this study, we separately account for this external contribution as
$C^\mathrm{lss}$, to be consistent with the procedure in \citet{Umetsu2014clash}.}
In the present work, 
we consider the diagonal form of the $C^\mathrm{int}$ matrix,
\begin{equation}
\label{eq:int}
 (C^\mathrm{int})_{ij} =
\alpha_\mathrm{int}^2
 \times
(\bs_{\mathrm{ML}})_i^2 
\delta_{ij},
\end{equation}
where the coefficient 
$\alpha_\mathrm{int}\approx\sqrt{C^\mathrm{int}_{ii}}/\kappa_{\infty,i}$
represents the fractional intrinsic scatter in $\kappa$.
On the basis of semi-analytical calculations calibrated by cosmological
numerical simulations \citep{Gruen2015}, we find that, for CLASH
clusters with a characteristic mass of 
$M_\mathrm{200c}\approx 10^{15}M_\odot\, h^{-1}$ \citep{Umetsu2014clash},
the diagonal part of the $C^\mathrm{int}$ matrix can be well
approximated by Equation (\ref{eq:int}) 
with $\alpha_\mathrm{int} \approx 0.2$ in the one-halo regime at
$R\simlt r_\mathrm{200m}$. 
In general,
the diagonal approximation to $C^\mathrm{int}$ can lead to an
underestimate of cluster parameters \citep[see][]{Gruen2015}, where the
degree of underestimation depends on the radial binning scheme for
cluster lensing measurements. 

We have tested the validity of this approximation in our radial binning
scheme by comparing against the semi-analytical model of
\citet{Gruen2015}. 
For a cluster halo of 
$M_\mathrm{200c}=10^{15}M_\odot\, h^{-1}$ 
at $z_\mathrm{l}=0.35$
($M_\mathrm{200m}\simeq 13\times 10^{14}M_\odot\, h^{-1}$) matching
approximately the average characteristics of the CLASH sample \citep{Umetsu2014clash},
we find that the total S/N
estimated using  
the diagonal approximation with $\alpha_\mathrm{int}=0.2$
is accurate to about $10\%$ in the regime of our cluster lensing
observations ($\mathrm{S/N}\sim 10$ per cluster), where the S/N is
defined as \citep{UB2008}\footnote{We note that this classical S/N
definition beaks down in the noise-dominated regime. In our analysis, the
binning scheme was chosen such that the per-pixel detection S/N is 
$\simgt 1$ for each cluster \citep{Umetsu2014clash}, and thus the noise 
contribution to this estimator is negligibly small.}
\begin{equation}
\label{eq:SNR}
(\mathrm{S/N})^2 = \sum_{i,j}s_i C^{-1}_{ij}s_j = \bs^t C^{-1} \bs,
\end{equation}
with $C$ the total
covariance matrix defined in Equation (\ref{eq:Ckappa}).
In the present study, we thus adopt $\alpha_\mathrm{int}=0.2$ in
Equation (\ref{eq:int}) to account for the effects of the intrinsic
profile variations in projection space.\footnote{Strictly speaking, when
simultaneously determining the mass and concentration for a given
individual cluster, the contribution from the intrinsic $c$--$M$
variance should be excluded from $C^\mathrm{int}$. 
To simplify the procedure, however,
we shall fix $\alpha_\mathrm{int}=0.2$ throughout this study.
We note that the effect of the $c$--$M$ variance becomes important only
at small cluster radii, $\theta\simlt 2\arcmin$ \citep{Gruen2015}.
}

\section{CLASH Individual Cluster Analysis}
\label{sec:profile} 

In this section we carry out a comprehensive analysis of strong-lensing,
weak-lensing shear  and magnification data sets for all 20 CLASH
clusters in our sample.

\subsection{CLASH Mass Profile Reconstruction}
\label{subsec:rec}


\begin{figure}[!htb] 
 \begin{center} 
  \includegraphics[width=0.45\textwidth,angle=0,clip]{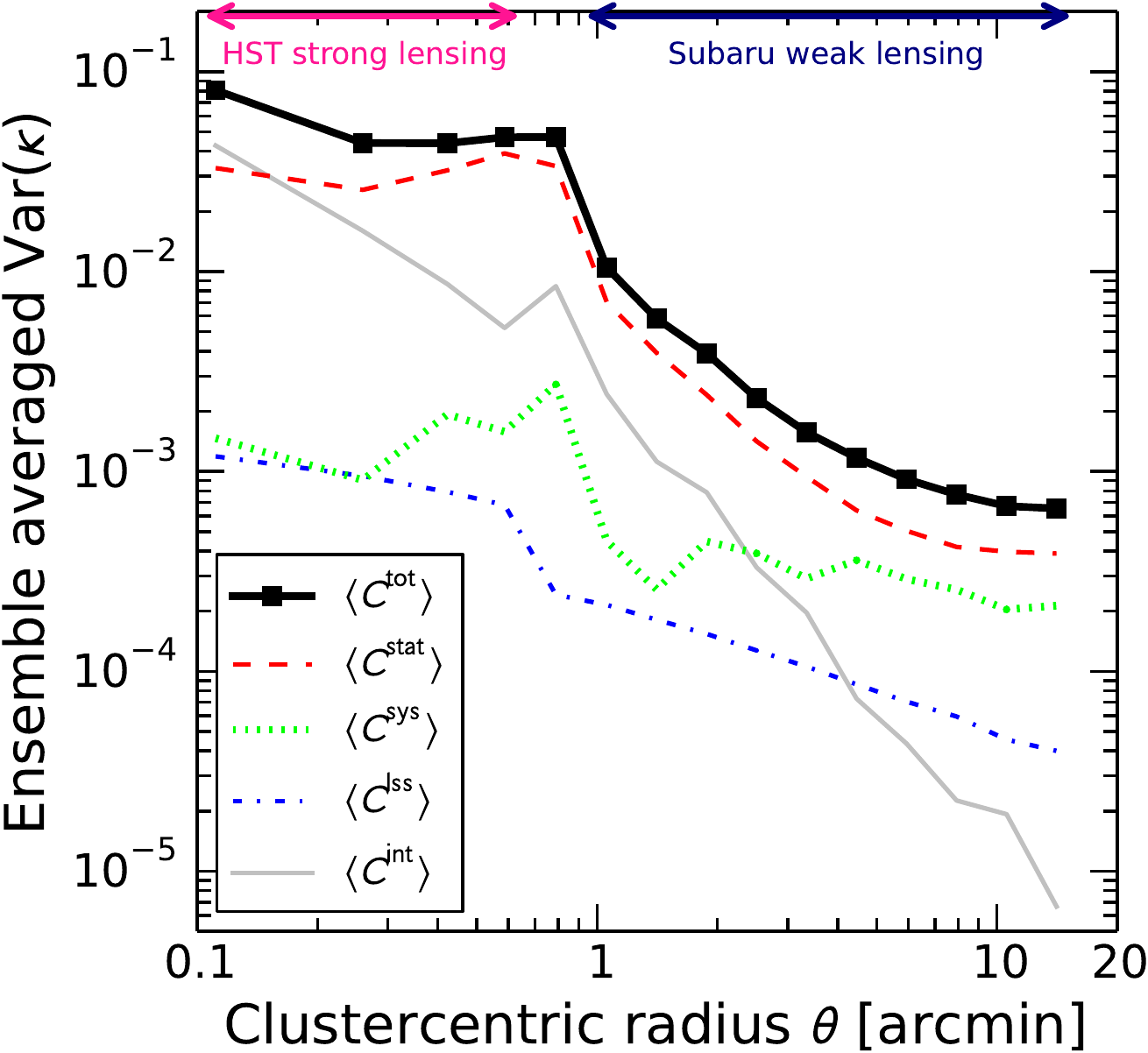}  
 \end{center}
 \caption{\label{fig:aveC}
Contributions to the total covariance matrix
 $C=C^\mathrm{stat}+C^\mathrm{sys}+C^\mathrm{lss}+C^\mathrm{int}$ (black
 line with squares) for the  binned cluster 
 convergence profile $\kappa(\theta)$ (see Figure \ref{fig:kappa})
 reconstructed from a joint likelihood analysis of strong-lensing,
 weak-lensing shear and magnification data sets.  
The results are obtained by averaging over 18 clusters observed with
 both {\em HST} and Subaru (i.e., all except RXJ1532.9$+$3021 without
 strong-lensing constraints and RXJ2248.7$-$4431 based on ESO/WFI
 data). 
The diagonal variance
 terms $\mathrm{Var}(\kappa)$, scaled to the mean depth of Subaru
 weak-lensing observations, are shown as a function of clustercentric
 radius $\theta$.  
}
\end{figure}


\begin{figure}[!htb] 
 \begin{center} 
  \includegraphics[width=0.48\textwidth,angle=0,clip]{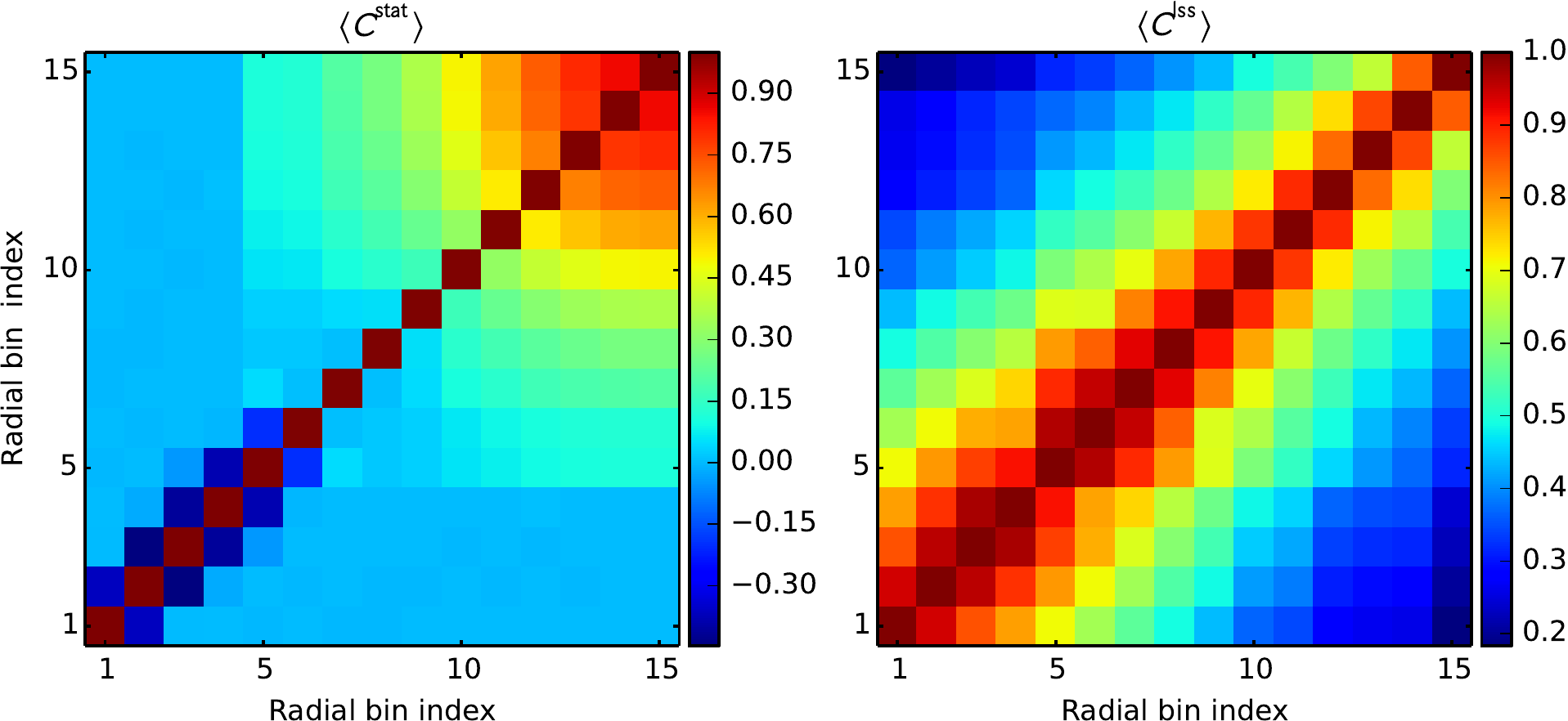}  
 \end{center}
 \caption{\label{fig:clss}
Cross-correlation coefficients of the ensemble-averaged
 $C^\mathrm{stat}$ (left panel) and $C^\mathrm{lss}$ (right panel)
 matrices, whose diagonal elements are shown in Figure \ref{fig:aveC}.  
}
\end{figure}

Following the methodology outlined in Section \ref{sec:method},
we analyze our weak- and strong-lensing data 
presented in \citet{Umetsu2014clash} and 
\citet{Zitrin2015clash}
to examine the underlying radial mass distribution
for a sample of 20 CLASH clusters (Table \ref{tab:sample}).
We have derived for each cluster a mass-profile solution
$\bSigma\equiv\Sigma_\mathrm{c,\infty}\bs=\{\Sigma_\mathrm{min},\Sigma_i\}_{i=1}^N$
from a joint likelihood analysis of our strong-lensing, weak-lensing
shear and magnification data.
We find that the minimum $\chi^2(=-2\ln{\cal L_\mathrm{GL}})$
values for the best-fit $\Sigma$ solutions range from $\chi^2=2.5$
(Abell 611)  to $14.8$ (MACSJ0429.6$-$0253)
for $9$ dof (a mean reduced $\chi^2$ of 0.95), indicating good
consistency between independent observations having different
systematics. 

In Appendix \ref{appendix:kappa} (Figure \ref{fig:kappa}) we show the
resulting mass-profile solutions, $\bSigma$ (black squares), obtained
for our sample along with those by 
\citet[][blue circles]{Umetsu2014clash} and
\citet[][red dots]{Merten2015clash}.
\citet{Umetsu2014clash} derived weak-lensing-only solutions for this
sample from a joint likelihood analysis of the shear and magnification
measurements.    
When the inner strong-lensing information (Table \ref{tab:sample}) is
combined with wide-field weak-lensing data, the central weak-lensing bin
$\Sigma(<0.9\arcmin)$ (Section \ref{subsubsec:gldata}) is resolved into
($N_\mathrm{SL}+1$) radial bins, hence improving the determination of
the inner mass profile.  
\citet{Merten2015clash} performed a two-dimensional {\sc SaWLens}
analysis of 19 CLASH X-ray-selected clusters,  
by combining the CLASH {\em HST} data with the shear catalogs of
\citet{Umetsu2014clash}.  
To break the mass-sheet degeneracy, \citet{Merten2015clash} assumed
outer boundary conditions such that the average convergence vanishes at
the edge of the reconstruction. 
This is a reasonable approximation in wide-field weak-lensing
observations entailing the full cluster field well beyond its virial
radius, $r_\mathrm{vir}$.  

We find overall good agreement between different reconstructions, except
for a few systems in the overlap sample, such as 
MACSJ1931.8$-$2635,
RXJ1347.5$-$1145, and
MACSJ0744.9$+$3927.
For these clusters,
the {\sc SaWLens} reconstructions are systematically lower
than those of this work and \citet{Umetsu2014clash} which include the
weak-lensing magnification data. 
Here 
MACSJ1931.8$-$2635 ($b=-20.09^\circ$) and 
MACSJ0744.9$+$3927 ($b=+26.65^\circ$) are the two lowest Galactic latitude
clusters of the overlap sample, implying higher stellar densities,
correspondingly large areas masked by bright saturated stars, 
and hence lower number densities of background galaxies usable for weak lensing
\citep[][see their Figure 2, Tables 3 and 4]{Umetsu2014clash}.
In fact, MACSJ1931.8$-$2635 has the lowest S/N of the
weak-lensing observations presented in \citet[][Table 5]{Umetsu2014clash}.
On the other hand, RXJ1347.5$-$1145  ($z=0.451$) and MACSJ0744.9$+$3927
($z=0.686$) represent the two highest-$z$ clusters of the overlap
sample, both of which exhibit complex mass distributions 
with a high degree of substructure 
\citep[see][]{Postman+2012CLASH,Merten2015clash}.
For clusters at lower Galactic latitudes and higher redshifts, the color
selection of background galaxies is correspondingly more difficult.  
Therefore, this discrepancy can be attributed in part to systematic
uncertainties in the present calibration of magnification measurements
for these low Galactic latitude clusters and high redshift clusters. 
Weak-lensing mass reconstructions are sensitive to the treatment of
boundary conditions if there are massive structures near the data
boundaries. Hence, mass profile reconstructions for clusters with high
degrees of substructure can be subject to a greater degree of mass-sheet
degeneracy.   
A more quantitative comparison with the {\sc SaWLens} results from
\citet{Merten2015clash} can be found in Sections \ref{subsec:mcalib} and
\ref{subsec:cM_discussion}.

Figure \ref{fig:aveC} shows the average contributions to the total 
covariance matrix $C$ (Equation (\ref{eq:Ckappa}))
for the convergence profile $\kappa(\theta)$ reconstructed from the
joint likelihood analysis of strong-lensing, weak-lensing shear and 
magnification data. 
The results are obtained by averaging over 18 clusters observed with
 both {\em HST} and Subaru (i.e., all except RXJ1532.9$+$3021 without
 strong-lensing constraints and RXJ2248.7$-$4431 based on ESO/WFI
 data). 
The diagonal variance terms $\mathrm{Var}(\kappa)$, scaled to the mean
 depth of Subaru weak-lensing observations, are shown as a function of
 clustercentric radius $\theta$.  
At all bins except the innermost bin, 
the reconstruction uncertainty is dominated by the
 observational statistical errors ($C^\mathrm{stat}$, red dashed).
The relative contribution from intrinsic variance ($C^\mathrm{int}$,
 gray solid) increases toward the cluster
 center and becomes important at small cluster radii, $\theta\simlt 2\arcmin$.
The $C^\mathrm{sys}$ term (green dotted) represents the level of residual
 variance (Equation (\ref{eq:Csys})) due primarily to the mass-sheet
 degeneracy.  In the regime of Subaru weak-lensing 
($\theta\ge 0.9\arcmin$), $C^\mathrm{sys}$ stays approximately constant
at $\sim (2-4)\times 10^{-4}$ with $\theta$, 
corresponding to a characteristic mass-sheet uncertainty of
 $\sigma_\kappa\sim (1-2)\times 10^{-2}$ per cluster.
A noticeable increase of the cosmic-noise contribution $C^\mathrm{lss}$
 (blue dotted--dashed) from projected uncorrelated LSS is seen at  
$\theta\le 40\arcsec$, within which the reconstruction is dominated by
 {\em HST} strong-lensing measurements with greater depth.   

In Figure \ref{fig:clss}, we show the cross-correlation coefficients of
the ensemble-averaged $C^\mathrm{stat}$ and $C^\mathrm{lss}$ matrices,
whose diagonal elements are shown in Figure \ref{fig:aveC}. 
We find that adjacent {\em HST} bins of the 
$\langle C^\mathrm{stat}\rangle$ matrix are anti-correlated at
around the 40\% level, where the first 4 bins in each panel of Figure
\ref{fig:clss} correspond to the {\em HST} data. This negative
covariance arises because they are to satisfy the observed cumulative 
mass constraints (Equation (\ref{eq:M2d})). We find small correlations
of $<1\%$ between the {\em HST} ($10\arcsec\le \theta\le 40\arcsec$) 
and Subaru ($0.9\arcmin\le \theta\le 16\arcmin$) radial bins. 
For the $\langle C^\mathrm{lss}\rangle$ matrix, 
since strongly lensed source galaxies at $z_\mathrm{s}\sim 2$ and
weakly lensed source galaxies at $z_\mathrm{s}\sim 1$ behind a given
cluster share the same mass overdensities at $z\simlt 1$ where the
geometrical lensing efficiency for the $z_\mathrm{s}\sim 2$ sources is
large, there are large positive correlations at the $\sim 70\%-20\%$
levels between the {\em HST} and Subaru bins, where the degree of
correlation increases with decreasing projected separation.

\subsection{Weighing CLASH Clusters}
\label{subsec:mass}

Tangential shear fitting with a spherical NFW profile is a standard
approach for measuring individual cluster masses from weak lensing 
\citep[][]{Umetsu+2009,Okabe+2010WL,WtG3}. 
Numerical simulations suggest that these mass estimates 
tend to be biased low \citep[by $\sim 5\%$--$10$\%; see][]{Meneghetti+2010a,Becker+Kravtsov2011,Rasia+2012} 
because local substructures that are abundant in cluster outskirts
dilute the shear tangential to the cluster center. 
%
This bias can be reduced if the fitting
range is restricted to within $\sim 2 r_\mathrm{500c} \sim r_\mathrm{vir}$
\citep{Becker+Kravtsov2011}. 
In the CLASH survey, the availability of multi-probe, multi-scale
lensing data allows us to combine weak shear lensing with magnification
and/or strong-lensing constraints 
\citep{Umetsu2014clash,Merten2015clash}. 
In particular, the complementary combination of shear and magnification
data enables to effectively break the
mass-sheet degeneracy and reconstruct the total mass distribution
$\Sigma$ \citep{Umetsu+2011}.

\begin{deluxetable}{lrrr}
\centering
\tabletypesize{\scriptsize}
\tablecolumns{4}
\tablecaption{
\label{tab:cm}
NFW halo parameters for individual CLASH clusters
}
\tablewidth{0pt}
\tablehead{
 \multicolumn{1}{c}{Cluster} &
 \multicolumn{1}{c}{$M_\mathrm{200c}$} &
 \multicolumn{1}{c}{$c_\mathrm{200c}$} &
 \multicolumn{1}{c}{$r_{-2}$} 
 \\ 
\colhead{} &
\multicolumn{1}{c}{($10^{14}M_{\odot}\,h_{70}^{-1}$)} &
\colhead{} &
\multicolumn{1}{c}{(Mpc\,$h_{70}^{-1}$)} 
}
\startdata
X-ray Selected:\\
         ~~Abell 383 & $7.98 \pm 2.66$ & $ 5.9 \pm  1.8$ & $0.31 \pm 0.13$\\
         ~~Abell 209 & $15.40 \pm 3.42$ & $ 2.7 \pm  0.6$ & $0.84 \pm 0.22$\\
        ~~Abell 2261 & $23.10 \pm 5.22$ & $ 3.7 \pm  0.9$ & $0.69 \pm 0.20$\\
    ~~RXJ2129.7+0005 & $6.14 \pm 1.79$ & $ 5.6 \pm  1.6$ & $0.30 \pm 0.11$\\
         ~~Abell 611 & $15.76 \pm 4.49$ & $ 3.9 \pm  1.2$ & $0.57 \pm 0.21$\\
       ~~MS2137-2353 & $13.56 \pm 5.27$ & $ 2.7 \pm  1.3$ & $0.80 \pm 0.45$\\
    ~~RXJ2248.7-4431 & $18.78 \pm 6.72$ & $ 3.6 \pm  1.4$ & $0.66 \pm 0.32$\\
  ~~MACSJ1115.9+0129 & $16.66 \pm 3.85$ & $ 3.0 \pm  0.8$ & $0.75 \pm 0.23$\\
  ~~MACSJ1931.8-2635 & $15.28 \pm 7.13$ & $ 4.4 \pm  1.9$ & $0.51 \pm 0.30$\\
    ~~RXJ1532.9+3021 & $5.98 \pm 2.32$ & $ 5.2 \pm  2.8$ & $0.29 \pm 0.18$\\
  ~~MACSJ1720.3+3536 & $14.50 \pm 4.30$ & $ 4.1 \pm  1.3$ & $0.51 \pm 0.20$\\
  ~~MACSJ0429.6-0253 & $9.76 \pm 3.50$ & $ 4.6 \pm  1.7$ & $0.40 \pm 0.19$\\
  ~~MACSJ1206.2-0847 & $18.17 \pm 4.23$ & $ 3.7 \pm  1.1$ & $0.60 \pm 0.21$\\
  ~~MACSJ0329.7-0211 & $8.65 \pm 1.97$ & $ 6.7 \pm  1.6$ & $0.26 \pm 0.08$\\
    ~~RXJ1347.5-1145 & $34.25 \pm 8.78$ & $ 3.2 \pm  0.9$ & $0.85 \pm 0.29$\\
  ~~MACSJ0744.9+3927 & $18.03 \pm 4.96$ & $ 3.5 \pm  1.2$ & $0.58 \pm 0.23$\\
\hline High Magnification:\\
  ~~MACSJ0416.1-2403 & $10.74 \pm 2.60$ & $ 2.9 \pm  0.7$ & $0.65 \pm 0.18$\\
  ~~MACSJ1149.5+2223 & $25.02 \pm 5.53$ & $ 2.1 \pm  0.6$ & $1.12 \pm 0.35$\\
  ~~MACSJ0717.5+3745 & $26.77 \pm 5.36$ & $ 1.8 \pm  0.4$ & $1.31 \pm 0.31$\\
  ~~MACSJ0647.7+7015 & $13.90 \pm 4.20$ & $ 4.1 \pm  1.5$ & $0.48 \pm 0.21$
\enddata
\tablecomments{Cluster parameters derived from single spherical NFW fits to individual surface mass density profiles (Figure \ref{fig:kappa}) reconstructed from combined strong-lensing, weak-lensing shear and magnification measurements. We adopt a concordance cosmology of $h=0.7$, $\Omega_\mathrm{m}=0.27$, and $\Omega_\Lambda=0.73$. The fitting radial range is restricted to $R\le 2$\,Mpc\,$h^{-1}\simeq 2.9$\,Mpc\,$h_{70}^{-1}$.}
\end{deluxetable}

\begin{deluxetable*}{lrrrrrrr}
\centering
\tabletypesize{\scriptsize}
\tablecolumns{8}
\tablecaption{
\label{tab:mass}
Mass estimates for individual CLASH clusters
}
\tablewidth{0pt}
\tablehead{
 \multicolumn{1}{c}{Cluster} &
 \multicolumn{1}{c}{$M_\mathrm{2500c}$} &
 \multicolumn{1}{c}{$M_\mathrm{1000c}$} &
 \multicolumn{1}{c}{$M_\mathrm{500c}$} &
 \multicolumn{1}{c}{$M_\mathrm{vir}$\tablenotemark{a}} &
 \multicolumn{1}{c}{$M_\mathrm{100c}$} &
 \multicolumn{1}{c}{$M_\mathrm{200m}$} &
 \multicolumn{1}{c}{$M(<1.5\mathrm{Mpc})$}
 \\ 
\colhead{} &
\multicolumn{1}{c}{($10^{14}M_{\odot}$)} &
\multicolumn{1}{c}{($10^{14}M_{\odot}$)} &
\multicolumn{1}{c}{($10^{14}M_{\odot}$)} &
\multicolumn{1}{c}{($10^{14}M_{\odot}$)} &
\multicolumn{1}{c}{($10^{14}M_{\odot}$)} &
\multicolumn{1}{c}{($10^{14}M_{\odot}$)} &
\multicolumn{1}{c}{($10^{14}M_{\odot}$)}  
}
\startdata
X-ray Selected:\\
         ~~Abell 383 & $ 2.78 \pm  0.63$ &  $ 4.43 \pm  1.16$ &$ 5.88 \pm  1.73$ & $ 9.41 \pm  3.33$ & $ 9.66 \pm  3.45$ & $10.34 \pm  3.78$ & $ 6.95 \pm  1.63$\\
         ~~Abell 209 & $ 2.95 \pm  0.68$ &  $ 6.18 \pm  1.25$ &$ 9.64 \pm  1.97$ & $19.60 \pm  4.61$ & $20.49 \pm  4.88$ & $22.36 \pm  5.44$ & $10.28 \pm  1.37$\\
        ~~Abell 2261 & $ 5.91 \pm  1.03$ &  $10.85 \pm  1.89$ &$15.65 \pm  3.05$ & $28.21 \pm  6.87$ & $29.39 \pm  7.26$ & $31.41 \pm  7.94$ & $14.22 \pm  1.79$\\
    ~~RXJ2129.7+0005 & $ 2.08 \pm  0.44$ &  $ 3.35 \pm  0.78$ &$ 4.48 \pm  1.16$ & $ 7.21 \pm  2.23$ & $ 7.48 \pm  2.34$ & $ 7.87 \pm  2.51$ & $ 5.75 \pm  1.23$\\
         ~~Abell 611 & $ 4.13 \pm  0.92$ &  $ 7.48 \pm  1.67$ &$10.73 \pm  2.65$ & $18.95 \pm  5.77$ & $19.99 \pm  6.21$ & $20.78 \pm  6.54$ & $11.24 \pm  1.95$\\
       ~~MS2137-2353 & $ 2.47 \pm  0.72$ &  $ 5.21 \pm  1.41$ &$ 8.28 \pm  2.57$ & $16.98 \pm  7.31$ & $18.25 \pm  8.12$ & $18.91 \pm  8.54$ & $ 9.67 \pm  2.12$\\
    ~~RXJ2248.7-4431 & $ 4.45 \pm  1.05$ &  $ 8.42 \pm  2.08$ &$12.45 \pm  3.62$ & $22.54 \pm  8.78$ & $24.12 \pm  9.68$ & $24.53 \pm  9.91$ & $12.67 \pm  2.50$\\
  ~~MACSJ1115.9+0129 & $ 3.50 \pm  0.82$ &  $ 7.02 \pm  1.43$ &$10.67 \pm  2.22$ & $20.30 \pm  4.97$ & $21.88 \pm  5.48$ & $22.24 \pm  5.60$ & $11.55 \pm  1.61$\\
  ~~MACSJ1931.8-2635 & $ 4.10 \pm  1.12$ &  $ 7.36 \pm  2.36$ &$10.51 \pm  4.05$ & $18.02 \pm  9.05$ & $19.18 \pm  9.88$ & $19.44 \pm 10.07$ & $11.23 \pm  3.07$\\
    ~~RXJ1532.9+3021 & $ 1.83 \pm  1.01$ &  $ 3.03 \pm  1.38$ &$ 4.17 \pm  1.71$ & $ 7.04 \pm  2.79$ & $ 7.51 \pm  3.02$ & $ 7.58 \pm  3.06$ & $ 5.81 \pm  1.63$\\
  ~~MACSJ1720.3+3536 & $ 3.92 \pm  0.86$ &  $ 7.00 \pm  1.59$ &$ 9.96 \pm  2.53$ & $17.04 \pm  5.39$ & $18.29 \pm  5.94$ & $18.30 \pm  5.94$ & $11.02 \pm  2.05$\\
  ~~MACSJ0429.6-0253 & $ 2.85 \pm  0.70$ &  $ 4.92 \pm  1.32$ &$ 6.85 \pm  2.10$ & $11.35 \pm  4.33$ & $12.15 \pm  4.76$ & $12.13 \pm  4.74$ & $ 8.40 \pm  2.03$\\
  ~~MACSJ1206.2-0847 & $ 4.62 \pm  1.01$ &  $ 8.47 \pm  1.63$ &$12.24 \pm  2.49$ & $21.35 \pm  5.29$ & $23.18 \pm  5.94$ & $22.82 \pm  5.81$ & $12.98 \pm  1.84$\\
  ~~MACSJ0329.7-0211 & $ 3.29 \pm  0.64$ &  $ 5.02 \pm  1.00$ &$ 6.51 \pm  1.37$ & $ 9.72 \pm  2.30$ & $10.35 \pm  2.50$ & $10.21 \pm  2.45$ & $ 7.94 \pm  1.36$\\
    ~~RXJ1347.5-1145 & $ 7.65 \pm  1.63$ &  $14.91 \pm  2.98$ &$22.33 \pm  4.89$ & $40.66 \pm 11.13$ & $44.50 \pm 12.60$ & $43.59 \pm 12.25$ & $19.33 \pm  2.61$\\
  ~~MACSJ0744.9+3927 & $ 4.32 \pm  1.02$ &  $ 8.12 \pm  1.76$ &$11.94 \pm  2.81$ & $20.57 \pm  5.98$ & $23.21 \pm  7.08$ & $21.32 \pm  6.29$ & $13.96 \pm  2.39$\\
\hline High Magnification:\\
  ~~MACSJ0416.1-2403 & $ 2.21 \pm  0.53$ &  $ 4.48 \pm  0.97$ &$ 6.85 \pm  1.52$ & $12.99 \pm  3.29$ & $14.13 \pm  3.66$ & $14.12 \pm  3.66$ & $ 8.74 \pm  1.37$\\
  ~~MACSJ1149.5+2223 & $ 3.73 \pm  1.11$ &  $ 8.74 \pm  1.98$ &$14.57 \pm  3.06$ & $30.45 \pm  7.06$ & $34.72 \pm  8.36$ & $32.62 \pm  7.71$ & $15.31 \pm  1.95$\\
  ~~MACSJ0717.5+3745 & $ 3.42 \pm  0.87$ &  $ 8.61 \pm  1.74$ &$14.98 \pm  2.85$ & $32.97 \pm  6.88$ & $37.95 \pm  8.17$ & $35.46 \pm  7.52$ & $15.63 \pm  1.75$\\
  ~~MACSJ0647.7+7015 & $ 3.72 \pm  0.98$ &  $ 6.65 \pm  1.65$ &$ 9.49 \pm  2.51$ & $15.91 \pm  5.07$ & $17.59 \pm  5.83$ & $16.64 \pm  5.40$ & $11.37 \pm  2.26$
\enddata
\tablecomments{Cluster mass estimates $M_\mathrm{3D}(<r)$ from single spherical NFW fits to individual surface mass density profiles (Figure \ref{fig:kappa}). All quantities in the table are given in physical units assuming a concordance cosmology of $h=0.7$, $\Omega_{\rm m}=0.27$, and $\Omega_{\Lambda}=0.73$. See Table \ref{tab:cm} for $M_\mathrm{200c}$. The fitting radial range is restricted to $R\le 2$\,Mpc\,$h^{-1}\simeq 2.9$\,Mpc\,$h_{70}^{-1}$.}
\tablenotetext{a}{Virial overdensity $\Delta_{\rm vir}$ based on the spherical collapse model (see Appendix A of \citet{Kitayama+Suto1996}). For our redshift range $0.187\le z\le 0.686$, $\Delta_{\rm vir}$ ranges approximately from $\simeq 110$ to 140 with respect to the critical density of the universe at the cluster redshift.}\end{deluxetable*}

Here we revisit the mass estimates for CLASH clusters using our improved
mass profile data set (Figure \ref{fig:kappa})   
and taking into account the intrinsic contribution ($C^\mathrm{int}$) to
the error covariance matrix (Section \ref{subsubsec:cmat}).  
We follow the Bayesian approach of \citet{Umetsu2014clash} to make
inference on the NFW halo parameters.  
We employ the radial dependence of the projected NFW profiles given
by \citet{2000ApJ...534...34W}, which provides a good
description of the projected mass distribution in the one-halo regime,
at least in an ensemble-average sense
\citep[][]{Oguri+Hamana2011,Umetsu+2011stack,Okabe+2013}.
We restrict the fitting range to $R\le 2\,$Mpc\,$h^{-1}$ 
\citep{Umetsu2014clash,Merten2015clash},
which is close to the virial radius for most CLASH clusters.
We specify the two-parameter NFW model using the halo mass
$M_\mathrm{200c}$ 
and the halo concentration
$c_\mathrm{200c}=r_\mathrm{200c}/r_{-2}$ with $r_{-2}$ the
characteristic radius at which the logarithmic density slope is -2 
(Section \ref{subsubsec:rho}).
We adopt uninformative log-uniform priors in the respective intervals, 
$0.1 \le M_\mathrm{200c}/(10^{15}M_\odot\,h^{-1}) \le 10$ 
and 
$0.1 \le c_\mathrm{200c}\le 10$ \citep{Umetsu2014clash}.
We check that the mass and concentration estimates for the CLASH 
sample are not sensitive to the choice of the 
priors as found by \citet[][see their Section 2.1]{Sereno2015cM}.
The $\chi^2$ function for our observations is
\begin{equation}
\chi^2(\bp)
=\sum_{i,j} 
\big[
  s_i-\hat{s}_i(\bp)
\big]
 C^{-1}_{ij}
\big[
  s_j-\hat{s}_j(\bp)
\big],
\end{equation}
where 
$\bp=(M_\mathrm{200c},c_\mathrm{200c})$, and
$\hat{s}_i(\bp)=\hat{\Sigma}_i(\bp)/\Sigma_{\mathrm{c},\infty}$ is
the predicted surface mass density averaged over the $i$th annulus, 
accounting for the effect of bin averaging \citep{Umetsu2014clash}.

In Table \ref{tab:cm}, we give marginalized constraints on the NFW halo
parameters $(M_\mathrm{200c}, c_\mathrm{200c})$ and the characteristic
radius $r_{-2}$.
Throughout this paper,
we employ the biweight estimators of
\citet{1990AJ....100...32B}
for the central location ($C_\mathrm{BI}$) and scale ($S_\mathrm{BI}$)
of the marginalized posterior distributions
\citep[][]{Sereno+Umetsu2011,Umetsu2014clash,Umetsu2015A1689}. 
From the posterior samples, we also derive marginalized constraints on the
total enclosed mass $M_\Delta=M_\mathrm{3D}(<r_\Delta)$ at several
characteristic interior overdensities $\Delta$ (see Section
\ref{sec:intro}). 
Table \ref{tab:mass} summarizes the results of our cluster mass estimates.
The median precision on our 
lensing mass measurements
is found to be
$\sim 28\%$ at $\Delta_\mathrm{c}=200$, 
$\sim 24\%$ at $\Delta_\mathrm{c}=500$,
$\sim 23\%$ at $\Delta_\mathrm{c}=1000$,
and
$\sim 24\%$ at $\Delta_\mathrm{c}=2500$.

In the CLASH weak-lensing study of \citet{Umetsu2014clash},
the $C^\mathrm{int}$ contribution (Equation \ref{eq:int}) to the total
covariance matrix (Equation (\ref{eq:Ckappa})) was not included in their
individual cluster mas measurements at $\Delta_\mathrm{c}\le 500$. 
Accordingly, their individual mass measurement errors were
underestimated by $\sim 20\%$ at $\Delta_\mathrm{c}=200$, on average 
\citep[see also Figure 5 of][]{Gruen2015}, even though $C^\mathrm{int}$
is less important at larger cluster radii (Figure \ref{fig:aveC}).   

On the other hand, the stacked cluster measurements of
\citet{Umetsu2014clash} empirically account for the cluster-to-cluster
profile variations from bootstrap resampling of the cluster sample.

\section{CLASH Stacked Mass Profile Analysis}
\label{sec:stack}

Stacking an ensemble of clusters helps average out the
projection effects of halo asphericity and substructure, as
well as the cosmic noise from projected uncorrelated LSS.
The statistical precision can be improved by stacking a large number of 
clusters, 
allowing a tighter comparison of the averaged lensing signal with
theoretical models
\citep{Okabe+2010WL,Okabe+2013,Umetsu+2011stack,Umetsu2014clash,Miyatake2015boss}.   

Hereafter, our analysis will focus on the X-ray-selected subsample of 16
CLASH clusters, which comprises a population of high-mass
X-ray regular clusters. 
The X-ray selection of this subsample is
optimized for radial profile measurements
\citep{Postman+2012CLASH}. 
Numerical simulations suggest that the CLASH X-ray-selected subsample is
prevalently composed of relaxed clusters ($\sim 70\%$) and largely free
of orientation bias \citep{Meneghetti2014clash}.\footnote{\citet{Meneghetti2014clash} showed that,
for this subsample, the median angle between the major axis of the halos
and the X-ray-selected sight lines is $\sim 57^\circ$, compared to 
$\sim 54^\circ$ expected for a distribution of random orientations.}   
The selection criteria based on X-ray morphology also
ensure well-defined cluster centers (Section \ref{subsec:sample}).
The four high-magnification clusters are thus excluded from
this part of the analysis (Sections \ref{sec:stack} and \ref{sec:cM}).

\subsection{Stacked Cluster Lensing Signal}
\label{subsec:stacklens}

 
\begin{figure*}[!htb] 
 \begin{center}
  \includegraphics[width=0.8\textwidth,angle=0,clip]{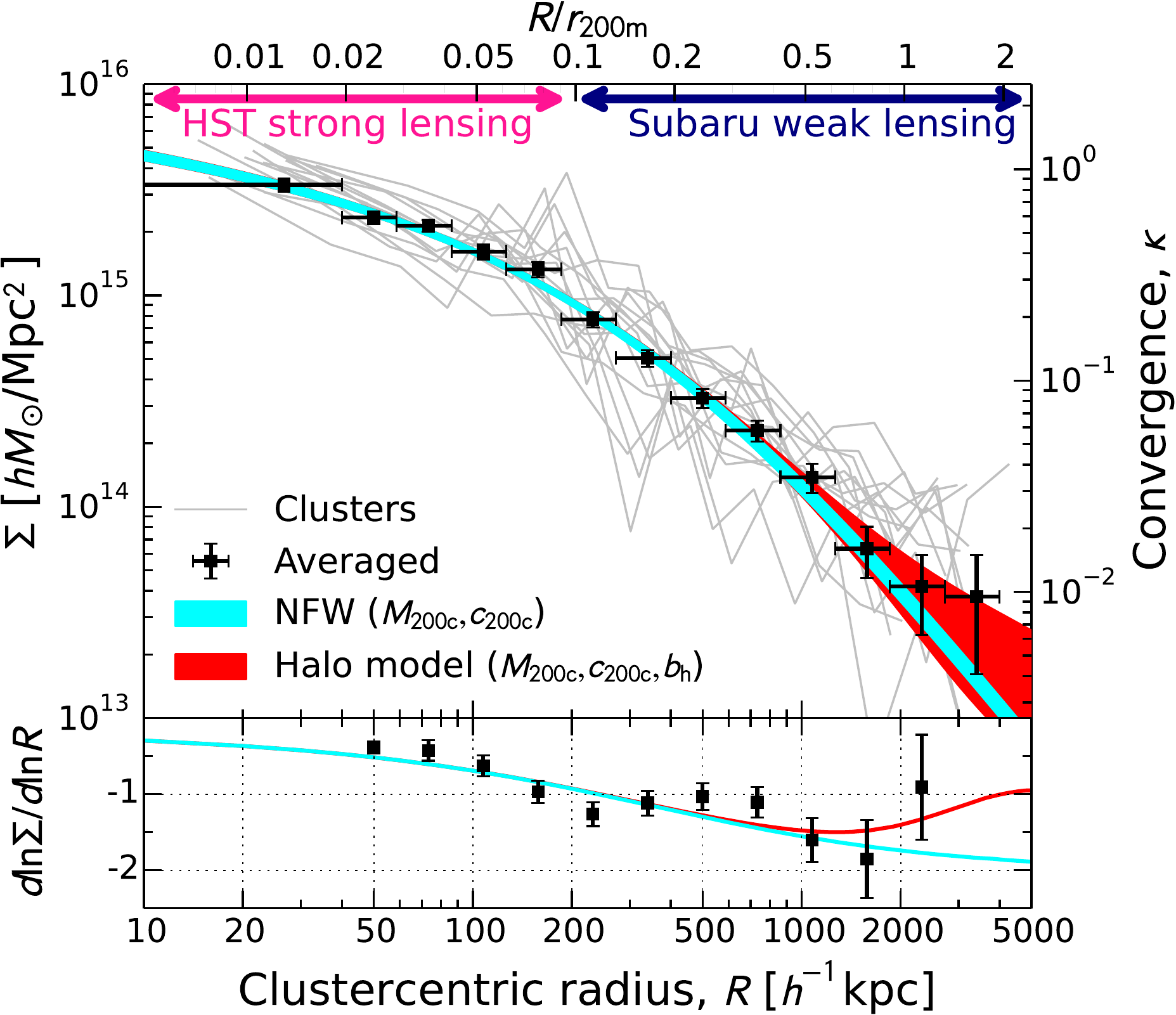} 
 \end{center} 
 \caption{\label{fig:kplot}
Upper panel: 
ensemble-averaged surface mass density $\llangle\Sigma\rrangle$
 (black squares) of the X-ray-selected subsample of 16 clusters,  
 which is obtained by stacking individual $\Sigma$ profiles 
(gray lines; Figure \ref{fig:kappa}) derived from the joint analysis of 
 {\em HST} and Subaru lensing data sets. 
The red-shaded area shows the $1\sigma$ confidence region of the
 three-parameter halo model fit (NFW+LSS (ii), Table \ref{tab:kmodel}).
The projected NFW model (cyan-shaded area, $1\sigma$)
 slightly underpredicts the total mass profile relative to the halo model
at $R\simgt r_\mathrm{200m}$. 
The scale on the right vertical axis indicates the corresponding lensing
 convergence
scaled to the mean depth of weak-lensing observations.
Lower panel: The logarithmic slope
 $d\ln\llangle\Sigma\rrangle/d\ln{R}$ (black squares) is shown along
 with the two best-fit models in the upper panel.
}
\end{figure*}

With a given set of surface mass density profiles for individual
clusters, we can stack them together to produce an ensemble-averaged
radial profile.   
Following the prescription of \citet{Umetsu+2011stack},
we re-evaluate the $\bSigma$ profiles of individual clusters 
in physical (proper) length units, using the same radial grid for all
clusters. 
Stacking an ensemble of clusters ($n=1,2,...$) is expressed as
\citep{Umetsu+2011stack}  
\begin{equation}
\label{eq:stack_mass}
  \llangle \bSigma \rrangle = 
  \left(\displaystyle\sum_n {\cal W}_n \right)^{-1}
  \,
  \left(
  \displaystyle\sum_n{ {\cal W}_n \bSigma_n}
  \right),
\end{equation}
where ${\cal W}_n$ is the sensitivity matrix of the $n$th cluster,
\begin{equation}
\label{eq:stack_weight}
 ({\cal W}_n)_{ij} \equiv \Sigma_{(\mathrm{c},\infty)n}^{-2} \,
  \left({C}^{-1}_n\right)_{ij},
\end{equation}
with $\Sigma_{(\mathrm{c,\infty}),n}$ the far-background critical
surface mass density and
$C_n$ the total covariance matrix (Equation
(\ref{eq:Ckappa})) for the $n$th cluster.\footnote{Since the covariance matrix $C$ is defined
for the far-background convergence $\kappa_\infty$, the associated
critical surface mass density too is a far-background quantity,
$\Sigma_{\mathrm{c},\infty} = \Sigma_\mathrm{c}(z\to\infty)$.}
The error covariance matrix for the stacked $\llangle\Sigma\rrangle$
profile is given by \citep{Umetsu+2011stack}
\begin{equation}
{\cal C} = \left(\sum_n {\cal W}_n\right)^{-1}.
\end{equation}
The weight matrix ${\cal W}_n$ is mass-independent when stacking in
physical length units 
\citep{Okabe+2010WL,Okabe+2013,Umetsu+2011stack,Umetsu2014clash},
which makes the effective halo mass extracted from the averaged lensing
signal a 
good proxy for the mean population value 
\citep[see Section
\ref{subsec:Meff};][]{Johnston+2007b,Okabe+2013,Umetsu2014clash,Sereno2015s8}. 
On the other hand, stacking in length units scaled to $r_\Delta$ weights
the contribution of each cluster to each radial bin in a nonlinear and
model-dependent manner \citep{Okabe+2013}, 
such that $\mathrm{tr}({\cal W}) \propto  r_\Delta^2 \propto M_\Delta^{2/3}$ 
when $C$ is dominated by the statistical noise contribution
$C^\mathrm{stat}$.

In Figure \ref{fig:kplot} we show the resulting
$\llangle\bSigma\rrangle$ profile averaged in $13$ radial bins. 
The innermost bin represents the mean density interior to
$R_\mathrm{min}=40$\,kpc\,$h^{-1}$, which corresponds approximately 
to the typical resolution limit of our strong-lensing data
($\theta_\mathrm{min}=10\arcsec$; Section \ref{subsubsec:likelihood}
and Figure \ref{fig:kappa}).
Since $R_\mathrm{min}$ is much larger than the rms offset between the
BCG and X-ray peak,  
$\sigma_\mathrm{off}\simeq 11$\,kpc\,$h^{-1}$ (Section \ref{subsec:sample}),
the miscentering effects on the $\llangle\bSigma\rrangle$
profile are expected to be insignificant for the X-ray-selected
subsample \citep{Umetsu2014clash}.
The other bins are logarithmically spaced over the range
$R=[R_\mathrm{min},R_\mathrm{max}]=[40,4000]$\,kpc\,$h^{-1}$
($0.02\simlt R/r_\mathrm{200m}\simlt 2$),
spanning two decades in radius.
For this sample, we find a sensitivity-weighted average redshift 
of $\llangle z_\mathrm{l}\rrangle\simeq 0.34$,
in close agreement with the median redshift of 
$\overline{z}_\mathrm{l}\simeq 0.35$.
We detect the stacked lensing signal at a total S/N of $\simeq 33$
using the total covariance matrix ${\cal C}$ including 
the statistical, systematic, projected uncorrelated LSS, and
intrinsic-variance contributions (Section \ref{subsubsec:cmat}).

\subsection{Modeling the Stacked Lensing Signal}

 
\begin{figure}[!htb] 
 \begin{center}
  \includegraphics[width=0.45\textwidth,angle=0,clip]{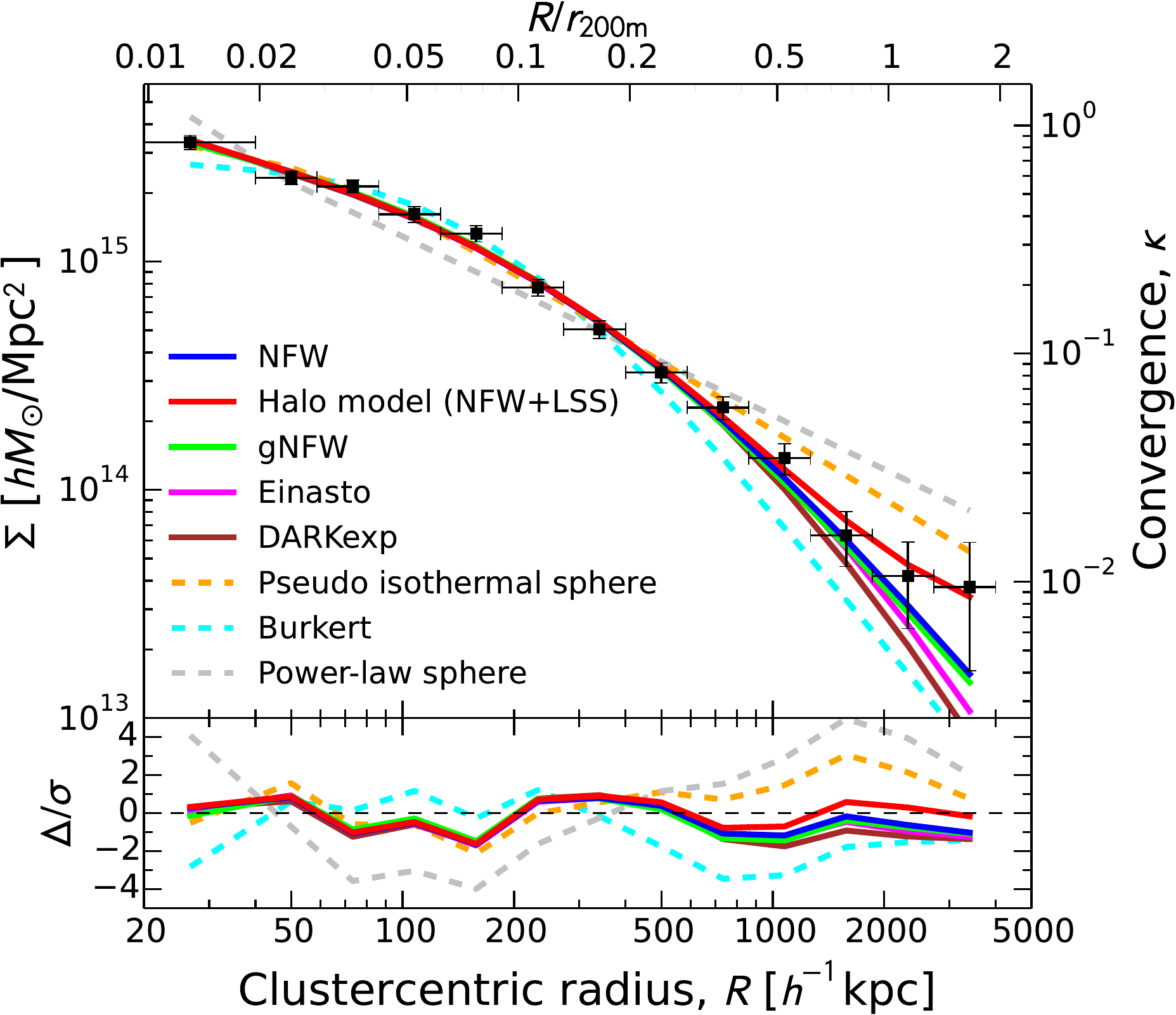} 
 \end{center}
 \caption{\label{fig:kmodel}
Upper panel: Comparison of models to the ensemble-averaged surface mass density 
 $\llangle\Sigma\rrangle$ (black squares) obtained for the
 X-ray-selected subsample of 16 clusters. Models with
 $\mathrm{PTE} > 0.05$ are shown with solid lines, while those with
 $\mathrm{PTE} < 0.05$ are shown with dashed lines (see Table
 \ref{tab:kmodel}).  
The red solid curve shows the best-fit two-parameter halo model (NFW+LSS
 (i), Table \ref{tab:kmodel}), including the effects of surrounding LSS
 as a two-halo term assuming the halo bias function of 
\citet{Tinker+2010} in the {\em WMAP} seven-year cosmology.
The lower panel shows the deviations $\Delta$ (in units of $\sigma$)
of the best-fit profiles with respect to the observed $\llangle
 \Sigma\rrangle$ profile.
}
\end{figure}

We quantify and characterize the ensemble-averaged mass
distribution of our X-ray-selected subsample using the
$\llangle\Sigma\rrangle$ profile (Section \ref{subsec:stacklens}). 
To interpret the observed averaged lensing signal,
we consider the line of sight projected surface mass density around the cluster center,
$\Sigma(R) = \int\!dl\,\Delta\rho(r)$,
with $\Delta\rho(r)=\rho(r)-\overline{\rho}_\mathrm{m}$ the mass
overdensity.
In the regime where $R\simlt r_\mathrm{200m}$, 
$\Sigma(R)$ is dominated by the cluster halo contribution
$\rho_\mathrm{h}(r)$, so that
$\Sigma(R)\simeq 2\int_0^\infty\!dl\,\rho_\mathrm{h}(r)$.

As shown in Figure \ref{fig:kplot}, our weak- and strong-lensing data
together cover a wide range of clustercentric distances $R$, extending
out to  
$R_\mathrm{max}=4000$\,kpc\,$h^{-1}\approx 2r_\mathrm{200m}$.
In the context of the $\Lambda$CDM model, the two outermost radial bins
lie in the transition between the one-halo and two-halo regimes
\citep{Cooray+Sheth2002},
$r_\mathrm{200m}\simlt R\simlt 2R_\mathrm{200m}$,
where the large-scale two-halo contribution 
to $\Sigma(R)$ is expected to become important
\citep{Oguri+Hamana2011,Silva+2013,Umetsu2014clash}.\footnote{On the
other hand, the tangential shear
$\gamma_+=\Delta\Sigma/\Sigma_\mathrm{c}$ is insensitive to the
projected two-halo term in the transition regime \citep{Oguri+Hamana2011}.}
In this work, we thus test models 
both with and without including the two-halo term.

\subsubsection{Halo Density Profiles}
\label{subsubsec:rho}

We give here a brief description of the halo profile models that we
consider. 
For each model $\rho_\mathrm{h}(r)$,
the halo mass is defined 
using a spherical overdensity $\Delta_\mathrm{c}=200$ 
as $M_\mathrm{200c}$.
We introduce the radius $r_{-2}$ at which the logarithmic density slope 
is {\em isothermal}, that is,
$d\ln{\rho_\mathrm{h}(r)}/d\ln{r}=-2$ at $r=r_{-2}$.
In analogy to the NFW concentration parameter, 
the degree of concentration is defined
by $c_\mathrm{200c}=r_\mathrm{200c}/r_{-2}$.  We use 
$M_\mathrm{200c}$ and $c_\mathrm{200c}$ as
fitting parameters, when possible.
\begin{enumerate}
\item Generalized NFW (gNFW) model
\citep[][]{Zhao1996}:
\begin{equation}
 \label{eq:gNFW}
  \begin{aligned}
   \rho_\mathrm{h}(r) &= \frac{\rho_\mathrm{s}}{(r/r_\mathrm{s})^{\gamma_\mathrm{c}}
   (1+r/r_\mathrm{s})^{3-\gamma_\mathrm{c}}},\\
   r_\mathrm{-2} &= (2-\gamma_\mathrm{c})r_\mathrm{s},
  \end{aligned}
\end{equation}
with  $\gamma_\mathrm{c}$ the central slope, $\rho_\mathrm{s}$ and
      $r_\mathrm{s}$ the characteristic density and 
      radius, respectively.
For $\gamma_\mathrm{c}=1$,
this reduces to the standard NFW model 
$\rho_\mathrm{NFW}(r)$ with $r_\mathrm{s}=r_{-2}$.
\item Einasto model \citep[][]{Einasto1965}:
\begin{equation}
\rho_\mathrm{h}(r)=\rho_{-2}
 \exp\left\{
      -\frac{2}{\alpha_\mathrm{E}}
      \left[
       \left(
	\frac{r}{r_\mathrm{-2}}
       \right)^{\alpha_\mathrm{E}}-1
      \right]
     \right\},
\end{equation}
with $\alpha_\mathrm{E}$ the shape parameter describing the degree of
      curvature and $\rho_{-2}=\rho_\mathrm{h}(r_{-2})$. An Einasto 
      profile with $\alpha_\mathrm{E}\approx 0.18$ 
      closely resembles the NFW profile over roughly two decades in radius
      \citep{Ludlow+2013}. 

\item DARKexp-$\gamma$ model.
      DARKexp is a theoretically derived model for collisionless
      self-gravitating systems with isotropic velocity distributions
      \citep{Hjorth+2010DARKexp,DARKexp2}. 
      We use Dehnen--Tremaine $\gamma$-models
      \citep{Dehnen1993,Tremaine1994} as an analytic fitting function
      for the DARKexp density profile \citep{Hjorth2015darkexp}:
\begin{equation}
 \begin{aligned}
  \rho_\mathrm{h}(r) &=
  \frac{\rho_\mathrm{s}}{(r/r_\mathrm{s})^{\gamma_\mathrm{c}}
  (1+r/r_\mathrm{s})^{4-\gamma_\mathrm{c}}},\\
  r_{-2} &= (1-\gamma_\mathrm{c}/2)r_\mathrm{s},\\
  \gamma_\mathrm{c} &\approx 3\log_{10}\phi_0-0.65 \ \ \ (1.7\le \phi_0 \le 6),\\
  \end{aligned}
\end{equation}
      where $\rho_\mathrm{s}$ and $r_\mathrm{s}$ are the scale density
      and radius, respectively, and
      $\phi_0$ represents the dimensionless depth of the halo potential
      describing the profile shape.
      The $\gamma$ models approximate DARKexp very well over nearly four decades
      in radius \citep{Hjorth2015darkexp}.\footnote{The DARKexp density
      profile is also well approximated by an Einasto profile at small halo
      radii \citep{Hjorth2015darkexp}.}

\item  Pseudo-isothermal (PI) sphere model:
\begin{equation}
\rho_\mathrm{h}(r)=\frac{\rho_\mathrm{c}}{1+(r/r_\mathrm{c})^2}
\end{equation}
with $\rho_\mathrm{c}$ and $r_\mathrm{c}$ the core density and radius, respectively.
The corresponding asymptotically flat circular velocity is 
$V_\mathrm{c}=(4\pi G\rho_\mathrm{c} r_\mathrm{c}^2)^{1/2}$ \citep{Shao2013}.

\item  Burkert model \citep[][]{Mori+Burkert2000}:
\begin{equation}
\rho_\mathrm{h}(r)=\frac{\rho_0}{(1+r/r_0)(1+r^2/r_0^2)}
\end{equation}
with $\rho_0$ and $r_0$ the core density and radius, respectively.

\item Power-law sphere model:
\begin{equation}
\rho_\mathrm{h}(r)\propto r^{-\gamma_\mathrm{c}}.
\end{equation}
This model includes the singular isothermal sphere model with $\gamma_\mathrm{c}=2$.

\end{enumerate}

The NFW, gNFW, and Einasto density profiles
represent a family of phenomenological models for cuspy DM halos
motivated by numerical simulations and observations.
The DARKexp model describes the distribution of particle energies
in finite, self-gravitating, collisionless, isotropic systems,
providing theoretical predictions for the structure of collisionless DM
halos.  
For radii accessible to $N$-body simulations, 
DARKexp allows for central slope values in the range
$-2\simlt d\ln{\rho_\mathrm{h}}/d\ln{r} \simlt 0$. 
The empirical PI and Burkert models describe cored density profiles.
The power-law model is often adopted as a lens model for its simplicity
\citep[e.g.,][]{Koopmans2009,Agnello2013}.

\subsubsection{Projected Halo Model}
\label{subsubsec:halomodel}

To interpret the averaged lensing signal in the context of the
standard $\Lambda$CDM model, we employ as our reference model the halo
model prescription of \citet{Oguri+Takada2011}. 
Specifically, we take into account the large-scale clustering
contribution $\rho_\mathrm{2h}(r)$ as
\begin{equation}
\label{eq:halomodel}
 \begin{aligned}
\Delta\rho(r) &= f_\mathrm{t}(r) \, \rho_\mathrm{h}(r) + \rho_\mathrm{2h}(r),\\
  f_\mathrm{t}(r) &= 
\left[
1+ \left(
  \frac{r}{r_\mathrm{t}}
  \right)^2
\right]^{-2},
  \end{aligned}
\end{equation}
where $f_\mathrm{t}(r)$ describes the steepening of the density profile
around a truncation radius $r_\mathrm{t}$
\citep{BMO,Diemer+Kravtsov2014}.
We fix the truncation parameter
$\tau_\mathrm{200c}\equiv r_\mathrm{t}/r_\mathrm{200c} =3$ 
\citep{Covone2014,Umetsu2014clash,Sereno2015s8},  
a typical value for cluster-sized halos in the $\Lambda$CDM
cosmology \citep{Oguri+Hamana2011}.
When the two-halo term is neglected, 
the standard NFW halo model \citep{Oguri+Hamana2011}
reduces to the Baltz--Marshall-Oguri (BMO) model
that describes a truncated NFW profile,
$f_\mathrm{t}(r)\rho_\mathrm{NFW}(r)$ \citep{BMO}.

For an ensemble of clusters with mass $M$ and redshift $z$,
the two-halo term is expressed as
$\rho_\mathrm{2h}(r) = \overline{\rho}_\mathrm{m} b_\mathrm{h}(M)\xi_\mathrm{m}^\mathrm{L}(r)$
with 
$\overline{\rho}_\mathrm{m}$ the mean background density of the universe,
$b_\mathrm{h}(M)$ the linear halo bias,
and
$\xi_\mathrm{m}^\mathrm{L}(r)$ the linear matter correlation function,
all evaluated at $z=\llangle z_\mathrm{l}\rrangle \simeq 0.34$ in the
{\em WMAP} seven-year cosmology (Section \ref{subsubsec:cmat}).
The two-halo term is proportional to the product
$b_\mathrm{h}\sigma_8^2$, 
where $\sigma_8$ is the rms amplitude of linear mass fluctuations in a
sphere of comoving radius $8$\,Mpc\,$h^{-1}$.  
In the adopted cosmology, $\sigma_8=0.81$ \citep{Komatsu+2011WMAP7}.
We compute the total surface mass density $\Sigma(R)$ by projecting
$\Delta\rho(r)=f_\mathrm{t}\rho_\mathrm{1h}+\rho_\mathrm{2h}$ along the
line of sight \citep[Section 2.2 of][]{Oguri+Hamana2011}.
To evaluate the halo bias factor $b_\mathrm{h}(M)$, we adopt the model
of \citet{Tinker+2010}, which is well calibrated using a large set of 
$N$-body simulations. We use their fitting formula
with a halo mass definition of $\Delta_\mathrm{c}=200$
\citep{Umetsu2014clash}, 
corresponding to $\Delta_\mathrm{m}\simeq 420$ in our adopted cosmology. 

\subsection{Model Comparison with Observations}

\begin{deluxetable*}{lccccccc}
\tablecolumns{8}
\tablecaption{
\label{tab:kmodel}
Best-fit models for the stacked mass profile of the CLASH X-Ray-selected subsample
}
\tablewidth{0pt}
\tablehead{
 \multicolumn{1}{c}{Model} &
 \multicolumn{1}{c}{$M_{200{\rm c}}$} &
 \multicolumn{1}{c}{$c_{200{\rm c}}$} &
 \multicolumn{1}{c}{Shape/Structural Parameters} &
 \multicolumn{1}{c}{$b_\mathrm{h}$} &
 \multicolumn{1}{c}{$\chi^2/\mathrm{dof}$} &
 \multicolumn{1}{c}{PTE\tablenotemark{a}} &
 \multicolumn{1}{c}{Notes}
 \\ 
\multicolumn{1}{c}{} &
\multicolumn{1}{c}{($10^{14}M_{\odot}\,h_{70}^{-1}$)} &
\multicolumn{1}{c}{} &
\multicolumn{1}{c}{} &
\multicolumn{1}{c}{} &
\multicolumn{1}{c}{} &
\multicolumn{1}{c}{} &
\multicolumn{1}{c}{}
}
\startdata
                 NFW & $14.4^{+1.1}_{-1.0}$ & $3.76^{+0.29}_{-0.27}$ & $\gamma_\mathrm{c}=1$ & --- & $11.3/11$ & $0.419$ & No truncation\\
                gNFW & $14.1^{+1.1}_{-1.1}$ & $4.04^{+0.53}_{-0.52}$ & $\gamma_\mathrm{c}=0.85^{+0.22}_{-0.31}$ & --- & $10.9/10$ & $0.366$ & No truncation\\
             Einasto & $14.7^{+1.1}_{-1.1}$ & $3.53^{+0.36}_{-0.39}$ & $\alpha_\mathrm{E}=0.232^{+0.042}_{-0.038}$ & --- & $11.7/10$ & $0.306$ & No truncation\\
DARKexp--$\gamma$\tablenotemark{b} & $14.5^{+1.2}_{-1.1}$ & $3.53^{+0.42}_{-0.42}$ & $\phi_0=3.90^{+0.41}_{-0.45}$ & --- & $13.5/10$ & $0.198$ & No truncation\\
   Pseudo isothermal & --- & --- & $V_\mathrm{c}=1762^{+  40}_{ -39}$\,km\,s$^{-1}$, $r_\mathrm{c}=69^{+ 7}_{-7}$\,kpc & --- & $23.6/11$ & $0.015$ & No truncation\\
             Burkert & $11.6^{+0.8}_{-0.8}$ & --- & $r_\mathrm{200c}/r_0=8.81^{+0.42}_{-0.41}$ & --- & $29.9/11$ & $0.002$ & No truncation\\
    Power-law sphere & $12.5^{+0.8}_{-0.8}$ & --- & $\gamma_\mathrm{c}=1.78^{+0.02}_{-0.02}$ & --- & $93.5/11$ & $0.000$ & No truncation\\
\hline Halo Model\tablenotemark{c}:\\
       ~~NFW+LSS (i) & $14.1^{+1.0}_{-1.0}$ & $3.79^{+0.30}_{-0.28}$ & $\gamma_\mathrm{c}=1$ & $ 9.3$ & $10.9/11$ & $0.450$ & $\Lambda$CDM $b_\mathrm{h}(M)$ scaling\\
      ~~NFW+LSS (ii) & $14.4^{+1.4}_{-1.3}$ & $3.74^{+0.33}_{-0.30}$ & $\gamma_\mathrm{c}=1$ & $ 7.4^{+ 4.6}_{-4.7}$ & $10.8/10$ & $0.377$ & $b_\mathrm{h}$ as a free parameter\\
   ~~Einasto+LSS (i) & $14.3^{+1.1}_{-1.1}$ & $3.69^{+0.36}_{-0.42}$ & $\alpha_\mathrm{E}=0.248^{+0.051}_{-0.047}$ & $ 9.3$ & $10.7/10$ & $0.385$ & $\Lambda$CDM $b_\mathrm{h}(M)$ scaling\\
  ~~Einasto+LSS (ii) & $14.5^{+1.9}_{-1.6}$ & $3.65^{+0.47}_{-0.61}$ & $\alpha_\mathrm{E}=0.245^{+0.061}_{-0.053}$ & $ 8.7^{+ 5.3}_{-5.6}$ & $10.6/9$ & $0.301$ & $b_\mathrm{h}$ as a free parameter\\
   ~~DARKexp+LSS (i) & $14.2^{+1.2}_{-1.1}$ & $3.64^{+0.44}_{-0.46}$ & $\phi_0=3.89^{+0.51}_{-0.54}$ & $ 9.3$ & $11.7/10$ & $0.308$ & $\Lambda$CDM $b_\mathrm{h}(M)$ scaling\\
  ~~DARKexp+LSS (ii) & $14.0^{+1.8}_{-1.6}$ & $3.69^{+0.53}_{-0.57}$ & $\phi_0=3.85^{+0.57}_{-0.61}$ & $10.1^{+ 4.9}_{-5.1}$ & $11.6/9$ & $0.235$ & $b_\mathrm{h}$ as a free parameter\
\enddata
\tablenotetext{a}{Probability to exceed the observed $\chi^2$ value.}\tablenotetext{b}{We use Dehnen--Tremaine $\gamma$-models with the central cusp slope $\gamma_\mathrm{c}=3\log_{10}\phi_0-0.65$ ($1.7\le \phi_0\le 6$) as an analytic fitting function for the DARKexp density profile.}
\tablenotetext{c}{For halo model predictions, we decompose the total mass overdensity $\Delta\rho(r)=\rho(r)-\overline{\rho}_\mathrm{m}$ as $\Delta\rho=f_\mathrm{t}\rho_\mathrm{h}+\rho_\mathrm{2h}$ where $\rho_\mathrm{h}(r)$ is the halo density profile, $\rho_\mathrm{2h}(r)=\overline{\rho}_\mathrm{m}b_\mathrm{h}\xi_\mathrm{m}^\mathrm{L}(r)$ is the two-halo term, and $f_\mathrm{t}(r)=(1+r^2/r_\mathrm{t}^2)^{-2}$ describes the steepening of the density profile in the transition regime around the truncation radius $r_\mathrm{t}$, which is assumed to be $r_\mathrm{t}=3r_\mathrm{200c}$.}
\end{deluxetable*}

We constrain model parameters using the $\llangle\bSigma\rrangle$
profile and its total covariance matrix  
${\cal C}$ (Section \ref{subsec:stacklens}).
The $\chi^2$ minimization is performed using the {\sc minuit} minimization
package from the CERN program libraries. The best-fit parameters are
reported in Table \ref{tab:kmodel}, along with the reduced $\chi^2$ and 
corresponding probability to exceed (PTE, hereafter) values. 

We first consider halo density profile models without including the
two-halo term. In this modeling approach, no truncation is applied
when calculating the surface mass density $\Sigma(R)$,
effectively accounting for the large-scale contribution.
Figure \ref{fig:kmodel} shows the best-fit model profiles along with the
observed $\llangle\Sigma\rrangle$ profile.
We find that two- or three-parameter cuspy models, namely,
the NFW, gNFW, Einasto, and DARKexp-$\gamma$ models, provide
satisfactory fits ($\mathrm{PTE}>0.05$) to the data.  
The best-fit gNFW model has a central cusp slope of
$\gamma_\mathrm{c}=0.85^{+0.22}_{-0.31}$, being consistent with the 
NFW model ($\gamma_\mathrm{c}=1$).
The cored PI and Burkert models provide poor fits
with $\chi^2/\mathrm{dof}$ (PTE) values of 
$23.6/11$ ($1.5\times 10^{-2}$) and 
$29.9/11$ ($1.7\times 10^{-3}$), respectively.
The power-law sphere model has a reduced $\chi^2$ value of
$\chi^2/\mathrm{dof}=93.5/11$ and is strongly disfavored by the
$\llangle\Sigma\rrangle$ profile having a pronounced radial
curvature.

Now we examine projected halo models following the prescription
described in Section \ref{subsubsec:halomodel}. 
In what follows, we will focus on our best models,
specifically, the NFW, Einasto, and DARKexp-$\gamma$
cuspy density profiles. 
We note that since the halo bias is given as a function of
$M_\mathrm{200c}$, the number of free parameters is unchanged for each
case.   
We find these models (NFW+LSS (i), Einasto+LSS (i), and DARKexp+LSS
(i) in Table \ref{tab:kmodel}) give statistically comparable fits.
In all cases,
including the two-halo term 
improves the fits,
while keeping the best-fit parameters essentially unchanged (the
difference in each parameter is much less than the $1\sigma$
uncertainty).
Independent of the chosen profile, we find 
$b_\mathrm{h}(M_\mathrm{200c})\sim 9.3$ ($b_\mathrm{h}\sigma_8^2\sim 6.1$)
for our best-fit models.

On the basis of the goodness-of-fit statistic, the highest-ranked model
among those considered is the NFW halo model (NFW+LSS (i); see the red
curve in Figure \ref{fig:kmodel}) with  
$M_\mathrm{200c}=14.1^{+1.0}_{-1.0}\times 10^{14}M_\odot\,h_{70}^{-1}$
and 
$c_\mathrm{200c}=3.79^{+0.30}_{-0.28}$,
followed by the Einasto and DARKexp halo models (Einasto+LSS
(i) and DARKexp+LSS (i)). 
The best-fit NFW halo model yields an Einstein radius of 
$\theta_\mathrm{Ein}=14.0^{+3.4}_{-3.2}$\,arcsec at $z_\mathrm{s}=2$,
which is about $1.7\sigma$ lower than
the median effective Einstein radius (see Section \ref{subsec:sl})
of $\overline{\theta}_\mathrm{Ein}=20.1\arcsec$ observed 
for the X-ray-selected subsample (Section \ref{subsec:sl}).
The best-fit NFW parameters of our projected halo model are in good
agreement with those from the stacked shear-only analysis of
\citet{Umetsu2014clash}, 
$M_\mathrm{200c}=13.4^{+1.0}_{-0.9}\times 10^{14}M_\odot\,h^{-1}$ and
$c_\mathrm{200c}=4.01^{+0.35}_{-0.32}$,
in which the two-halo term can be safely neglected as the stacked
tangential-shear signal, $\llangle \Delta\Sigma\rrangle$, is only
sensitive to the intra-halo mass distribution in our radial range. 
The Einasto shape parameter is constrained to be
$\alpha_\mathrm{E}=0.248^{+0.051}_{-0.047}$ from an Einasto halo-model  
fit to the $\llangle \Sigma\rrangle$ profile (Einasto+LSS (i)),
in agreement with $\alpha_\mathrm{E}=0.191^{+0.071}_{-0.068}$
from the stacked shear-only analysis of \citet{Umetsu2014clash}.
Our measurements agree well with predictions from $\Lambda$CDM numerical
simulations, 
$\alpha_\mathrm{E}=0.21\pm 0.07$ 
\citep[][$\alpha_\mathrm{E}=0.24\pm 0.09$ when fitted to surface mass density profiles]{Meneghetti2014clash}.
The fitting formula given by \citet{Gao+2008} yields
$\alpha_\mathrm{E}\simeq 0.29$ for our X-ray-selected subsample.
This is consistent with our results at the $1\sigma$ level. 

The halo-model interpretation we have adopted is a crude
approximation in the transition between the intra-halo and two-halo
regimes, neglecting nonlinear effects
\citep{Baldauf2010,Diemer+Kravtsov2014}.  
In order to effectively account for possible modifications due to
nonlinear effects, here we allow the halo bias factor $b_\mathrm{h}$
to be a free parameter.
We find that allowing $b_\mathrm{h}$ to be free does not improve the
fits and that the resulting best-fit parameters remain unchanged
within the errors (Table \ref{tab:kmodel}). The {\em effective} halo
bias $b_\mathrm{h}$ is detected at $1.6\sigma$--$2.0\sigma$ 
significance, where the exact level of significance depends on the
details of the halo density profile in the transition regime.
In Figure \ref{fig:kplot}, we compare the resulting NFW halo model profile
(NFW+LSS (ii), red shaded area) and the NFW profile (cyan-shaded
area) along with the $\llangle \Sigma\rrangle$ profile.


\subsection{Interpreting the Effective Halo Mass}
\label{subsec:Meff}


\begin{figure}[!htb] 
 \begin{center}
  \includegraphics[width=0.45\textwidth,angle=0,clip]{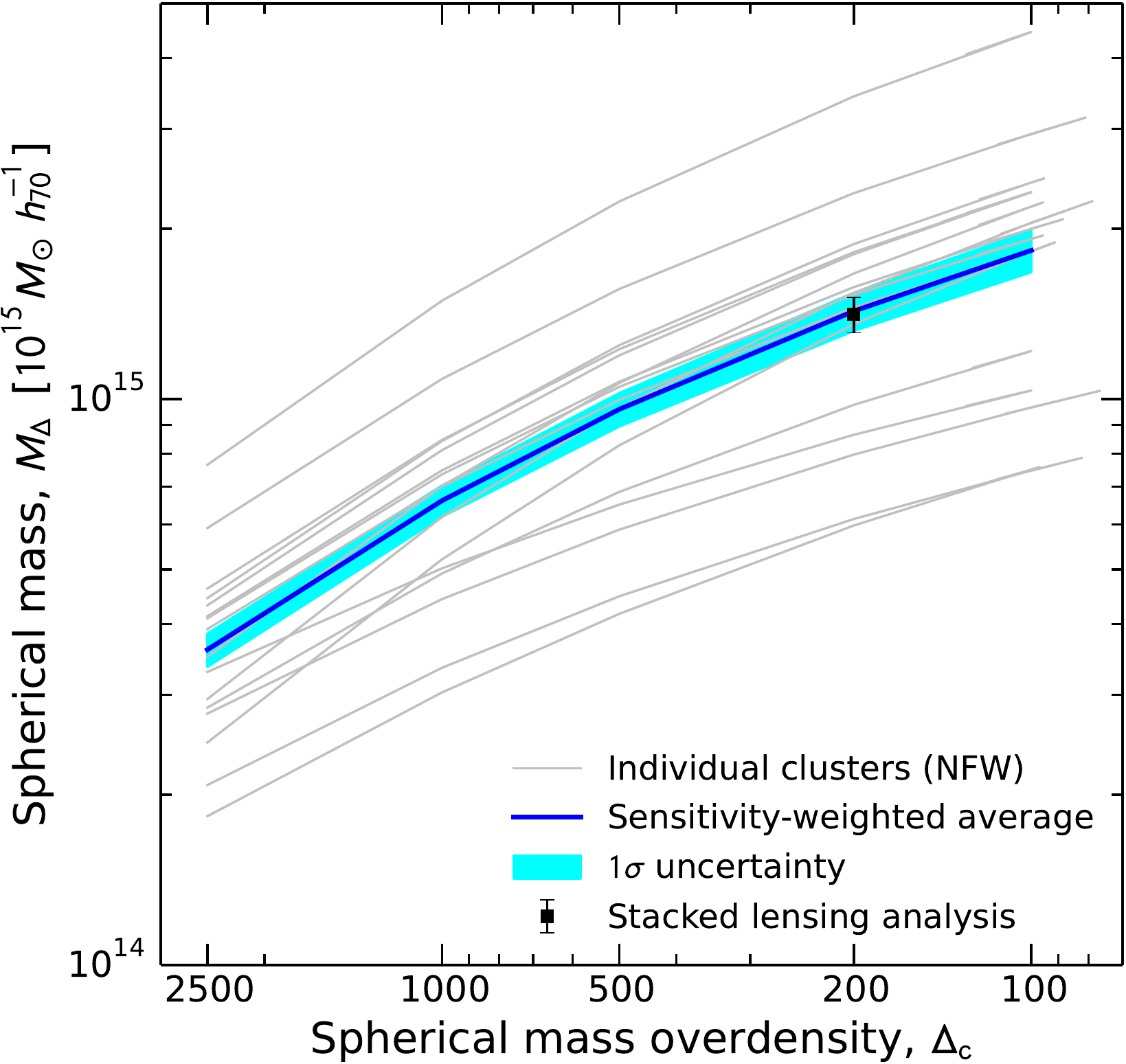}
 \end{center}
\caption{\label{fig:Meff}
Cumulative mass profiles $M_\Delta=M_\mathrm{3D}(<r_\Delta)$ for
 the X-ray-selected subsample of 16 clusters,
shown at several characteristic values of the spherical mass overdensity
 $\Delta_\mathrm{c}=\rho_\mathrm{h}(<r)/\rho_\mathrm{c}$.
The gray lines show the $M_\Delta$ profiles for individual clusters
obtained from NFW fits to their $\Sigma$ profiles (Tables \ref{tab:cm}
 and \ref{tab:mass}; Figure \ref{fig:kappa}). 
The blue line and the cyan-shaded area show the mean and its $1\sigma$ error
 of the composite-halo profile $\llangle M_\Delta\rrangle$
from a sensitivity-weighted average of the individual $M_\Delta$ profiles.
The filled square at $\Delta_\mathrm{c}=200$
marks the best-fit halo mass 
$M_\mathrm{200c}=14.1^{+1.0}_{-1.0}\times 10^{14}M_\odot\,h_{70}^{-1}$ 
obtained from a halo-model fit
(NFW+LSS (i) in Table \ref{tab:kmodel}) to the stacked
 $\llangle\Sigma\rrangle$ profile (Figure \ref{fig:kmodel}).
This is in excellent agreement with the mean sample mass 
$\llangle M_\mathrm{200c}\rrangle = (14.3 \pm 1.1)\times 10^{14}M_\odot\,h_{70}^{-1}$ 
averaged using $\mathrm{tr}({\cal W})$ (Equation
 (\ref{eq:stack_weight})) as weights.  
}
\end{figure}

Interpreting the effective mass from stacked lensing requires caution
when the cluster sample spans a broad range of masses and redshifts
\citep{Mandelbaum2005halomodel,Niikura2015}.  
If there is a significant scatter in the luminosity--mass relation of
halos, then the halo mass distribution of a luminosity-selected sample 
becomes significantly broader, with the width and asymmetry of the
distribution typically increasing for higher luminosity
bins. Accordingly, there is no typical mass, and in general the
effective mass determination from NFW fits falls between the mean and
the median masses of the halo population
\citep{Mandelbaum2005halomodel}.
Alternatively, if a halo sample has a broad redshift distribution, the
effective lensing mass determined from stacking with redshift-dependent
weights (Equation (\ref{eq:stack_weight})) is generally different from
the mean population mass \citep{Umetsu2014clash}.

To assess this possibility, we compare the stacked lensing results
with the individual cluster masses (Tables \ref{tab:cm} and
\ref{tab:mass}).
For the former, we consider our best model (NFW+LSS (i) in Table
\ref{tab:kmodel}) for the X-ray-selected subsample,
namely the NFW halo model using the $b_\mathrm{h}$--$M$ relation of
\citet{Tinker+2010}.

Following \citet{Umetsu2014clash}, we construct a composite-halo
mass profile $\llangle M_\Delta\rrangle$ \citep[see
also][]{Coupon+2013,Ford2014cfhtlens} from a sensitivity-weighted
average of NFW fits to individual cluster $\bSigma$ profiles as 
\begin{equation}
 \llangle M_\Delta\rrangle = \frac{\sum_n \mathrm{tr}({\cal W}_n)\,
  M_{\Delta,n}}{\sum_n \mathrm{tr}({\cal W}_n)}
\end{equation}
using $\mathrm{tr}({\cal W}_n)\propto \Sigma_{(\mathrm{c},\infty)n}^{-2}$
(Equation (\ref{eq:stack_weight}))
as an effective sensitivity weight for each cluster. 
Hence, the weight depends on cluster redshift $z_\mathrm{l}$
through $\Sigma_{(\mathrm{c},\infty)}$. 
The error variance for $\llangle M_\Delta\rrangle$ is given by
$\sigma^2_\Delta=[\sum_n\mathrm{tr}({\cal W}_n)]^{-2}
\sum_n \sigma_{\Delta,n}^2 [\mathrm{tr}({\cal W}_n)]^2$
with $\sigma_{\Delta,n}$ the $1\sigma$ uncertainty for $M_{\Delta,n}$.
It was shown by \citet{Umetsu2014clash} that this composite-halo approach 
using the trace approximation for the sensitivity matrix can give an
adequate description of the observed stacked lensing signal.

The results of this comparison are summarized in Figure \ref{fig:Meff}.
At $\Delta_\mathrm{c}=200$, we find a sensitivity-weighted average of
$\llangle M_\mathrm{200c}\rrangle= (14.3 \pm 1.1)\times 10^{14}M_\odot\,h_{70}^{-1}$
for the X-ray-selected subsample, in excellent agreement with
the best-fit halo mass of 
$M_\mathrm{200c}=14.1^{+1.0}_{-1.0}\times 10^{14}M_\odot\,h_{70}^{-1}$ 
extracted from a halo-model fit to the $\llangle\Sigma\rrangle$ profile.
On the other hand, 
the unweighted median masses of the clusters are 
$\sim 3\%$--$8\%$ higher than the sensitivity-weighted masses
$\llangle M_\Delta\rrangle$ at each overdensity.
This difference is not due to the asymmetry of the distribution of the
cluster masses because the unweighted median and mean masses agree to
within $\sim 2\%$ at each overdensity.
Our results appear to be robust against different ensemble-averaging
techniques once the effects of sensitivity weighting are consistently
taken into account, supporting the findings of \citet{Umetsu2014clash}.  

\section{CLASH Concentration--Mass Relation}
\label{sec:cM}

\subsection{Bayesian Linear Regression}
\label{subsec:regression}

We examine the relationship between halo mass and concentration for our  
CLASH X-ray-selected subsample.  
Here we closely follow the Bayesian regression approach of
\citet{Sereno2015cM}, taking into account 
the correlation between the errors on mass and concentration
\citep[e.g.,][]{Du+Fan2014} 
and the effects of nonuniformity of the mass probability distribution
\citep[e.g.,][]{Kelly2007,Andreon+Berge2012}.
We consider a power-law function of the following form:
\begin{equation}
\label{eq:cMz}
c_\mathrm{200c}
= 10^\alpha
\left(\frac{M_\mathrm{200c}}{M_\mathrm{piv}}\right)^\beta
\left(
\frac{1+z}{1+z_\mathrm{piv}}
\right)^\gamma,
\end{equation}
with $M_\mathrm{piv}$ 
and
$z_\mathrm{piv}$ the pivot mass and redshift, respectively.
We set $M_\mathrm{piv}=10^{15}M_\odot\, h^{-1}$ 
and $z_\mathrm{piv}=0.34$, approximately the effective mean values of the 
cluster mass and redshift, respectively (Section \ref{sec:stack}). 
In the present analysis, we fix $\gamma =-0.668$ \citep{Meneghetti2014clash}
as predicted for the CLASH X-ray-selected population (for details, see
Section \ref{subsec:cM_CLASH}) because our data do not have sufficient
statistics to constrain the redshift evolution of the $c$--$M$ relation
\citep{Merten2015clash}.

To perform regression analysis we define new dependent and independent
variables in logarithmic form as 
\begin{equation}
\label{eq:XY}
 \begin{aligned}
  Y &\equiv
\log_{10}
\left[
\left(
\frac{1+z}{1+z_\mathrm{piv}}
\right)^{-\gamma}
c_\mathrm{200c}
\right],\\
 X &\equiv \log_{10}\left(M_\mathrm{200c}/M_\mathrm{piv}\right).
 \end{aligned}
\end{equation}
Equation (\ref{eq:cMz}) is then expressed by a linear relation,
$Y=\alpha+\beta X$. The observed values of latent variables are denoted
with lowercase letters.
If the selected clusters in our sample are not uniformly distributed in
logarithmic mass $X$, this can lead to biased estimates of regression
parameters \citep{Kelly2007, Sereno2015cM}. We properly 
account for these effects in Bayesian regression 
\citep{Kelly2007, CoMaLit2, CoMaLit4, Sereno2015cM}. 

The clusters in our X-ray-selected subsample are selected to be
X-ray hot ($>5$\,keV) and regular in X-ray morphology
\citep{Postman+2012CLASH}. 
In general, 
if we select a cluster sample by imposing certain thresholds on
cluster observables, the steepening at the high end of the
intrinsic mass function, combined with the selection effects, makes the
resulting mass probability distribution of selected clusters
approximately lognormal \citep[Appendix A of][]{CoMaLit4}.

In this study, we model the intrinsic probability distribution
${\cal P}(X)$ of logarithmic mass $X$ with a single Gaussian function 
characterized by two parameters, namely the mean $\mu$ and the
dispersion $\tau$ \citep{Kelly2007}.
This approach alleviates the problem of assuming a uniform prior
distribution on the independent variable $X$
and, in general, provides a good approximation for a regular unimodal
distribution \citep{Kelly2007,Andreon+Berge2012,CoMaLit4}.  
A uniform prior distribution can be
recovered in the limit of $\tau\to\infty$.

The conditional probability distribution ${\cal P}(\by|\bx)=\prod_n{\cal P}(y_n|x_n)$ 
of $\by=\{y_n\}$ given $\bx=\{x_n\}$ is then written as \citep{Kelly2007}
\begin{equation}
  \ln{\cal P}(\by|\bx) = -\frac{1}{2}\sum_{n}
\left[
\ln{(2\pi \sigma_n^2)} +
 \left(
 \frac{y_n-\langle y_n|x_n\rangle}{\sigma_n}
 \right)^2
\right],
\end{equation}
where $\langle y_n|x_n\rangle$ and $\sigma_n^2\equiv \mathrm{Var}(y_n|x_n)$ are 
the conditional mean and variance of $y_n$ given $x_n$, respectively:
\begin{equation}
 \begin{aligned}
 \langle y_n|x_n \rangle &= \alpha + \beta \mu + \frac{\beta\tau^2+C_{xy,n}}{\tau^2+C_{xx,n}}(x_n-\mu),\\
 \sigma_n^2 &=
  \beta^2\tau^2+\sigma_{Y|X}^2+C_{yy,n}
  -\frac{(\beta\tau^2+C_{xy,n})^2}{\tau^2+C_{xx,n}},
 \end{aligned}
\end{equation}
where $\sigma_{Y|X}$ is the intrinsic scatter in the $Y$--$X$
relation;
$C_{xx,n}$, $C_{yy,n}$, and $C_{xy,n}=C_{yx,n}$ are the elements of the
covariance matrix between the observables $x_n$ and $y_n$ of the $n$th
cluster. 
If ${\cal P}(X_n)$ is assumed to be uniform, we have 
$\langle y_n|x_n\rangle=\alpha +\beta x_n$ and
$\sigma_n^2=\sigma^2_{Y|X} + C_{yy,n} + \beta^2 C_{xx,n} -2\beta
C_{xy,n}$ \citep[e.g.,][]{Kelly2007,Okabe+Smith2015}.

We use uninformative priors for all parameters.
We adopt uniform priors for the intercept $\alpha$, the slope
$\beta$, and the mean $\mu$. For the variance parameters
$\sigma_{Y|X}^2$ and $\tau^2$, we use the inverse Gamma
distribution, $\mathrm{I\Gamma}(\epsilon,\epsilon)$, with
$\epsilon \ll 1$ a small number \citep{Andreon+Hurn2010,CoMaLit1,CoMaLit4}. 
In our analysis, we choose $\epsilon=10^{-3}$ \citep{CoMaLit1,CoMaLit4}.


\begin{figure}[!htb] 
 \begin{center}
  \includegraphics[width=0.48\textwidth,angle=0,clip]{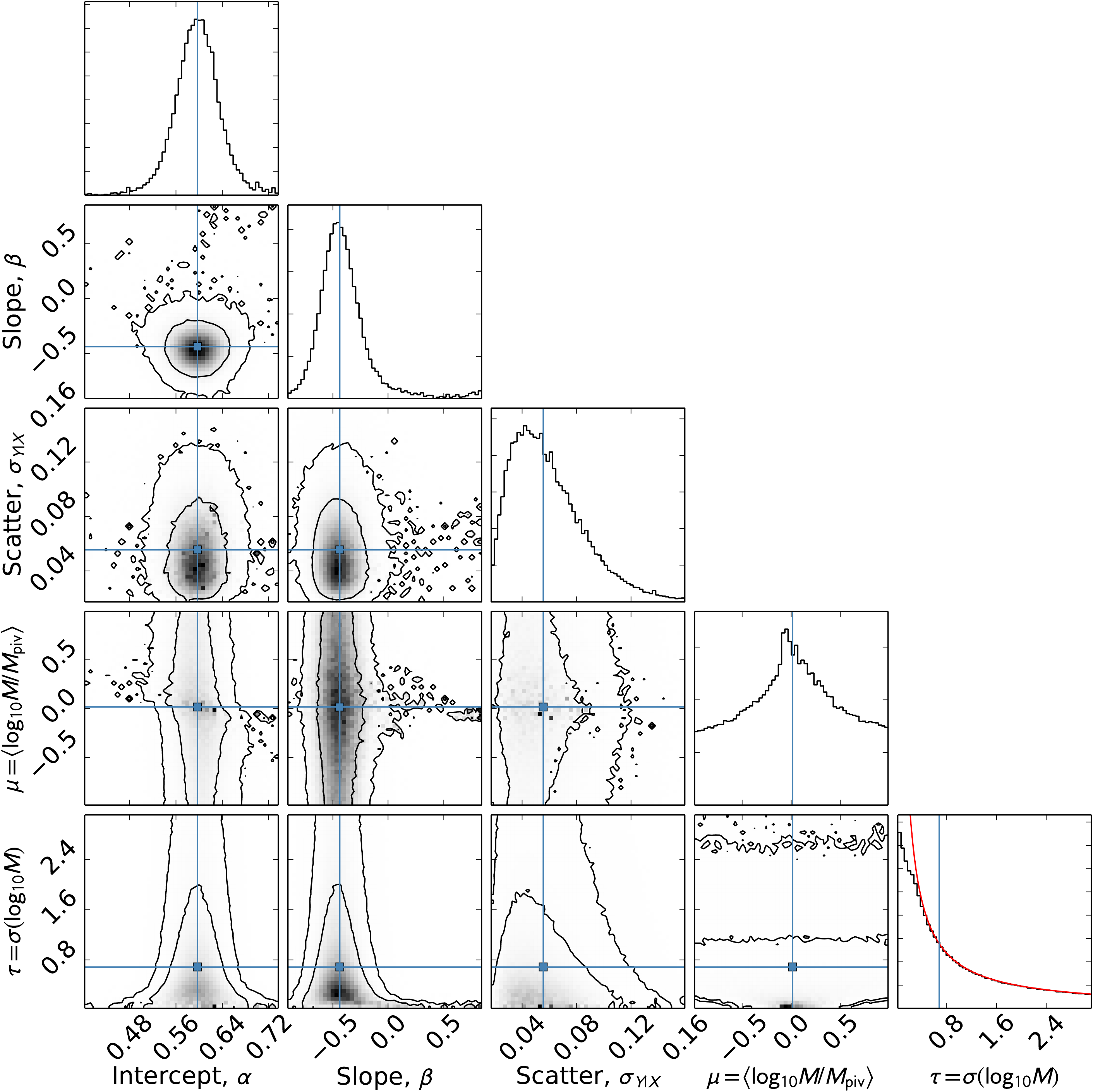} 
 \end{center}
 \caption{\label{fig:cMmodel}
Constraints on the regression parameters
 $(\alpha,\beta,\sigma_{Y|X},\mu,\tau)$ 
of the $c$--$M$ relation (Section \ref{subsec:regression}) obtained from
 {\em HST} and Subaru lensing observations of 16 CLASH X-ray-selected
 clusters, showing marginalized one-dimensional distributions and
 two-dimensional 68\% and 95\% limits. For each parameter, the blue
 solid line shows the biweight central location ($C_\mathrm{BI}$) of the
 marginalized one-dimensional posterior distribution.
For the $\tau$ parameter, the prior probability distribution function is
 shown by the red line.
}
\end{figure}

Figure \ref{fig:cMmodel} shows the two-dimensional marginalized
posterior distributions derived for all pairs of the regression
parameters $(\alpha,\beta,\sigma_{Y|X}, \mu,\tau)$.
The marginalized constraints ($C_\mathrm{BI}\pm S_\mathrm{BI}$; 
Section \ref{subsec:mass}) on the intercept, slope, and intrinsic
scatter are 
$\alpha=0.60 \pm 0.04$, 
$\beta=-0.44 \pm 0.19$,
and
$\sigma_{Y|X}=0.056\pm 0.026$ (or, a natural logarithmic scatter
of $0.13\pm 0.06$).
The posterior distributions show a long tail extending toward 
positive values of $\beta$, corresponding to shallower slopes
for the $c$--$M$ relation. This tail is associated with small values of
the $\tau$ parameter that describes the width of the mass probability
distribution ${\cal P}(X)$. Hence, accounting for the nonuniformity of  
the mass probability distribution is crucial for making 
inference of the underlying $c$--$M$ relation for the CLASH sample. 
We have checked that the tail in the posterior distribution of $\beta$
disappears if ${\cal P}(X)$ is assumed to be uniform.
We see from Figure \ref{fig:cMmodel} that the constraint on the $\tau$
parameter is dominated by the prior especially at $\tau\simgt 0.5$ and
that the data are not informative enough to determine the dispersion of
the intrinsic mass distribution.

\subsection{Comparison with Predictions for the CLASH Survey}
\label{subsec:cM_CLASH}


\begin{figure*}[!htb] 
 \begin{center}
  \includegraphics[width=0.7\textwidth,angle=0,clip]{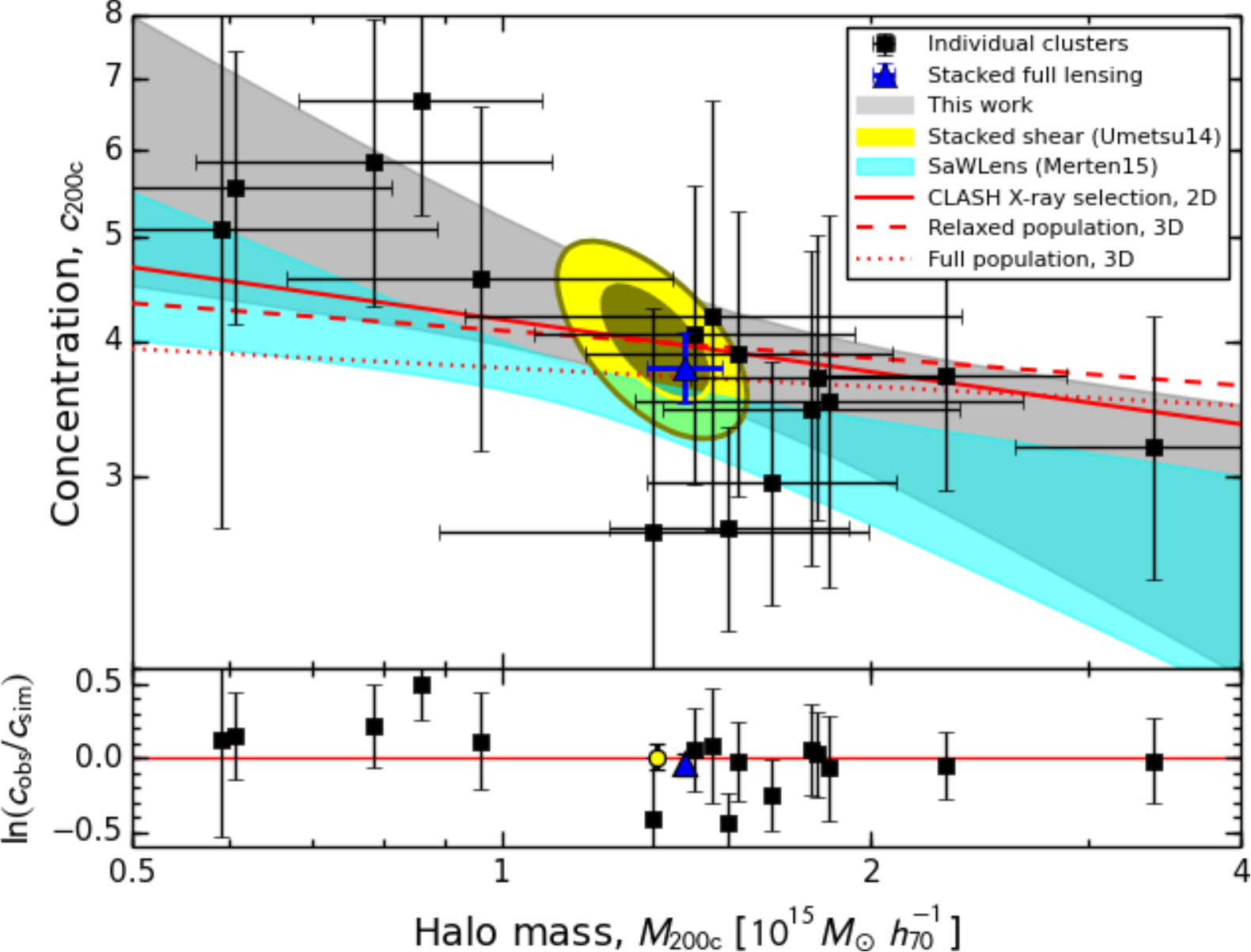} 
 \end{center}
 \caption{\label{fig:cMplot}
Upper panel: concentration--mass relation for the CLASH X-ray-selected
 subsample of 16 clusters derived from a joint analysis of {\em HST} and
 Subaru lensing data sets.
The black squares with error bars represent the measured parameters and their
 $1\sigma$ uncertainties for individual clusters. 
The gray shaded region shows the $1\sigma$ confidence region of the
 CLASH $c$--$M$ relation ($z=0.34$) from our Bayesian regression.
The blue triangle shows the best-fit parameters from a halo-model fit
 (NFW+LSS (i) in Table \ref{tab:kmodel}) to the ensemble-averaged
 surface mass density profile, $\llangle \bSigma\rrangle$ 
(Figure \ref{fig:kmodel}). 
The yellow contours represent the $1\sigma$ and $2\sigma$ confidence
 regions determined from the stacked shear-only analysis of the same sample 
\citep{Umetsu2014clash}. 
The cyan-shaded band shows the $1\sigma$ uncertainty on the CLASH
 $c$--$M$ relation obtained by \citet{Merten2015clash}. 
The red-solid line represents the theoretical expectation from numerical
 simulations accounting for the projection effects and the CLASH selection
 function based on X-ray morphology \citep{Meneghetti2014clash}.
The red-dashed and red-dotted lines show the intrinsic three-dimensional
 $c$--$M$ relations for the relaxed and full populations, respectively 
 \citep{Meneghetti2014clash}. 
The lower panel shows, for each case, the ratio between the measured
 concentration and the predicted value using the $c(M,z)$ relation for
 the CLASH X-ray-selected population (red-solid line in the upper panel). 
}
\end{figure*} 

In Figure \ref{fig:cMplot}, we summarize in the $c$--$M$ plane 
our regression results obtained for our 16 CLASH X-ray-selected
clusters. 
A detailed comparison with the {\sc SaWLens} results
\citep{Merten2015clash} is presented in Section
\ref{subsubsec:cM_SaWLens}.

Understanding the selection function and observational biases arising
from projection effects is crucial when interpreting the $c$--$M$
relation derived from lensing observations.  For the CLASH 
sample, this has been addressed in detail by
\citet[][]{Meneghetti2014clash} using a sample of  
$\sim 1400$ cluster-sized halos ($0.25\le z\le 0.67$) selected from the
MUSIC-2 nonradiative hydrodynamical $N$-body simulations
\citep{Sembolini2013music2} in a $\Lambda$CDM universe
($h=0.7,\Omega_\mathrm{m}=0.27,\Omega_\Lambda=1-\Omega_\mathrm{m},\sigma_8=0.82$).

\citet{Meneghetti2014clash} characterized a sample of halos that follows
the CLASH selection function based on X-ray morphological regularity.
Their results suggest that the CLASH X-ray-selected subsample is
prevalently composed of relaxed clusters ($\sim 70\%$) and largely free
of orientation bias (Section \ref{sec:stack}).
Another important implication is that this subsample is expected to have
a very small scatter in concentration because of the high degree of
regularity in their X-ray morphology.  

\citet{Meneghetti2014clash} find that the mean two-dimensional
concentration of the CLASH X-ray-selected sample is expected to be $\sim
11\%$ higher than that for the full population of clusters
\citep{Meneghetti2014clash}. 
According to their simulations, we expect that the NFW concentrations of
this sample determined from 
noise-free $\Sigma$ profiles are in the range 
$3\simlt c_\mathrm{200c} \simlt 6$,
with a mean (median) value of 3.87 (3.76) and a standard deviation of
0.61.  
The distribution follows a power-law relation of the form 
$c\propto M^{-0.160\pm 0.108} \times (1+z)^{-0.668\pm 0.341}$
with a normalization of 
$c|_{z=0.34}= 3.96\pm 0.14$ at $10^{15}M_\odot\,h^{-1}$.
This model, shown as the red solid line in Figure \ref{fig:cMplot}, is
in excellent agreement with our regression results (gray shaded area) with
$c|_{z=0.34}=3.95\pm 0.35$ at $10^{15}M_\odot\,h^{-1}$,
as well as with the stacked lensing measurements:
$c_\mathrm{200c}=3.79^{+0.30}_{-0.28}$ (blue triangle; Section
\ref{sec:stack}) from our stacked 
lensing analysis
and
$c_\mathrm{200c}=4.01^{+0.35}_{-0.32}$ (yellow contours) from the
stacked shear-only analysis of \citet{Umetsu2014clash}.
The derived slope of $\beta=-0.44\pm 0.19$ is somewhat steeper than,
but consistent within errors with, the predicted values of 
$\beta=-0.160\pm 0.108$ \citep{Meneghetti2014clash}.
The intrinsic scatter, $\sigma_{Y|X}=0.056\pm 0.026$, is in agreement
with the expected standard deviation of $0.61$ around the mean 
$3.87$ \citep{Meneghetti2014clash}, corresponding to
$\sigma(\log_{10}c_\mathrm{200c})\sim 0.07$, or
$\sigma(\ln c_\mathrm{200c})\sim 0.16$.
This is significantly smaller than the typical intrinsic scatter
predicted for the full (relaxed) population of halos,
$\sigma(\log_{10} c_\mathrm{200c})\sim 0.14$--$0.16$ 
($0.1$--$0.12$) \citep{2007MNRAS.381.1450N,Duffy+2008,Bhatt+2013}.


\section{Discussion}
\label{sec:discussion}

\subsection{Systematic Errors}
\label{subsec:sys}

As described in Section \ref{sec:method}, we have accounted for various
sources of errors associated with our strong-lensing, weak-lensing shear
and magnification measurements.   
All these errors are encoded in the measurement uncertainties
$(\sigma_+,\sigma_\mu,\sigma_M)$ that enter the joint likelihood
analysis (Section \ref{subsec:bayesian}) and contribute to the posterior
covariance matrix $C^\mathrm{stat}$ of the mass profile solution
$\bs=\bSigma/\Sigma_\mathrm{\infty,c}$.  
In particular, our magnification bias analysis (Section
\ref{subsec:magbias}) accounted for spurious large-scale variations of
the red galaxy counts 
($\sigma_\mu^\mathrm{sys}$), as well as the angular clustering
 ($\sigma_\mu^\mathrm{int}$) and Poisson error
 ($\sigma_\mu^\mathrm{stat}$) contributions. 
The fractional area $f_\mathrm{mask}$ lost to cluster members,
foreground objects, and defects was calculated as a function of
clustercentric radius and corrected for in the source counts according
to Equation (\ref{eq:nb}). In our mass profile analysis, 
the total covariance matrix $C$ includes additional contributions from
the residual mass-sheet uncertainty $C^\mathrm{sys}$, 
the cosmic noise $C^\mathrm{lss}$, and
the intrinsic variations $C^\mathrm{int}$ of the cluster
signal due to the $c$--$M$ variance,  halo asphericity, and the presence
of correlated halos (Section \ref{subsubsec:cmat}).   

Additionally, we quantified potential sources of systematic errors in
our weak-lensing shear+magnification measurements as follows (Section
\ref{sec:method}):  
(1) dilution of the weak-lensing signal by cluster members
($2.4\%$, Section \ref{subsec:back}),
(2) photometric-redshift bias in the mean depth estimates $\langle\beta\rangle$
($0.27\%$, Section \ref{subsec:back}),
and
(3) shear calibration uncertainty ($5\%$, Section \ref{subsec:gt}).
These systematic errors scale approximately linearly with the cluster mass and add
to $5.6\%$
in quadrature.
For the absolute calibration of cluster masses, one needs to take into 
account the systematic bias due to mass model uncertainties.
\citet{Meneghetti2014clash} find that spherical NFW masses of
 cluster-sized halos estimated from their surface mass densities are
 biased low on average by 5\% and that the bias is significantly reduced
 for relaxed halos (1\%--2\%) because they are more spherical than
 unrelaxed ones. According to their simulations, the degree of negative
 bias expected for our sample dominated by relaxed clusters is $3\%$
 (Massimo Meneghetti, private communication).  
Hence, the systematic
 uncertainty in the absolute mass calibration is estimated to be 
$\simeq 6\%$. This implies that the total uncertainty in the absolute mass
 calibration with the sample of 20 clusters is
$\sqrt{0.28^2/20+0.06^2}\simeq 9\%$ at $\Delta_\mathrm{c}=200$ (see
 Section \ref{subsec:mass}).

\begin{figure}[!htb] 
 \begin{center}
  \includegraphics[width=0.45\textwidth,angle=0,clip]{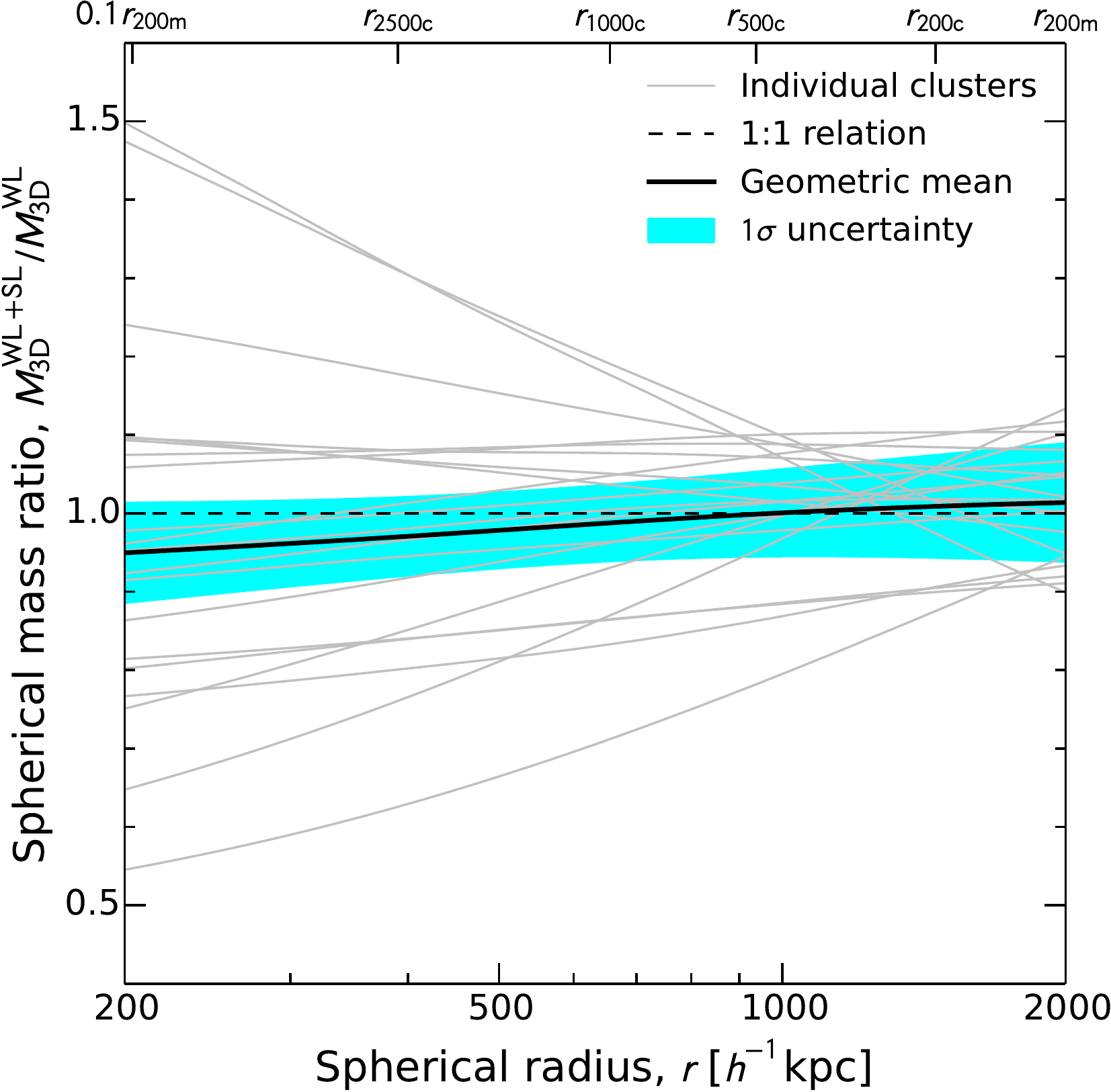} 
 \end{center}
 \caption{\label{fig:mratio}
Ratio of cluster masses $M_\mathrm{3D}(<r)$ from NFW fits to the
 $\Sigma$ profile obtained in our weak+strong lensing analysis (Figure 
 \ref{fig:kappa}) and to that from the weak-lensing analysis by
 \citet{Umetsu2014clash}.  
The results are shown for our full sample of 20 clusters (gray lines). 
The black line and the cyan-shaded area show the geometric-mean mass
 ratio and its $1\sigma$ uncertainty, respectively. The dashed
 horizontal line marks the 1:1 relation. 
This comparison shows that, adding the inner strong-lensing information
 to the weak-lensing constraints has a substantial impact on the
 individual cluster mass determinations at $r\simlt r_\mathrm{2500c}$.
When averaged over the 20 clusters, our ensemble analysis shows no
 significant evidence for a systematic bias between weak-lensing-only
 and weak+strong-lensing measurements.  
} 
\end{figure}

\subsection{Impact of Adding HST Lensing Data}
\label{subsec:msys}

A joint analysis of multi-scale, multiple lensing probes makes it
possible to improve the precision of cluster mass profile
reconstructions over a wide range of clustercentric radii and to
self-calibrate systematics as well as observational parameters that
describe the intrinsic properties of source populations 
\citep{Rozo+Schmidt2010,Umetsu2013}.
When the wide-field CLASH weak-lensing data are combined with the inner
{\em HST} lensing constraints ($N_\mathrm{SL}=4$), the central
weak-lensing bin $\Sigma(<0.9\arcmin)$ of \citet{Umetsu2014clash} is
resolved into $5$ radial bins of the surface mass density. 
We find a significant improvement of $\sim 45\%$ on average in terms of
the total S/N (Equation (\ref{eq:SNR})) from adding the {\em HST}
lensing information to the wide-field weak-lensing data. 

The improved sensitivity and resolution at $10\arcsec$--$40\arcsec$
have also allowed us to determine the inner characteristic radius
$r_{-2}$ and the halo concentration   
$c_\mathrm{200c}=r_\mathrm{200c}/r_{-2}$ for each individual cluster 
(Tables \ref{tab:cm}). 
\citet{Umetsu2014clash} did not attempt to determine $c_\mathrm{200c}$
for each cluster because the weak-lensing profiles of individual 
clusters are highly degenerate in $M_\mathrm{200c}$ and
$c_\mathrm{200c}$, which can potentially 
bias the slope of the $c$--$M$ relation determined from 
weak lensing \citep{Hoekstra+2011,Du+Fan2014}.
This is particularly the case for high-$z$, low-mass systems, 
for which the characteristic profile curvature around 
$r_{-2}$ is unconstrained by noisy and sparse weak-lensing
data. For such clusters, the constraints on $c_\mathrm{200c}$ are
essentially imposed by prior information. 
We note again that the $C^\mathrm{int}$ contribution to the total
covariance matrix $C$ was not included in the individual cluster mass
measurements of \citet[][Table 6]{Umetsu2014clash}, so that their mass
measurement errors were underestimated with respect to this work
(Section \ref{subsec:mass}).

A multi-probe approach combining complementary probes of cluster lensing
 allows us to test the consistency and robustness of cluster mass
 measurements \citep{Umetsu2013}.
Now we compare our weak+strong lensing mass estimates derived for our
full sample of 20 CLASH clusters with the weak-lensing masses of
 \citet{Umetsu2014clash} to 
assess the level of systematic uncertainties in the ensemble mass
calibration.   

In Figure \ref{fig:mratio}, we show for each cluster the 
weak+strong to weak lensing mass ratio,
$M^\mathrm{WL+SL}_\mathrm{3D}/M^\mathrm{WL}_\mathrm{3D}$,
 as a function of spherical radius $r$.
The results are shown in the range $r=[200,2000]$\,kpc\,$h^{-1}$ 
($0.1\simlt r/r_\mathrm{200m}\simlt 1$)
where the weak-lensing mass measurements were obtained by
\citet[][see their Figure 4]{Umetsu2014clash}.  
At each cluster radius, we compute the unweighted average of the mass
ratios for our sample using geometric averaging \citep{Donahue2014clash,Umetsu2014clash}.\footnote{
The geometric mean $\langle X\rangle$ is defined as
$\langle X\rangle=(\prod_{n=1}^N X_n)^{1/N}=\exp\left({\frac{1}{N}\sum_{n=1}^N\ln{X_n}}\right)$, so that
$\langle Y/X\rangle=1/\langle X/Y\rangle$ for the ratios of samples
$X$ and $Y$.} 
Figure \ref{fig:mratio} shows that the ensemble-averaged mass ratio 
$\langle M^\mathrm{WL+SL}_\mathrm{3D}/M^\mathrm{WL}_\mathrm{3D}\rangle$
is consistent with unity within the
errors at all cluster radii. 
In particular, the mass offset is within $2\%$ at 
$r\simgt 600$\,kpc\,$h^{-1}\sim 1.5r_\mathrm{2500c}$.
We see a trend for the average ratio to decrease toward the center,  
reaching $0.95\pm 0.06$ at $r=200$\,kpc\,$h^{-1}$.
This indicates that the {\em HST} analysis \citep{Zitrin2015clash}
favors relatively low central masses and hence low halo concentrations
although the difference is not significant,
$|\langle M_\mathrm{3D}^\mathrm{WL+SL}/M_\mathrm{3D}^\mathrm{WL}\rangle-1| \le 5\%\pm 6\%$,
compared to the sensitivity limit with 20 clusters.
This level of mass offset is smaller than, but consistent with,
the value $8\%\pm 9\%$ based on the shear--magnification consistency
test of \citet{Umetsu2014clash}. 

On the basis of these consistency checks,
the residual systematic uncertainty in the ensemble mass calibration 
is estimated to be of the order $\sim 5\%$--$8\%$ in the one-halo regime,
$r=[200,2000]$\,kpc\,$h^{-1}$.
This however implies that, on an individual cluster basis, there is a
large scatter of $\sim 20\%$--$40\%$ between different reconstruction
methods that use different combinations of data. 
The level of uncertainty in the ensemble mass calibration (5\%--8\%)
empirically estimated using different combinations of lensing probes is
in agreement with the systematic uncertainty in the absolute mass
calibration (6\%) assessed in Section \ref{subsec:sys}. 

\subsection{Ensemble Calibration of Cluster Masses}
\label{subsec:mcalib}


\begin{figure*}[!htb] 
 \begin{center}
 $
 \begin{array}
  {c@{\hspace{.1in}}c@{\hspace{.1in}}c}
  \includegraphics[width=0.40\textwidth,angle=0,clip]{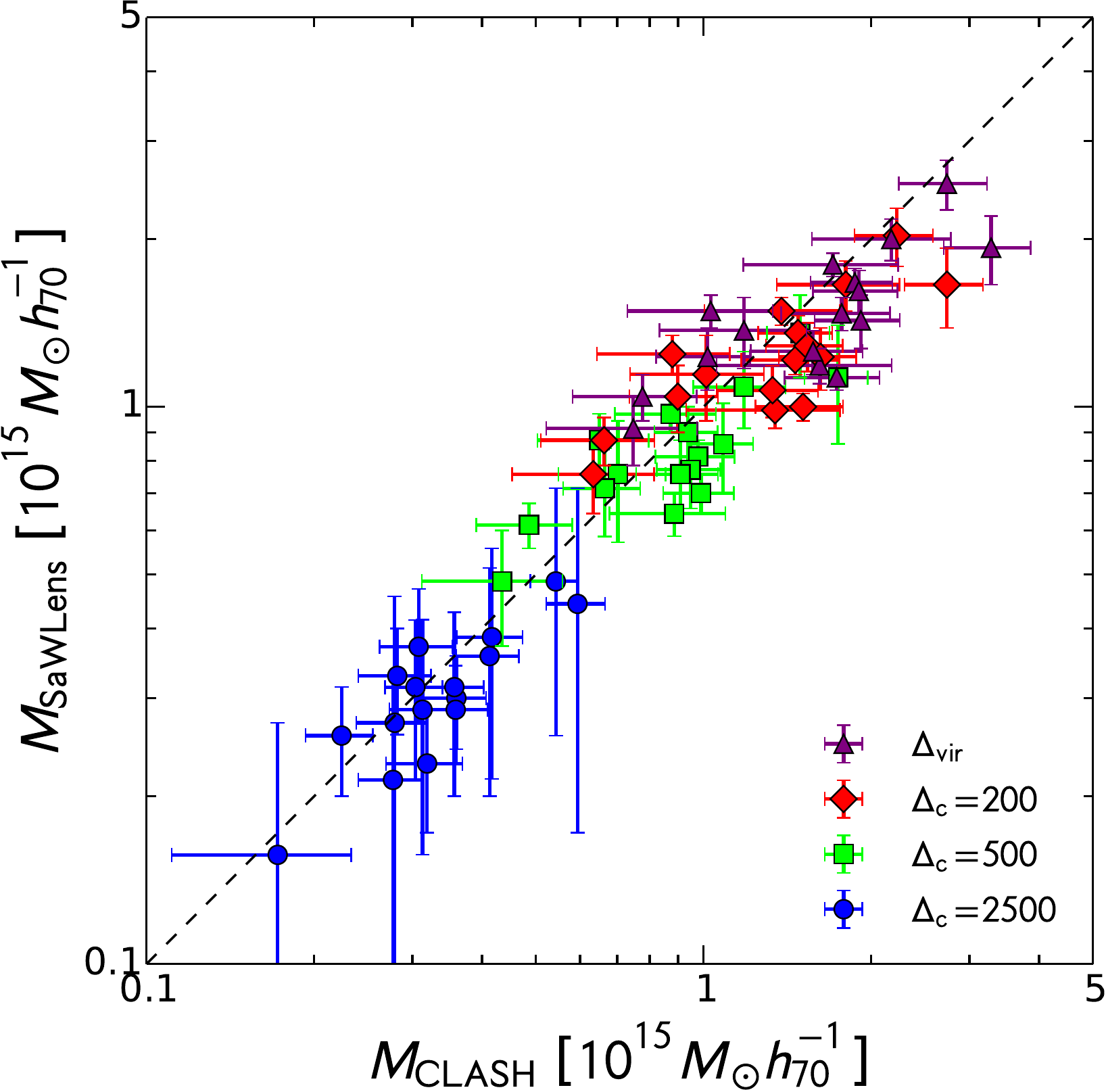} &
  \includegraphics[width=0.40\textwidth,angle=0,clip]{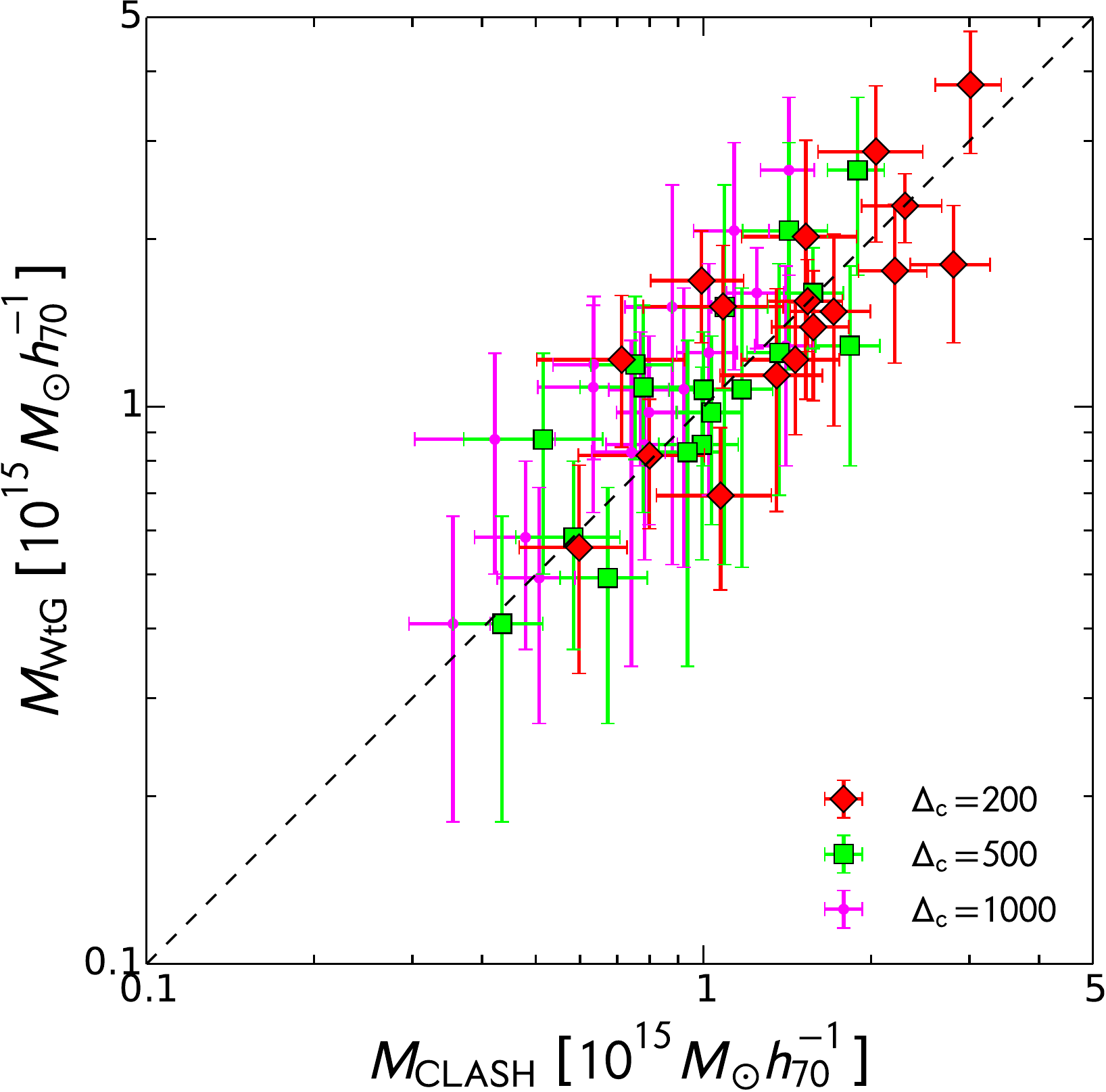} \\
 \end{array}
 $
 $
 \begin{array}
  {c@{\hspace{.1in}}c@{\hspace{.1in}}c}
  \includegraphics[width=0.40\textwidth,angle=0,clip]{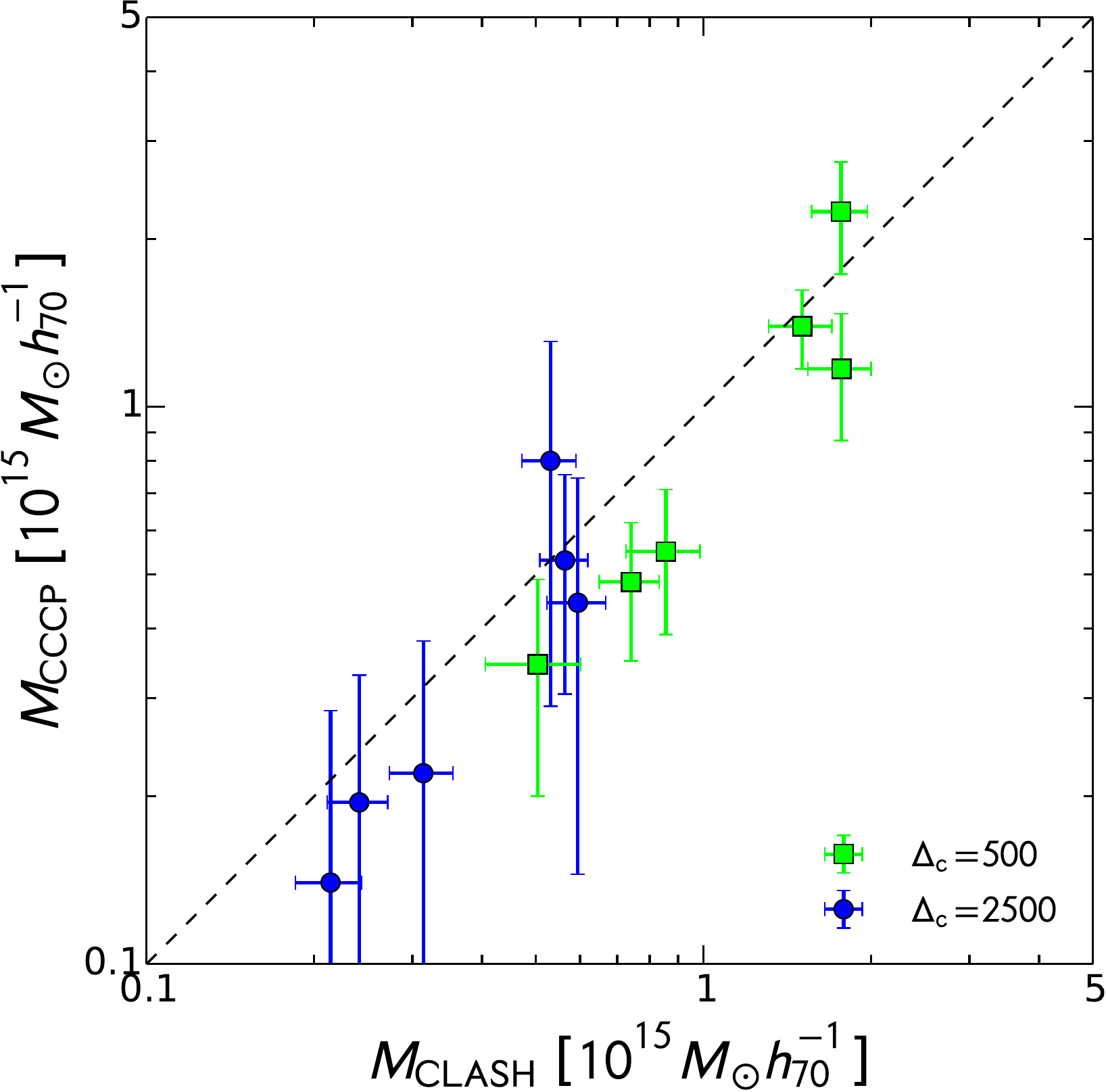}&
  \includegraphics[width=0.40\textwidth,angle=0,clip]{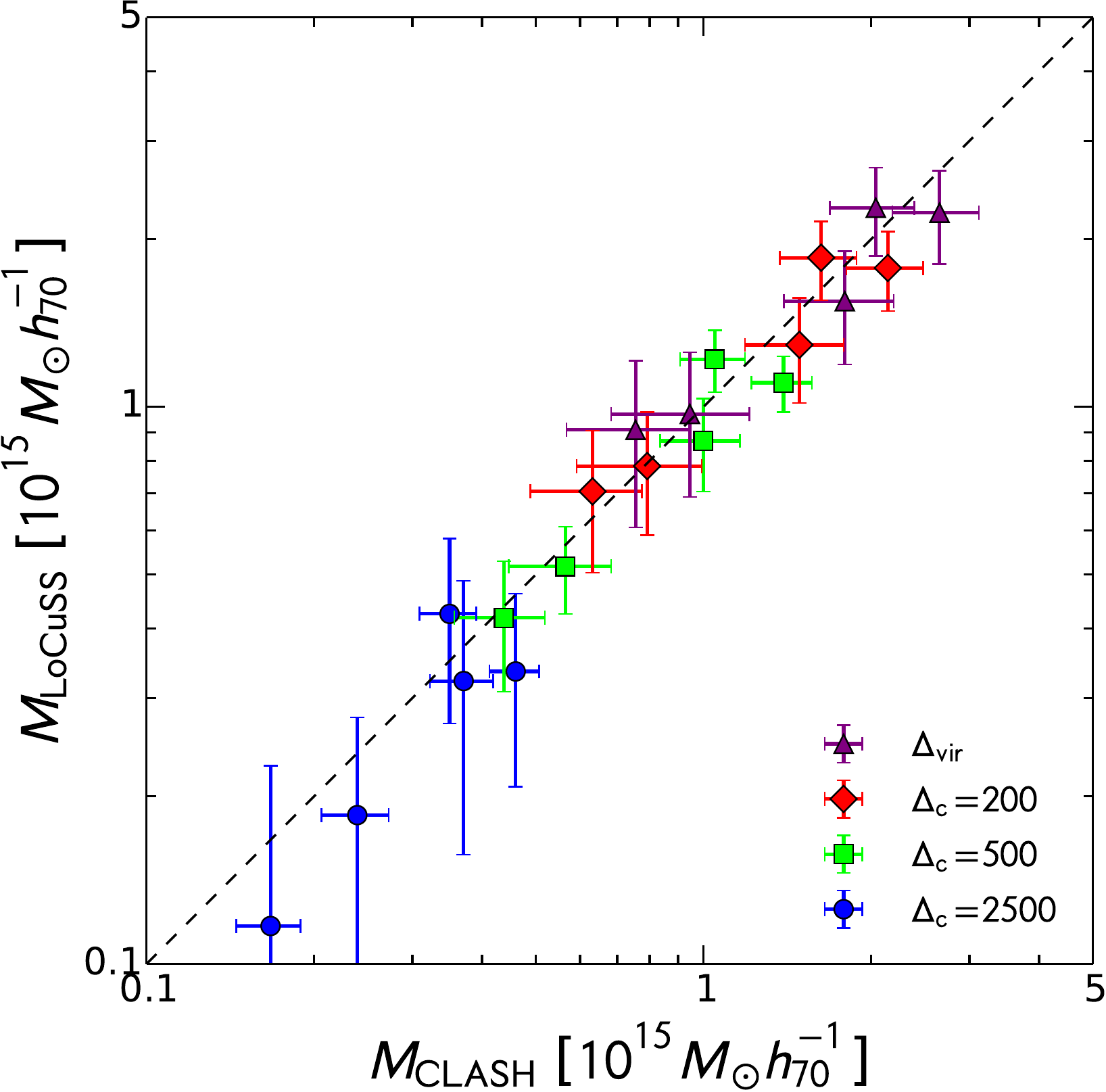}
 \end{array}
 $
 \end{center}
\caption{\label{fig:mcomp}
Comparison of our 
weak+strong lensing mass measurements
 ($M_\mathrm{CLASH}$) of 20 clusters to results from
 \citet[][{\sc SaWLens}]{Merten2015clash}, 
 \citet[][WtG]{WtG3}, \citet[][CCCP]{Hoekstra2015CCCP}, and 
 \citet[][LoCuSS]{Okabe+Smith2015}.
For each comparison, we measure the mass of clusters
within characteristic overdensity radii $r_\Delta$ of the respective
 work assuming the spherical NFW density profile (Section \ref{subsec:mcalib}). 
The dashed line shows the one-to-one relation.
}
\end{figure*}

CLASH provides a sizable sample 
(20 clusters at $0.19<z<0.69$, $\overline{z}=0.377$) 
for the calibration of the high end of the cluster mass function. 
In principle, weak lensing can yield unbiased mass estimates (assuming
sphericity) for a sample of clusters that is largely free of orientation 
bias \citep{Meneghetti2014clash}.
In practice, however, lensing mass measurements can be subject to
various (known and unknown) systematic effects, 
as discussed in Section \ref{subsec:sys}.
In our early work \citep[e.g.,][]{BTU+05,UB2008}, we established
that the dominant source of systematic effects in cluster weak lensing
is the contamination of background galaxy samples by cluster members,
which can lead to a substantial underestimation of the true lensing
signal. 
This point has been supported by recent observations
\citep{Okabe+2013,WtG3,Hoekstra2015CCCP} through systematic mass
comparisons between different surveys that use different approaches to
measuring weak lensing.

Here we compare our cluster mass estimates 
(Tables \ref{tab:cm} and \ref{tab:mass}) with 
those obtained from other cluster lensing surveys that overlap with our
sample, namely 
the {\em  Weighing the Giants} program \citep[WtG;][]{WtG3},
the Canadian Cluster Cosmology Project \citep[CCCP;][]{Hoekstra2015CCCP},
and the LoCuSS \citep[][LoCuSS]{Okabe+Smith2015},\footnote{\href{http://www.sr.bham.ac.uk/locuss}{http://www.sr.bham.ac.uk/locuss}} 
as well as with those from the {\sc SaWLens} analysis of
\citet{Merten2015clash}.    
The WtG sample is a representative X-ray luminous subset of the {\em
ROSAT} All-sky Survey (RASS) clusters at $0.15 < z < 0.7$, 
with a median redshift of $\overline{z}=0.387$.
The WtG clusters span a wide range of dynamical states, as well as of
redshifts \citep{WtG1}. 
The CCCP sample is a mixture of 
X-ray luminous clusters for which archival $B$- and $R$-band
observations made with the CFH12k camera on the 3.6\,m
Canada--France--Hawaii Telescope (CFHT) were available \citep{Hoekstra2007}
and
a temperature-selected subset of clusters drawn from the {\em ASCA}
survey \citep[$k_\mathrm{B}T_X>5$\,keV,][]{Hoekstra2012CCCP}, 
spanning the range $0.15 < z < 0.55$
($\overline{z} = 0.233$). 
The LoCuSS sample is drawn from the RASS catalogs at $0.15<z<0.3$  
($\overline{z} = 0.229$) 
and is approximately X-ray luminosity limited \citep{Okabe+Smith2015}.
The observed X-ray temperatures of the LoCuSS clusters are 
$k_\mathrm{B}T_X\simgt 5$\,keV \citep[][see their Figure 5]{Martino2014locuss}.
The LoCuSS sample is selected purely on the X-ray luminosity,
ignoring other physical parameters and relaxation properties
\citep{Smith2015locuss}.   

In all cases, the cluster masses are measured assuming a spherical NFW
halo (Section \ref{subsec:mass}). 
The WtG and LoCuSS mass measurements are based on weak-lensing
observations with Subaru/Suprime-Cam, whilst the CCCP survey uses
weak-lensing data taken with CFHT.  
The CCCP and LoCuSS surveys used the BCG as the cluster center as done
in our work (Section \ref{subsec:sample}), whereas the WtG survey
adopted the X-ray centroid as the cluster center \citep{WtG1}.
We note that, for all clusters in our sample, the mass measurements
presented in Tables \ref{tab:cm} and \ref{tab:mass} are insensitive to
the choice of the cluster center as discussed in Section 
\ref{subsec:sample}.
Another key difference is that the LoCuSS and CLASH surveys controlled
contamination of their background galaxy samples at the $\simlt 2\%$
level by imposing stringent color cuts (Section \ref{subsec:back}),
albeit with increased shot noise, whereas the WtG and CCCP surveys
boosted the diluted shear signal to correct statistically for
contamination, 
by assuming that the observed number density profile of a pure
background galaxy sample is flat. This correction, referred to as a
boost factor, is not valid in general as it ignores the depletion or
enhancement of the number density of background galaxies due to
magnification bias \citep{Umetsu2014clash,Okabe+Smith2015,Ziparo2015locuss,Chiu2015magbias}. 

In the following, we compare cluster masses between two studies 
by using the same aperture radii to avoid aperture mismatch problems.
These comparisons are limited to those overdensity radii ($r_\Delta$)
where the fitting ranges typically overlap.
The results of the comparisons are shown in Figure \ref{fig:mcomp}.
For each case, we calculate the mass ratios for the overlap sample using
the unweighted geometric mean  (Section \ref{subsec:msys}), unless
otherwise noted.

\subsubsection{CLASH: \citet{Merten2015clash}}
\label{subsubsec:sawlens}
 
There are 16 clusters in common with the CLASH {\sc SaWLens} analysis of
\citet[][]{Merten2015clash}. These are all CLASH X-ray-selected
clusters. 
\citet{Merten2015clash} measure masses by reconstructing two-dimensional
convergence maps of individual clusters, binning the maps into $\Sigma$
profiles, and fitting these profiles with an NFW model within
$2\,$Mpc\,$h^{-1}$ ($R\simlt r_\mathrm{vir}$),
closely following the procedure suggested by \citet[][]{Meneghetti2014clash}.  
An important difference between the data used by \citet{Merten2015clash}
and the data used here 
is the availability of azimuthally integrated magnification
constraints \citep{Umetsu2014clash}.\footnote{\citet{Merten2015clash} and
\citet{Zitrin2015clash} use identical sets of {\em HST} lensing 
constraints (i.e., {\em HST} shear catalogs plus locations and redshifts
of multiple images) as input for respective mass reconstructions.
\citet{Merten2015clash} simultaneously combine the {\em HST} lensing
constraints and ground-based shear catalogs of
\cite{Umetsu2014clash}. In this work, lensing constraints are combined
a posteriori in the form of radial profiles according to the procedure
described in Section \ref{subsec:bayesian}.}  

This comparison shows that, on average, 
the {\sc SaWLens} masses are $7\% \pm 6\%$ lower than our masses at 
$\Delta_\mathrm{c}=\Delta_\mathrm{vir}$, $200$, and $500$; 
their masses are $9\% \pm 11\%$ lower than our masses at
$\Delta_\mathrm{c}=2500$. 
This difference is smaller than the systematic mass offset of 
$\sim 10\%\pm 5\%$ at $r=0.5$\,Mpc, $1$\,Mpc, and $r_\mathrm{500c}$ 
found between the CLASH weak-lensing \citep{Umetsu2014clash} and
{\sc SaWLens} \citep{Merten2015clash} mass profiles
\citep[see][Table 6]{Donahue2014clash}.
Hence, combining the central {\em HST}-lensing and outer weak-lensing
data has reduced the discrepancy with respect to the {\sc SaWLens}
results.   
We find that this remaining discrepancy can be attributed to the three
outliers discussed in Section \ref{subsec:rec},
which correspond to clusters at the lowest Galactic latitudes of the
overlap sample (MACSJ1931.8¡Ý2635, MACSJ0744.9+3927) and those at the
highest redshifts (RXJ1347.5¡Ý1145, MACSJ0744.9+3927) which exhibit
complex mass distributions with a high degree of substructure
\citep{Postman+2012CLASH,Merten2015clash}.
When excluding the three outliers, we find that the {\sc SaWLens} mass
estimates ($M_\mathrm{SaWLens}$) are statistically in excellent
agreement with our results ($M_\mathrm{CLASH}$) at all overdensities: 
$\langle M_\mathrm{SaWLens}/M_\mathrm{CLASH}\rangle = 1.01\pm 0.07,1.00\pm 0.07, 0.99\pm 0.07$,  
and $0.95 \pm 0.12$ at 
$\Delta_\mathrm{c}=\Delta_\mathrm{vir}, 200, 500$, and $2500$,
respectively. 
We note that this agreement is achieved in spite of using
substantially different reconstruction methods even though the data used  
are largely common to the two analyses.

\subsubsection{The WtG Project}
\label{subsubsec:wtg}

The {\em WtG} collaboration conducted weak-lensing shear mass
measurements for 51 X-ray-selected luminous clusters at  
$0.15\simlt z\simlt 0.7$ ($\overline{z}=0.387$)
using deep multi-color Subaru/Suprime-Cam and
CFHT/MegaPrime optical imaging \citep{WtG1,WtG3}. 
Their cluster sample includes the majority of the CLASH clusters.
There are 17 clusters in common between the two studies, both of which
use Subaru data.
The overlap sample includes 
14 CLASH X-ray-selected clusters and 3 high-magnification clusters.
\citet{WtG3} derived
cluster masses from NFW fits to reduced tangential shear profiles over 
the radial range $0.75$--$3$\,Mpc\,$h_{70}^{-1}$ 
($R\simgt 0.8r_\mathrm{1000c}$)
assuming a fixed concentration parameter of $c_\mathrm{200c}=4$ for all
clusters. 
In contrast, we have measured masses from NFW fits to surface mass
density profiles within $R\le 2$\,Mpc\,$h^{-1}\simeq 2.9$\,Mpc$\,h_{70}^{-1}$,
allowing both $M_\mathrm{200c}$ and
$c_\mathrm{200c}$ as free parameters (see Section
\ref{subsec:mass}). This fitting procedure is the same as that adopted
by \citet{Umetsu2014clash} and \citet{Merten2015clash}.

In the upper-right panel of Figure \ref{fig:mcomp} we compare
shear-only masses ($M_\mathrm{WtG}$) of \citet{WtG3}
and
our full-lensing masses ($M_\mathrm{CLASH}$), obtaining good agreement.
We find that the WtG masses are 
$3\% \pm 9\%$, $7\% \pm 12\%$, and $7\% \pm 12\%$ higher than 
our masses at $\Delta_\mathrm{c}=200$, $500$, and $1000$, respectively. 
We see a trend of increasing mass offset with increasing overdensity
(or, decreasing overdensity radius $r_\Delta$).
However, the offsets are well within the $1\sigma$ errors and not
statistically significant.

\subsubsection{The CCCP}
\label{subsubsec:cccp}

The CCCP conducted weak-shear lensing mass measurements for a sample of
52 clusters at $0.15<z<0.55$ 
($\overline{z} = 0.233$) 
on the basis of CFHT observations. We have 6 clusters in common with
the CCCP project, including 5 CLASH X-ray-selected clusters (Abell
383, Abell 209, Abell 2261, Abell 611, RXJ1347.5$-$1145) and one
high-magnification cluster (MACSJ0717.5$+$3745).
\citet{Hoekstra2015CCCP} measure NFW masses from CFHT 
reduced tangential shear 
measurements within $0.5$--$2$\,Mpc\,$h_{70}^{-1}$ 
($r_\mathrm{2500c}\simlt R \simlt 1.5 r_\mathrm{500c}$),
assuming the $c$--$M$ relation of \citet{Dutton+Maccio2014} (see Section
\ref{subsubsec:cM_LCDM}).  

We find that this mass comparison is somewhat sensitive to weighting
schemes because of the large observational uncertainties in the
CFHT-based CCCP masses
(with a typical uncertainty of $\sim 34\%$)
and of the small number of clusters. Taking the
error-weighted geometric mean (down-weighting clusters with noisy mass
measurements), we find that the CCCP masses are on average 
$16\% \pm 10\%$ and $9\% \pm 24\%$ lower than our
masses at $\Delta_\mathrm{c}=500$ and $2500$,
respectively. With limited statistics, we do not find statistically
significant differences between the CCCP and our masses.
We note that \citet{Hoekstra2015CCCP} 
obtained an excellent
agreement between the Subaru weak-lensing masses of
\citet{Umetsu2014clash} and their masses  measured from the Subaru
imaging data processed by the CLASH collaboration.\footnote{\href{https://archive.stsci.edu/prepds/clash/}{https://archive.stsci.edu/prepds/clash/}}
\citet{Hoekstra2015CCCP} found that the Subaru-based CCCP masses
are on average only $\sim 2.4\%$ lower than the CLASH weak-lensing
masses of \citet{Umetsu2014clash}.
Since our mass calibration is highly consistent
with that of \citet{Umetsu2014clash} (Section \ref{subsec:msys}), a
similar improvement is expected when the Subaru-based CCCP masses are
compared to our weak+strong lensing masses.

\subsubsection{The LoCuSS}
\label{subsubsec:locuss}

The LoCuSS has carried out a
systematic weak-shear lensing analysis of a sample of 50 X-ray luminous
clusters at $0.15<z<0.3$ ($\overline{z} = 0.229$) 
based on deep two-band imaging with
Subaru/Suprime-Cam \citep{Okabe+Smith2015}.
There are 5 clusters in common between the LoCuSS and our samples
(Abell 383, Abell 209, Abell\ 2261, RXJ2129.7$+$0005, Abell 611).
Both studies use Subaru weak-shear lensing data.
To reduce biases due to noisy inner profiles, \citet{Okabe+Smith2015}
optimize the binning scheme (i.e., the fitting range and the number of bins) for
each individual cluster. 
They perform NFW fits to a suite of reduced tangential shear profiles 
that span inner radii in the range $50$--$300$\,kpc\,$h^{-1}$,
outer radii in the range $2000$--$3000$\,kpc\,$h^{-}$,
and number of bins in the range 4--8.
The NFW concentration parameter is treated as a
free parameter in their fits.

For this comparison, we obtain excellent agreement between the LoCuSS
masses ($M_\mathrm{LoCuSS}$) and our masses, even on an individual
cluster basis (see the lower-right panel of Figure \ref{fig:mcomp}).
Accordingly, 
the comparison results are insensitive to the choice of weighting
schemes.
The LoCuSS masses are on average 
$0\% \pm 15\%$, $2\%\pm 13\%$, $7\% \pm 10\%$, and $16\%\pm 22\%$ 
lower than our masses at
$\Delta_\mathrm{c}=\Delta_\mathrm{vir}$, $200$, $500$, and $2500$,
respectively.  
A better agreement is seen for lower overdensities where the constraints
are dominated by the Subaru weak-lensing measurements.
A possible explanation for this excellent agreement is the fact that
both surveys controlled contamination of the background samples at the
$\simlt 2\%$ level without employing a boost factor (Section
\ref{subsec:mcalib}) and
both used independent but very similar shape measurement algorithms.
In particular, \citet{Okabe+Smith2015} adopted a shear calibration
procedure that is nearly identical to the one developed by 
\citet{Umetsu+2010CL0024} and adopted by \citet{Umetsu2014clash} (see
Section \ref{subsec:gt}).
\citet{Okabe+Smith2015} obtained a shear calibration bias of 
$m\sim -3\%$, which is close to the value derived by
\citet[][$m\sim -5\%$]{Umetsu+2010CL0024}.
We note that adding the inner strong-lensing information 
\citep{Zitrin2015clash} to the weak-lensing constraints
\citep{Umetsu2014clash} has resulted in, on average, a decrease of the
mass estimates, especially at $r\simlt 1.5r_\mathrm{2500c}$ (see Figure
\ref{fig:mratio}; Section  \ref{subsec:msys}).


\subsection{Cluster $c$--$M$ Relation}
\label{subsec:cM_discussion}

\subsubsection{Comparison with $\Lambda$CDM Predictions from the Literature}
\label{subsubsec:cM_LCDM}

\begin{deluxetable*}{lcccrrrr}
\centering
\tabletypesize{\scriptsize}
\tablecolumns{8}
\tablecaption{
\label{tab:LCDM}
Comparison of measured and predicted concentrations for the
 CLASH X-ray-selected subsample
}
\tablewidth{0pt}
\tablehead{
 \multicolumn{1}{c}{Author} &
 \multicolumn{1}{c}{Sample} &
 \multicolumn{1}{c}{3D/2D} &
 \multicolumn{1}{c}{Function\tablenotemark{a}} &
 \multicolumn{2}{c}{$c^\mathrm{(obs)}/c^\mathrm{(pred)}$} &
 \multicolumn{1}{c}{$\chi^2$} &
 \multicolumn{1}{c}{PTE\tablenotemark{b}} 
\\
 \multicolumn{1}{c}{} &
 \multicolumn{1}{c}{} &
 \multicolumn{1}{c}{} &
 \multicolumn{1}{c}{} &
 \multicolumn{1}{c}{Average\tablenotemark{c}} &
 \multicolumn{1}{c}{$\sigma$\tablenotemark{d}} &
 \multicolumn{1}{c}{} &
 \multicolumn{1}{c}{} 
}
\startdata
Theory:\\
~~\citet{Duffy+2008} &         full & 3D & $c$--$M$ & $1.331 \pm 0.108$ & 0.334 & 22.6 & 0.046\\
~~\citet{Duffy+2008} &      relaxed & 3D & $c$--$M$ & $1.165 \pm 0.094$ & 0.290 & 13.6 & 0.399\\
~~\citet{Prada2012} &         full & 3D & $c$--$\nu$ & $0.733 \pm 0.065$ & 0.244 & 24.6 & 0.026\\
~~\citet{Bhatt+2013} &         full & 3D & $c$--$\nu$ & $1.169 \pm 0.095$ & 0.292 & 14.1 & 0.369\\
~~\citet{Bhatt+2013} &      relaxed & 3D & $c$--$\nu$ & $1.131 \pm 0.092$ & 0.277 & 12.4 & 0.494\\
~~\citet{Dutton+Maccio2014} &         full & 3D & $c$--$M$ & $1.061 \pm 0.086$ & 0.262 & 10.4 & 0.659\\
~~\citet{Meneghetti2014clash} &         full & 3D & $c$--$M$ & $1.061 \pm 0.089$ & 0.279 & 10.2 & 0.675\\
~~\citet{Meneghetti2014clash} &      relaxed & 3D & $c$--$M$ & $0.990 \pm 0.083$ & 0.249 &  9.2 & 0.760\\
~~\citet{Diemer+Kravtsov2015} & full (median) & 3D & $c$--$\nu$ & $1.021 \pm 0.083$ & 0.330 & 14.4 & 0.349\\
~~\citet{Diemer+Kravtsov2015} &  full (mean) & 3D & $c$--$\nu$ & $1.060 \pm 0.086$ & 0.326 & 13.8 & 0.391\\
~~\citet{Meneghetti2014clash} &         full & 2D & $c$--$M$ & $1.087 \pm 0.092$ & 0.336 & 13.5 & 0.413\\
~~\citet{Meneghetti2014clash} &      relaxed & 2D & $c$--$M$ & $1.040 \pm 0.086$ & 0.283 & 10.8 & 0.628\\
~~\citet{Meneghetti2014clash} &        CLASH & 2D & $c$--$M$ & $0.988 \pm 0.078$ & 0.227 &  9.6 & 0.730\\
Observations:\\
~~\citet{Merten2015clash} &        CLASH & 2D & $c$--$M$ & $1.133 \pm 0.087$ & 0.209 &  9.2 & 0.754
\enddata
\tablenotetext{a}{$c$--$M$: power-law $c(M,z)$ relation; $c$--$\nu$:
 halo concentration given as a function of peak height $\nu(M,z)$.}
\tablenotetext{b}{Probability to exceed the measured $\chi^2$ value assuming the standard $\chi^2$ probability distribution function.}
\tablenotetext{c}{Weighted geometric average of observed-to-predicted
 concentration ratios.}
\tablenotetext{d}{Standard deviation of the distribution of
 observed-to-predicted concentration ratios.} 
\end{deluxetable*}

We compare our individual cluster mass and concentration measurements
(Table \ref{tab:cm}; Figure \ref{fig:cMplot}) with predictions from
numerical simulations in the literature. 
To statistically quantify the level of agreement with a given predicted 
$c$--$M$ relation, we use frequentist measures of goodness of fit.
Specifically, for a given fixed $c(M,z)$ function, 
we evaluate the $\chi^2$ goodness of fit 
to the null hypothesis that the sample data are derived
from the model population. The $\chi^2$ statistic is then defined as 
\begin{equation}
   \chi^2 =
  \sum_n\left[\frac{\log_{10}{c_n}-\log_{10}{c(M_n,z_n)}}{\sigma_n}\right]^2,
\end{equation}
where $\sigma_n$ is the total statistical uncertainty of the $n$th cluster,
\begin{equation}
  \sigma_n^2 = \sigma^2_{\mathrm{int},0} + C_{yy,n} + \beta_0^2 C_{xx,n} -
   2\beta_0  C_{xy,n}
\end{equation}
with $\beta_0$ the mass slope of the intrinsic $c$--$M$ relation under
consideration and $\sigma_{\mathrm{int},0}$ the intrinsic scatter in 
$\log_{10}{c_\mathrm{200c}}$ at fixed mass and redshift.  For all
models, we fix 
$\sigma_{\mathrm{int}, 0}=\sigma_{Y|X}\simeq 0.057$ according to our
regression results (Section \ref{subsec:regression})
and assume that the effective number of parameters for the null model is
three (i.e., the intercept, mass slope, and redshift evolution).  

Table \ref{tab:LCDM} lists for each model the values of $\chi^2$ and PTE, 
along with the weighted geometric average\footnote{Specifically, the
weighted geometric average $\langle Y/X\rangle$ is defined as
$\langle Y/X\rangle = \exp\left\{\left[\sum_{n} u_n\ln{(Y_n/X_n)}\right]\,\left(\sum_{n} u_n\right)^{-1} \right\}$ with $u_n$ 
the inverse variance weight for the $n$th cluster, 
$u_n^{-1}=\sigma_{X,n}^2/X_n^2+\sigma_{Y,n}^2/Y_n^2$.} 
 and the standard deviation of observed-to-predicted concentration ratios
$c^\mathrm{(obs)}/c^\mathrm{(pred)}$. 
The theoretical predictions from \citet{Meneghetti2014clash} are based
on nonradiative simulations of DM and baryons, and those from the others
are based on DM-only simulations. 
In all cases, halo masses and concentrations are
defined using the overdensity $\Delta_\mathrm{c}=200$ and measured
assuming the spherical NFW profile, either in projection (2D) or
in three-dimensions (3D).  
For peak-height-based $c(\nu)$ relations
\citep{Prada2012,Diemer+Kravtsov2015}, we assume the {\em WMAP}
seven-year cosmology of \citet{Komatsu+2011WMAP7} to calculate the
relationship between peak height and halo mass at each
redshift.\footnote{We find that the average mass of the CLASH
X-ray-selected subsample corresponds to a halo peak height of 
$\nu\simeq 3.8$ in the adopted cosmology.}

We first consider models for the full population of halos based on the
three-dimensional characterization of the halo density profile.
We find that recent $c$--$M$ relations from 
\citet{Bhatt+2013},
\citet{Dutton+Maccio2014},
\citet{Meneghetti2014clash},
and
\citet{Diemer+Kravtsov2015} 
are in satisfactory agreement with the data.
The \citet{Meneghetti2014clash} model ($\mathrm{PTE}=0.675$), which is
calibrated for a cosmology with a relatively high normalization  
($h=0.7$, $\Omega_\mathrm{m}=0.27$, $\sigma_8=0.82$; Section
\ref{subsec:cM_CLASH}), yields the highest goodness of fit among those
considered here, followed by the \citet{Dutton+Maccio2014} model
($\mathrm{PTE}=0.659$) calibrated for the {\em Planck} cosmology
($h=0.671$, $\Omega_\mathrm{m}=0.3175$, $\sigma_8=0.8344$).   
Our measurements are $33\% \pm 11\%$ higher than the predictions of
\citet{Duffy+2008} based on the {\em WMAP} five-year cosmology
($h=0.742, \Omega_\mathrm{m}=0.258, \sigma_8=0.796$). 
This is in line with the findings of \citet{Dutton+Maccio2014}, who
showed that the $c$--$M$ relation in the {\em WMAP} five-year cosmology
has a 20\% lower normalization at $z=0$ than in the {\em Planck}
cosmology.
On the other hand, the observed concentrations are $27\% \pm 7\%$ lower
than the predictions of \citet{Prada2012}.
Their model exhibits a flattening and upturn of the $c$--$M$ relation in
the high $\nu$ regime, so that their concentrations
derived for cluster halos are substantially higher than those of
others. We refer the reader to \citet{Meneghetti+Rasia2013} and
\citet{Diemer+Kravtsov2015} for detailed
discussions on the possible origin of this discrepancy.

Next, we consider models derived for relaxed populations of
cluster halos  \citep{Duffy+2008,Bhatt+2013,Meneghetti2014clash}.
Numerical simulations suggest that relaxed subsamples have
concentrations that are on average $\sim 10\%$ higher than 
for the full samples. 
In all cases, we find improved agreement with the data compared to the
full-sample comparison (Table \ref{tab:LCDM}). This is consistent with the 
expectation that the CLASH X-ray-selected subsample is largely composed
of relaxed clusters (Section \ref{subsec:cM_CLASH}).  
For the \citet{Meneghetti2014clash} model, the agreement is particularly
excellent, with a PTE of $0.760$.

Finally, we examine the two-dimensional $c$--$M$ relations of
\citet{Meneghetti2014clash} obtained from fitting $\Sigma$ profiles of
simulated halos (Section \ref{subsec:cM_CLASH}).\footnote{We refer to
\citet{Meneghetti2014clash} and \citet{Giocoli2012bias} for general
discussions on the effects of projection bias in measuring the $c$--$M$
relation from lensing.}  
As summarized in Table \ref{tab:LCDM}, we find a better agreement with
the model that explicitly takes into account the CLASH selection
function based on X-ray morphology ($\mathrm{PTE}=0.730$).

In summary, we find that, overall, cluster $c$--$M$ relations that are 
calibrated for recent cosmologies yield good agreement with the
observations
\citep{Bhatt+2013,Dutton+Maccio2014,Meneghetti2014clash,Diemer+Kravtsov2015}. 
An improved agreement is achieved when selection effects are 
taken into account in the models
\citep{Duffy+2008,Bhatt+2013,Meneghetti2014clash}, matching the
characteristics of the CLASH clusters in terms of the overall degree of 
relaxation and X-ray morphological regularity.

\subsubsection{Comparison with the {\sc SaWLens} Results}
\label{subsubsec:cM_SaWLens}

The observational $c$--$M$ relation of \citet[][]{Merten2015clash} is
 derived from their {\sc SaWLens} analysis of a sample of 19 CLASH
 X-ray-selected clusters 
(Section \ref{subsec:rec}; Figure \ref{fig:kappa}).
Their sample includes all 16 clusters in our X-ray-selected subsample. 
Their NFW concentrations scale with halo mass and redshift as  
$c\propto M^{-0.32\pm 0.18}$ and  
$c\propto (1+z)^{0.14\pm 0.52}$
with a normalization of $c|_{z=0.37}=3.66\pm 0.16$ at 
$M_\mathrm{200c}=8\times 10^{14}M_\odot\,h^{-1}$.
Their slope is in good agreement with our results, while their
normalization is somewhat lower than our measurement (Sections
\ref{subsec:regression} and \ref{subsec:cM_CLASH}).  
The best-fit $c$--$M$ relation of \citet{Merten2015clash} is compared
with our regression results in Figure \ref{fig:cMplot}.
We see that the confidence regions overlap well at the $1\sigma$ level.
The \citet{Merten2015clash} relation is also in agreement with the
stacked lensing results of \citet{Umetsu2014clash} and those of this
work. As discussed by \citet{Merten2015clash}, their normalization is  
slightly lower than that of the \citet{Meneghetti2014clash} relation
predicted for the CLASH-like X-ray-selected halos and in better
agreement with that of the full-sample relation (including both relaxed
and unrelaxed halos).

A quantitative comparison of the \citet{Merten2015clash} $c$--$M$
relation with our individual cluster measurements
is summarized in Table
\ref{tab:LCDM}. The observed-to-predicted concentration ratio has a
small scatter of $0.209$ around the mean value of $1.133 \pm 0.087$ and
shows an excellent goodness of fit, with $\mathrm{PTE}=0.754$.
A direct comparison for the 16 clusters in common shows 
that our NFW concentrations (Table \ref{tab:cm}) are
$9.7\%\pm 10.3\%$ higher than those obtained by \citet{Merten2015clash}.  
We emphasize here that this agreement comes in spite of using
substantially different reconstruction methods even though the data and
sample used are largely common to the two studies.  
A multi-probe approach for cluster lensing is one of the great
advantages of the CLASH survey, providing consistency checks between
different lensing methods and different data sets 
\citep[][]{Coe+2012A2261,Umetsu+2012,Umetsu2014clash,Medezinski+2013}.

\subsubsection{Revisiting the Overconcentration Problem}
\label{subsubsec:overcon}


\begin{figure*}[!htb] 
 \begin{center}
  \includegraphics[width=0.7\textwidth,angle=0,clip]{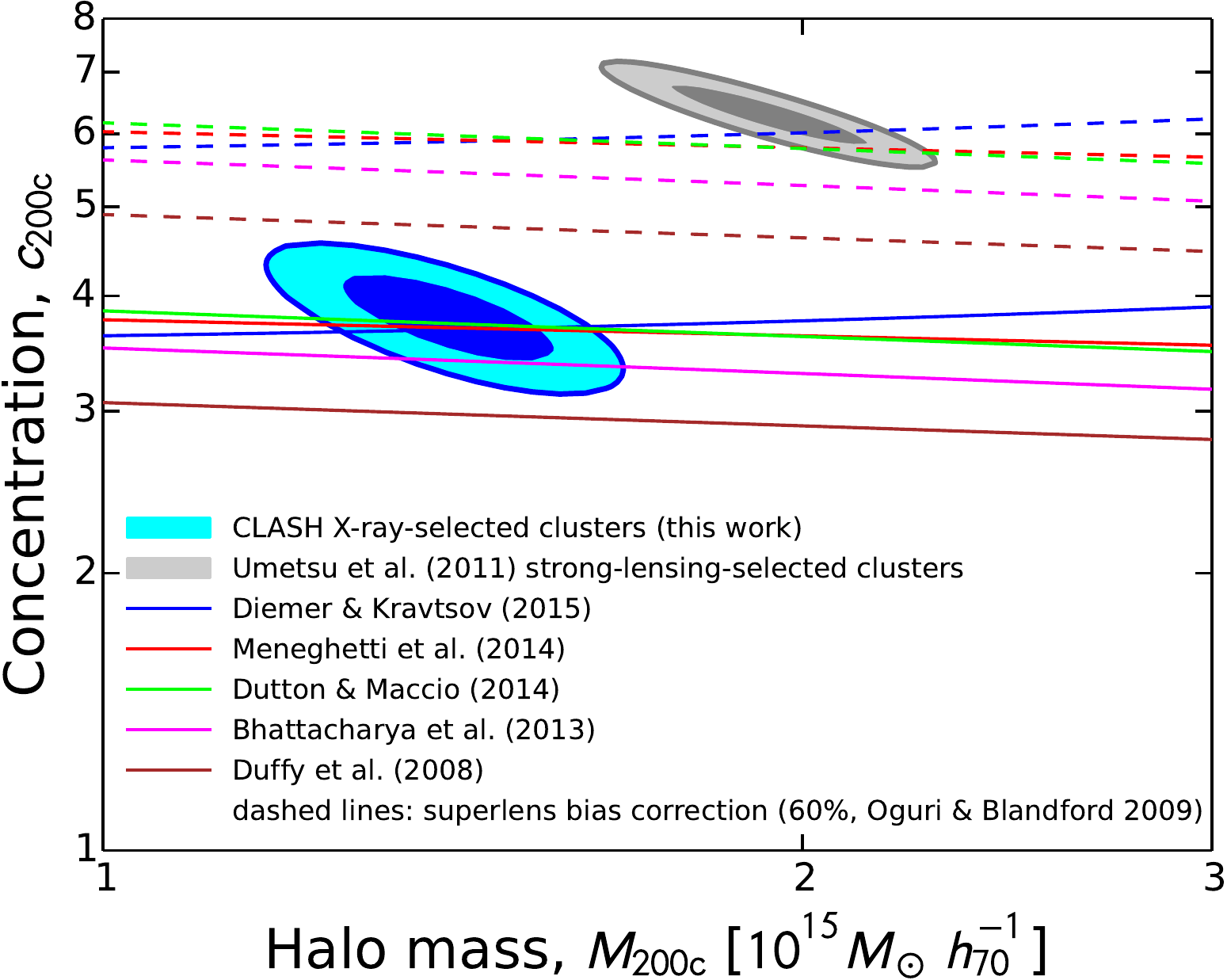} 
 \end{center}
 \caption{\label{fig:cM_U11}
Joint constraints on the mass and concentration parameters
 ($M_\mathrm{200c}, c_\mathrm{200c}$) for our CLASH X-ray-selected
 subsample ($\llangle z_\mathrm{l}\rrangle\simeq 0.34$; blue contours) and the
 strong-lensing-selected sample of \citet{Umetsu+2011stack} 
($\llangle z_\mathrm{l}\rrangle\simeq 0.32$; gray contours) derived from
 the respective 
lensing analyses of {\em HST}+Subaru observations.
For each case, the contours show the 68.3\% and 95.4\% confidence levels
 ($\Delta\chi^2=2.3$ and $6.17$).
The results are compared to theoretical $c$--$M$ relations (solid lines)
 from numerical simulations of $\Lambda$CDM cosmologies
 \citep{Duffy+2008,Bhatt+2013,Dutton+Maccio2014,Meneghetti2014clash,Diemer+Kravtsov2015},
 all evaluated at $z\simeq 0.32$ for the full population of halos. 
The dashed lines show 60\% superlens corrections to the solid lines,
 accounting for the effects of selection and orientation bias expected
 for a strong-lensing cluster population
 \citep[60\%;][]{Oguri+Blandford2009}.  
Once the effects of superlens bias are taken into account, the
stacked-lensing constraints on the \citet{Umetsu+2011stack} sample come
 into line with the models of  
\citet{Dutton+Maccio2014}, 
\citet{Meneghetti2014clash}, and
\citet{Diemer+Kravtsov2015}, the three most recent $c$--$M$ models
 studied in this work.
 }
\end{figure*}

In contrast to the CLASH X-ray-selected subsample,
clusters selected to have large
Einstein radii ($\theta_\mathrm{Ein}$) represent a highly biased
population with their major axis preferentially aligned with the line of sight.
A population of such {\em superlenses} might also be biased toward halos with
intrinsically higher concentrations \citep{2007ApJ...654..714H}.
In the context of $\Lambda$CDM,
the projected mass distributions of superlens clusters
have $\sim 40\%$--$60\%$ higher concentrations than typical
clusters with similar masses and redshifts \citep{Oguri+Blandford2009}.

Prior to the CLASH survey,
\citet[][hereafter U11]{Umetsu+2011stack} performed a
strong-lensing, weak-lensing shear and magnification analysis of four
strong-lensing-selected clusters of similar masses (A1689, A1703, A370, and 
Cl0024+17) at $\llangle z_\mathrm{l}\rrangle \simeq 0.32$ using 
high-quality {\em HST} and Subaru observations.
These clusters display prominent
strong-lensing features, characterized by 
$\theta_\mathrm{Ein}\simgt 30\arcsec$ ($z_\mathrm{s}=2$).
U11 show that the stacked $\llangle\Sigma\rrangle$ profile of the four
clusters is 
well 
described by a single NFW profile in the one-halo
regime, with an effective concentration of 
$c_\mathrm{vir}=7.68^{+0.42}_{-0.40}$
($c_\mathrm{200c}\simeq 6.29$) at 
$M_\mathrm{vir}=22.0^{+1.6}_{-1.4}\times 10^{14}M_\odot\,h_{70}^{-1}$
($M_\mathrm{200c}\simeq 19.4\times 10^{14}M_\odot\,h_{70}^{-1}$),
corresponding to an Einstein radius of 
$\theta_\mathrm{Ein}\simeq 36\arcsec$ ($z_\mathrm{s}=2$).
Semianalytical simulations of $\Lambda$CDM incorporating idealized
triaxial halos yield a $\sim 40\%$--$60\%$ bias correction for a
strong-lensing cluster population \citep{Oguri+Blandford2009}. 
After applying a 50\% superlens correction,
U11 found a discrepancy of $\sim 2\sigma$ with
respect to the \citet{Duffy+2008} $c$--$M$ relation based on the 
{\em WMAP} five-year cosmology.
U11 conclude that there is no significant tension
between the concentrations of their clusters and those of CDM halos if
large lensing biases are coupled to a sizable intrinsic scatter in the
$c$--$M$ relation. 

Figure \ref{fig:cM_U11} compares in the $c$--$M$ plane the stacked
full-lensing constraints for our CLASH X-ray-selected subsample (blue
contours; NFW+LSS (i) in Table \ref{tab:kmodel}) and those for the
strong-lensing-selected sample of U11 (gray contours).  
In the figure we overplot theoretical $c$--$M$ relations of
\citet{Duffy+2008}, \citet{Bhatt+2013}, \citet{Dutton+Maccio2014},
\citet{Meneghetti2014clash}, and 
\citet[][their mean relation; see Table \ref{tab:LCDM}]{Diemer+Kravtsov2015}, all 
evaluated for the full population of halos at the mean redshift 
$\llangle z_\mathrm{l}\rrangle\simeq 0.32$ of the U11 sample.
Figure \ref{fig:cM_U11} demonstrates that $c$--$M$ relations that are
calibrated for more recent cosmologies and simulations provide better
agreement with our CLASH measurements (see Section
\ref{subsubsec:cM_LCDM}).

To account for the superlens bias in the U11 sample,
we plot each of the $c$--$M$ models with a maximal 60\% correction 
\citep{Oguri+Blandford2009}. 
We find that, once the effects of selection and
orientation bias are taken into account, the full-lensing results of U11
come into line with the models of 
\citet{Dutton+Maccio2014}, 
\citet{Meneghetti2014clash}, and
\citet{Diemer+Kravtsov2015}, the three most recent $c$--$M$ models
studied in this work.
Hence, the discrepancy found by U11 can be fully reconciled by the higher
normalization of the $c$--$M$ relation as favored by recent {\em WMAP}
and {\em Planck} cosmologies.

\section{Summary and Conclusions}
\label{sec:summary}

We have presented a comprehensive analysis of strong-lensing,
weak-lensing shear and magnification data for a sample of
16 X-ray-selected and 4 high-magnification-selected galaxy clusters  
at $0.19\simlt z\simlt 0.69$ (Table \ref{tab:sample}) targeted in the
CLASH survey \citep{Postman+2012CLASH}. 
Our analysis combines constraints from 16-band {\em HST} observations
\citep{Zitrin2015clash} and wide-field multi-color imaging
\citep{Umetsu2014clash}
taken primarily with Subaru/Suprime-Cam, spanning a
wide range of cluster radii, $\theta=10\arcsec$--$16\arcmin$. 
We have carefully taken into account several major sources of
uncertainties in our error analysis (Section \ref{sec:method}).

We have reconstructed surface mass density profiles $\bSigma$ of
 all clusters from a joint analysis of strong-lensing, weak-lensing
 shear and magnification constraints (Section \ref{sec:profile}; Figure 
 \ref{fig:kappa}), 
 providing a unique cluster mass profile data set.
We find a significant improvement of $\sim 45\%$ on average in terms of
 the total S/N of the cluster mass profile measurement from adding the
 central {\em HST} lensing constraints ($10\arcsec$--$40\arcsec$) to
 the wide-field weak-lensing data ($0.9\arcmin$--$16\arcmin$). 

With the improved sensitivity and resolution at
$10\arcsec$--$40\arcsec$,
we have measured masses and concentrations from these mass profiles for
individual  clusters assuming a spherical NFW halo (Tables \ref{tab:cm}
and \ref{tab:mass}).     
The median precision on individual cluster mass measurements
is found to be
$\sim 28\%, 24\%, 23\%$, and $24\%$
at
$\Delta_\mathrm{c}=200, 500, 1000$, and $2500$, respectively.
We assessed the internal consistency of cluster mass measurements
across the multiple probes of cluster lensing effects (Section
\ref{subsec:msys}). 
We find internal consistency of the ensemble mass calibration to be
$\le 5\%\pm 6\%$ 
in the one-halo regime, $r=200$--$2000$\,kpc\,$h^{-1}$ 
($0.01\simlt r/r_\mathrm{200m}\simlt 1$; Figure \ref{fig:mratio}),
by comparison with the CLASH weak-lensing-only measurements of
\citet{Umetsu2014clash}.
This level of uncertainty in the ensemble mass calibration,
empirically estimated using different combinations of lensing probes, is 
in agreement with the systematic uncertainty in the absolute mass
calibration of $\simeq 6\%$ assessed in Section \ref{subsec:sys}.
This implies that the total uncertainty in the absolute mass calibration
with the sample of 20 clusters is 
$\sqrt{0.28^2/20+0.06^2}\simeq 9\%$ at $\Delta_\mathrm{c}=200$ 
(Section \ref{subsec:mass}).

The CLASH X-ray-selected sample was selected to have a high degree of
regularity in their X-ray morphology \citep{Postman+2012CLASH}.
Numerical simulations suggest that this sample is prevalently 
composed of relaxed clusters ($\sim 70\%$) and largely free of
orientation bias \citep{Meneghetti2014clash}.  
An important effect of the CLASH selection function is to significantly
reduce the scatter in concentration  because of their X-ray regularity
(Section \ref{subsec:cM_CLASH}). 
For a lensing-unbiased subsample of 16 CLASH X-ray-selected clusters, 
we have examined the mean $c$--$M$ relation and its
 intrinsic scatter using Bayesian regression methods (Section
 \ref{sec:cM}; Figures \ref{fig:cMmodel} and \ref{fig:cMplot}).
Our analysis takes into account the correlation between the errors on
mass and concentration and the effects of nonuniformity of the intrinsic
mass probability distribution.
Our model yields a mean concentration of $c|_{z=0.34}=3.95\pm 0.35$
 at 
$M_\mathrm{200c}=10^{15}M_\odot\,h^{-1}\simeq 14\times 10^{14}M_\odot\,h_{70}^{-1}$ and an
 intrinsic scatter of 
$\sigma(\log_{10}c_\mathrm{200c})=0.056\pm 0.026$, or
$\sigma(\ln c_{\mathrm{200c}})=0.13\pm 0.06$.
The normalization, slope, and scatter of the observed $c$--$M$ relation
 are all consistent with 
$\Lambda$CDM predictions \citep{Meneghetti2014clash} when the projection
effects and the CLASH selection function based on X-ray morphological
regularity are taken into account.  Our regression results are in
 agreement with the {\sc SaWLens} analysis of \citet{Merten2015clash}
 and the stacked shear-only analysis of \citet{Umetsu2014clash}. 
This multi-probe approach for cluster lensing is one of the
key advantages of the CLASH survey, providing consistency checks between 
different lensing methods and different data sets.

We have derived an ensemble-averaged surface mass density profile
 $\llangle \bSigma\rrangle$ at an average redshift of 
$\llangle z_\mathrm{l}\rrangle \simeq 0.34$ 
for the X-ray-selected subsample of 16 clusters by stacking their
 individual $\bSigma$ profiles (Section \ref{sec:stack}; Figure \ref{fig:kplot}).
The stacked lensing signal is detected at $33\sigma$ significance over the
 entire radial range, $R\le 4000$\,kpc\,$h^{-1}$, 
accounting for the effects of intrinsic profile variations and
 uncorrelated LSS along the line of sight (Figure \ref{fig:aveC}).

Our CLASH lensing determination of the cluster mass distribution
provides a firm basis for a detailed comparison with theoretical models.
We show that the $\llangle\bSigma\rrangle$ profile is well described by
 a family of density profiles predicted for DM-dominated halos in
 gravitational equilibrium (Table \ref{tab:kmodel}; Figure
 \ref{fig:kmodel}), namely,  
  the NFW, Einasto, and DARKexp models. Of these, the first two are
 phenomenological models and the last is a theoretically derived model
 \citep{Hjorth+2010DARKexp,DARKexp2}. 
The single power-law, cored isothermal and Burkert density
profiles are disfavored by the observed mass profile having a pronounced
radial curvature.  

We find that cuspy halo models that include the large-scale two-halo 
 contribution using the $b_\mathrm{h}$--$M$ relation of \citet{Tinker+2010}
provide improved agreement with the data. 
Independent of the chosen halo density profile, we find 
$b_\mathrm{h}(M_\mathrm{200c})\sim 9.3$ 
($b_\mathrm{h}\sigma_8^2\sim 6.1$).
For the NFW halo model (NFW+LSS (i)), we measure a mean concentration of 
 $c_\mathrm{200c}=3.79^{+0.30}_{-0.28}$
at $M_\mathrm{200c}=14.1^{+1.0}_{-1.0}\times 10^{14}M_\odot\,h_{70}^{-1}$,
demonstrating 
consistency between complementary analysis methods
 (Figure \ref{fig:cMplot}). 
This model yields an Einstein radius of 
$\theta_\mathrm{Ein}=14.0^{+3.4}_{-3.2}$\,arcsec at $z_\mathrm{s}=2$,
which agrees within $2\sigma$
with the observed median Einstein radius of
$\overline{\theta}_\mathrm{Ein}=20.1\arcsec$ ($z_\mathrm{s}=2$) for this
subsample (Section \ref{subsec:sl}).   
The Einasto shape parameter is constrained to be
 $\alpha_\mathrm{E}=0.248^{+0.051}_{-0.047}$ 
 ($1/\alpha_\mathrm{E}=4.03^{+0.93}_{-0.69}$), 
 which is in good agreement
 with predictions from $\Lambda$CDM numerical simulations
 \citep{Gao+2008,Meneghetti2014clash}. 

A systematic comparison between different cluster lensing surveys that
use different approaches to measuring the masses of clusters 
allows us to identify (known and unknown) systematic effects.
In the last few years, a substantial effort has been devoted to
establishing an accurate mass calibration from cluster lensing
\citep{vonderLinden2014calib,Hoekstra2015CCCP,Israel+2015},  
in light of the apparent tension in cosmological constraints from {\em
Planck} primary CMB data and SZE cluster counts
\citep{Planck2014XX,Planck2015XXIV}.  
We compared our CLASH lensing masses (Tables \ref{tab:cm} and 
\ref{tab:mass}) with weak-lensing masses from other surveys  (WtG, CCCP,
LoCuSS). 
Our mass measurements are in excellent agreement within $1\sigma$
with the WtG \citep{WtG3} and LoCuSS \citep{Okabe+Smith2015} surveys,
with which we have 17 and 5 clusters in common, respectively.
At higher mass overdensities $\Delta_\mathrm{c}\simgt 500$ where
weak-lensing measurements are subject to various systematics, the
agreement appears to be less impressive (Sections \ref{subsubsec:wtg}
and \ref{subsubsec:locuss}).  
We find that our measurements are on average 
$16\%$ and $9\%$
higher than the CCCP measurements at $\Delta_\mathrm{c}=500$ and $2500$,
respectively. 
With limited statistics, however, we do not find statistically
significant differences between the CCCP and CLASH results.

Our mass measurements are found to be $\sim 7\%$--$9\%$ higher than the 
CLASH {\sc SaWLens} results of \citet{Merten2015clash}, with which we
have 16 clusters in common. 
This difference is smaller than the systematic mass offset of 
$\sim 10\% \pm 5\%$ 
found between the CLASH weak-lensing \citep{Umetsu2014clash}
and {\sc SaWLens} \citep{Merten2015clash} mass profiles.
Hence, combining the central {\em HST}-lensing and outer weak-lensing
data has reduced the discrepancy with respect to the {\sc SaWLens}
results. 
We find that this remaining discrepancy can be attributed to three
outliers (Section \ref{subsec:rec}), which correspond to
clusters at the lowest Galactic latitudes ($b<30^\circ$) of the overlap
sample  
and those at the highest redshifts ($z\simgt 0.45$)
which exhibit complex mass distributions with a high degree of
substructure \citep{Postman+2012CLASH,Merten2015clash}.
Since the data used are largely common to the two analyses except for
the inclusion of weak-lensing magnification data in this work,
this discrepancy likely arises from systematics in the present
calibration of magnification measurements for these low Galactic
latitude clusters and high redshift clusters. 
Weak lensing reconstructions are sensitive to the treatment of
boundary conditions if there are massive structures near the data
boundaries. Hence, mass profile reconstructions for clusters with 
high degrees of substructure can be subject to a greater degree
of mass-sheet degeneracy.
Excluding the three outliers brings the two results into agreement
within $1\%$ at $\Delta_\mathrm{c}=\Delta_\mathrm{vir}$, $200$, and
$500$ and within $5\%$ at $\Delta_\mathrm{c}=2500$.



In the CLASH survey \citep{Postman+2012CLASH}, 
we have demonstrated the power of multi-probe, multi-scale data sets
available from a space telescope cluster survey  
\citep{Zitrin2015clash,Merten2015clash} combined with 
X-ray \citep{Donahue2014clash},
SZE \citep{Sayers2013pressure,Czakon2015}, and 
wide-field imaging plus spectroscopic 
\citep{Umetsu2014clash,Rosati2014VLT} follow-up observations.
Extending this type of cluster survey to a large sky area, as planned
for the WFIRST and Euclid missions,
will be a
significant step forward in obtaining a comprehensive picture of
the evolution of clusters over cosmic time and across populations, as
well as in understanding the evolutionary and tidal
effects of surrounding LSS on the mass distribution of the central
cluster. 

%




\acknowledgments
We thank the anonymous referees for their careful reading of the
manuscript and useful suggestions.
This work was made possible by the availability of high-quality lensing
data produced by the CLASH team. We express our gratitude to all
members of the CLASH team who enabled us to carry out the work.
We thank all authors of \cite{Umetsu2014clash} and \cite{Zitrin2015clash}
for their valuable contributions to the lensing analyses used here.  
We thank John Moustakas for his assistance in the early stages of this work.
We thank Massimo Meneghetti and Elinor Medezinski for providing useful
information and suggestions.
We acknowledge very fruitful discussions with
Mauro Sereno,
Nobuhiro Okabe,
Bau-Ching Hsieh,
Benedikt Diemer,
Nicole Czakon,
Tom Broadhurst,
Masahiro Takada,
Jens Hjorth,
Liliya L.R. Williams,
and
Graham Smith.
This work is partially supported by 
the Ministry of Science and Technology of Taiwan under the grants
MOST 103-2112-M-001-030-MY3
and
MOST 103-2112-M-001-003-MY3.
AZ acknowledges  support by
NASA through Hubble Fellowship Grant HST-HF2-51334.001-A awarded by
STScI. 
JM is supported by the People Programme (Marie Curie Actions) of the 
European Union's Seventh Framework Programme (FP7/2007-2013) under REA
grant agreement number 627288.
DG was supported by SFB-Transregio 33 ``The Dark Universe'' by the
Deutsche Forschungsgemeinschaft (DFG), the DFG cluster of excellence
``Origin and Structure of the Universe'', and NASA through the Einstein
Fellowship Program, grant PF5-160138.

\appendix

\section{Discretized Expressions for Cluster Lensing Profiles}
\label{appendix:estimators}

First, we derive a discrete expression 
for the mean interior convergence $\kappa_\infty(<\theta)$
as a function of clustercentric radius $\theta$
using the azimuthally averaged convergence $\kappa_\infty(\theta)$.
For a given set of $(N+1)$ concentric radii $\theta_i$
defining $N$ radial bands in the range 
$\theta_\mathrm{min}\equiv\theta_1\le\theta\le \theta_{N+1}\equiv \theta_\mathrm{max}$,
a discretized estimator for $\kappa_\infty(<\theta)$
can be written as
\begin{equation}
\label{eq:avkappa_d}
\kappa_\infty(<\theta_i)=
\left(\frac{\theta_\mathrm{min}}{\theta_i}\right)^2
\kappa_\infty(<\theta_\mathrm{min})+
\frac{2}{\theta_i^2}\sum_{j=1}^{i-1}
\Delta\ln\theta_{j}\,
\overline\theta_{j}^2
\kappa_\infty(\overline\theta_{j}),
\end{equation}
with
$\Delta\ln\theta_i \equiv (\theta_{i+1}-\theta_i)/\overline\theta_i$
and $\overline\theta_i$
the area-weighted center of the $i$th 
bin defined by $[\theta_i,\theta_{i+1}]$.
In the continuous limit, 
\begin{equation}
\label{eq:medianr}
\overline{\theta}_i
=
2\int_{\theta_i}^{\theta_{i+1}}\!d\theta'\theta'^2/
(\theta_{i+1}^2-\theta_{i}^2)\nonumber\\ 
=
\frac{2}{3}
\frac{\theta_{i+1}^3-\theta_{i}^3}
{\theta_{i+1}^2-\theta_{i}^2}. 
\end{equation}

Next, we derive discretized expressions for the tangential
reduced shear $g_{+}(\theta)$ and 
the inverse magnification $\mu^{-1}(\theta)$ in terms of the binned
convergence $\kappa_\infty(\overline{\theta}_i)$, using the following relations:
\begin{eqnarray}
g_+(\overline\theta_i) &=&
\frac{
\langle W\rangle_g \left[
 \kappa_\infty(<\overline\theta_i)-\kappa_\infty(\overline\theta_i)
\right]
}{1- f_{W,g}\langle W\rangle_g\kappa_\infty(\overline\theta_i)},\\
\mu^{-1}(\overline\theta_i)&=&
\left[1-\langle W\rangle_\mu\kappa_\infty(\overline\theta_i)\right]^2
-
\langle W\rangle_\mu^2\left[
 \kappa_\infty(<\overline\theta_i)-\kappa_\infty(\overline\theta_i)
\right]^2,
\end{eqnarray}
where both depend on the mean convergence interior to $\overline\theta_i$, 
$\kappa_\infty(<\overline{\theta}_i)$.
By assuming a constant density in each radial band,
we find the following expression for 
$\kappa_\infty(<\overline\theta_i)$:
\begin{equation} 
\kappa_\infty(<\overline\theta_i) = 
\frac{1}{2}\Big[
\left(\theta_{i} / \overline{\theta}_i\right)^2
\kappa_\infty(<\theta_i) 
+
\left(\theta_{i+1} / \overline{\theta}_i\right)^2
\kappa_\infty(<\theta_{i+1})
\Big],
\end{equation}
where $\kappa_\infty(<\theta_i)$ and $\kappa_\infty(<\theta_{i+1})$
can be computed using Equation (\ref{eq:avkappa_d}).

Accordingly, all relevant cluster lensing observables can be uniquely
specified by the binned convergence profile 
$\{\kappa_{\infty,\mathrm{min}},\kappa_{\infty,i}\}_{i=1}^{N}$
with $\kappa_{\infty,\mathrm{min}}\equiv \kappa_\infty(<\theta_\mathrm{min})$ and 
$\kappa_{\infty,i}\equiv \kappa_\infty(\overline{\theta}_i)$.

\section{Comparison of Surface Mass Density Profiles}
\label{appendix:kappa}

Appendix B includes the surface mass density profiles for our cluster
sample obtained in this study, along with those of
\citet{Umetsu2014clash} and \citet{Merten2015clash}.


\begin{figure*}[!htb] 
 \begin{center}
 $
 \begin{array}
  {c@{\hspace{.1in}}c@{\hspace{.1in}}c}
  \includegraphics[width=0.39\textwidth,angle=0,clip]{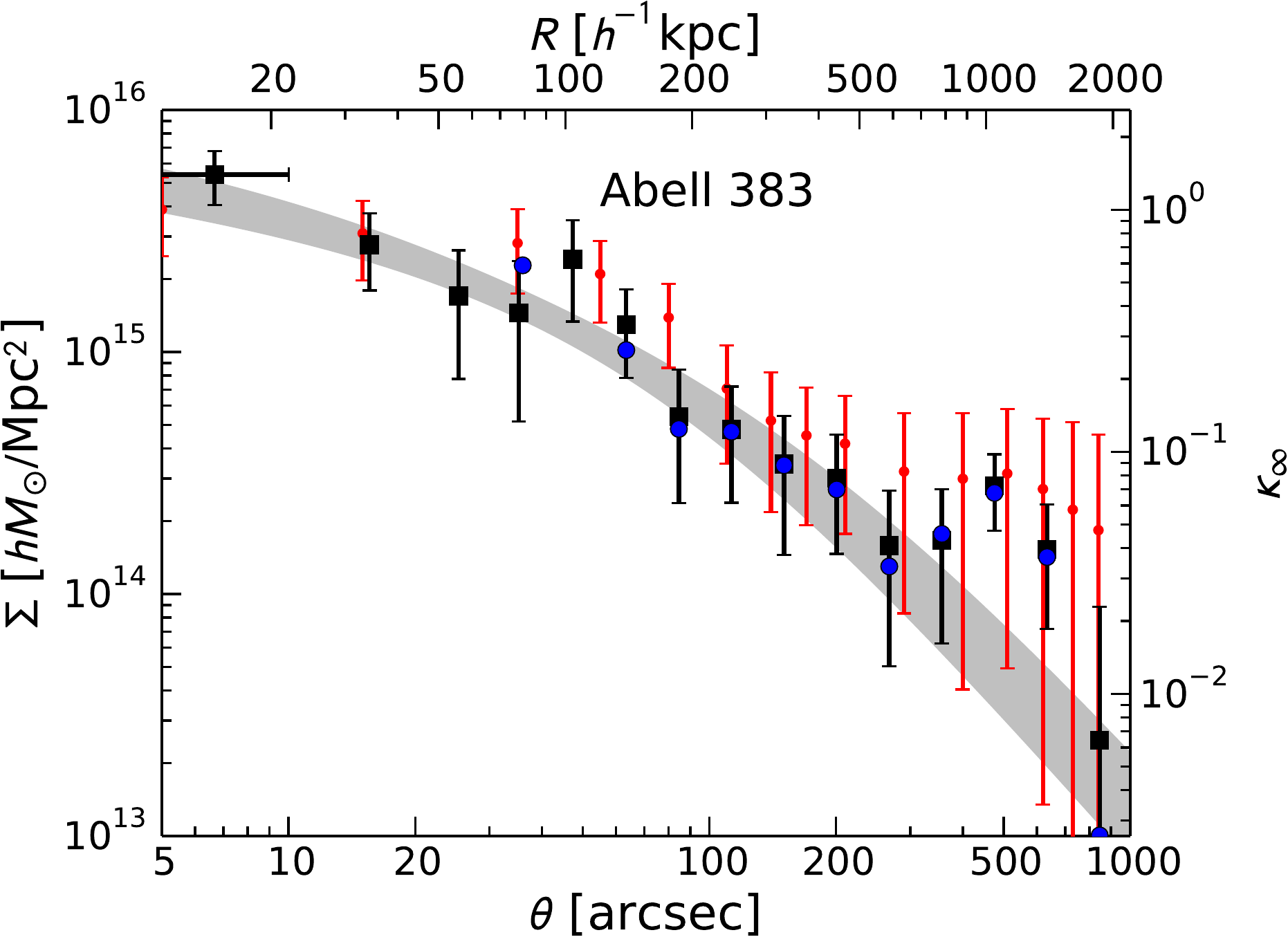} &
  \includegraphics[width=0.39\textwidth,angle=0,clip]{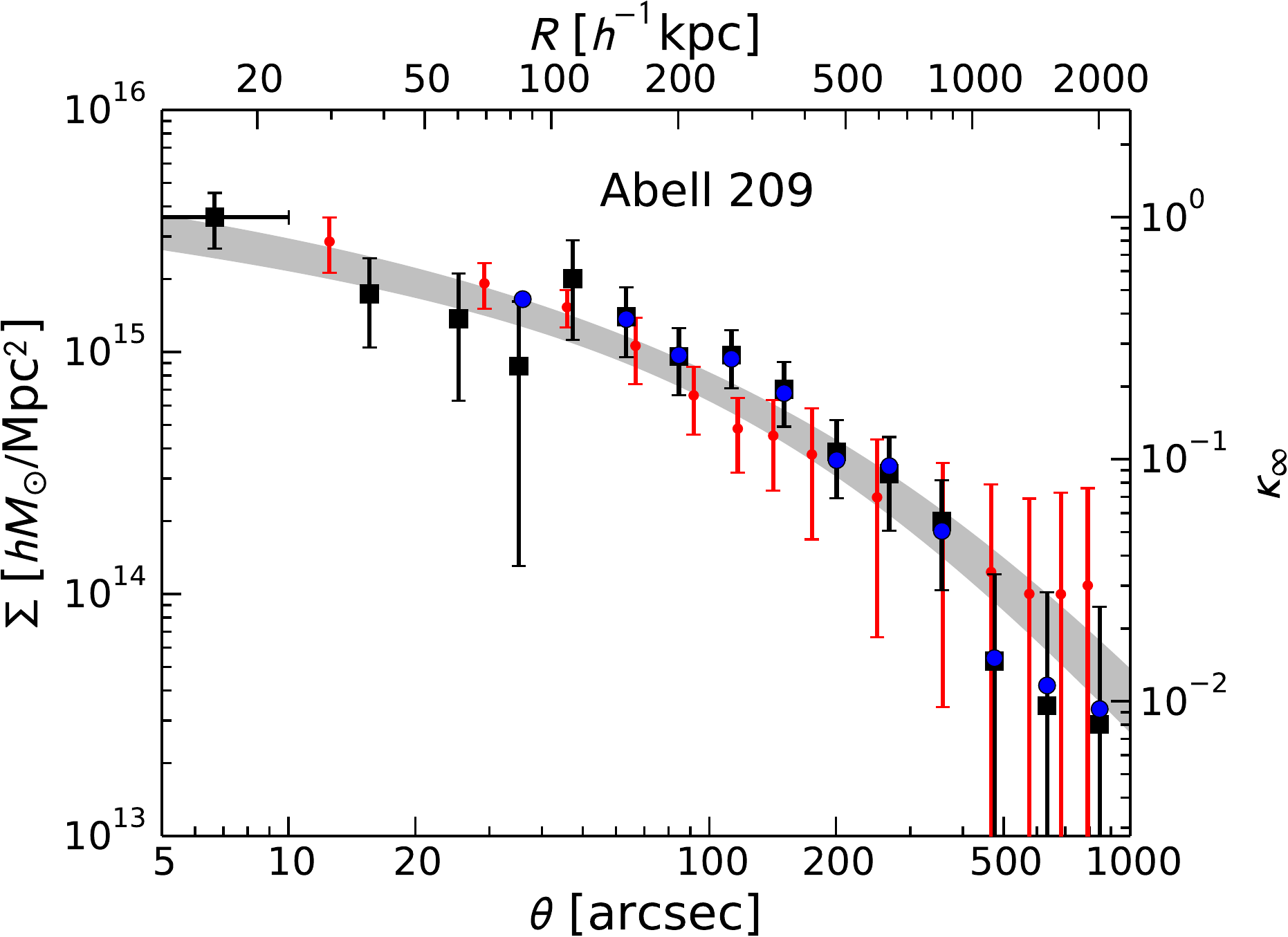} \\
 \end{array}
 $
 $
 \begin{array}
  {c@{\hspace{.1in}}c@{\hspace{.1in}}c}
  \includegraphics[width=0.39\textwidth,angle=0,clip]{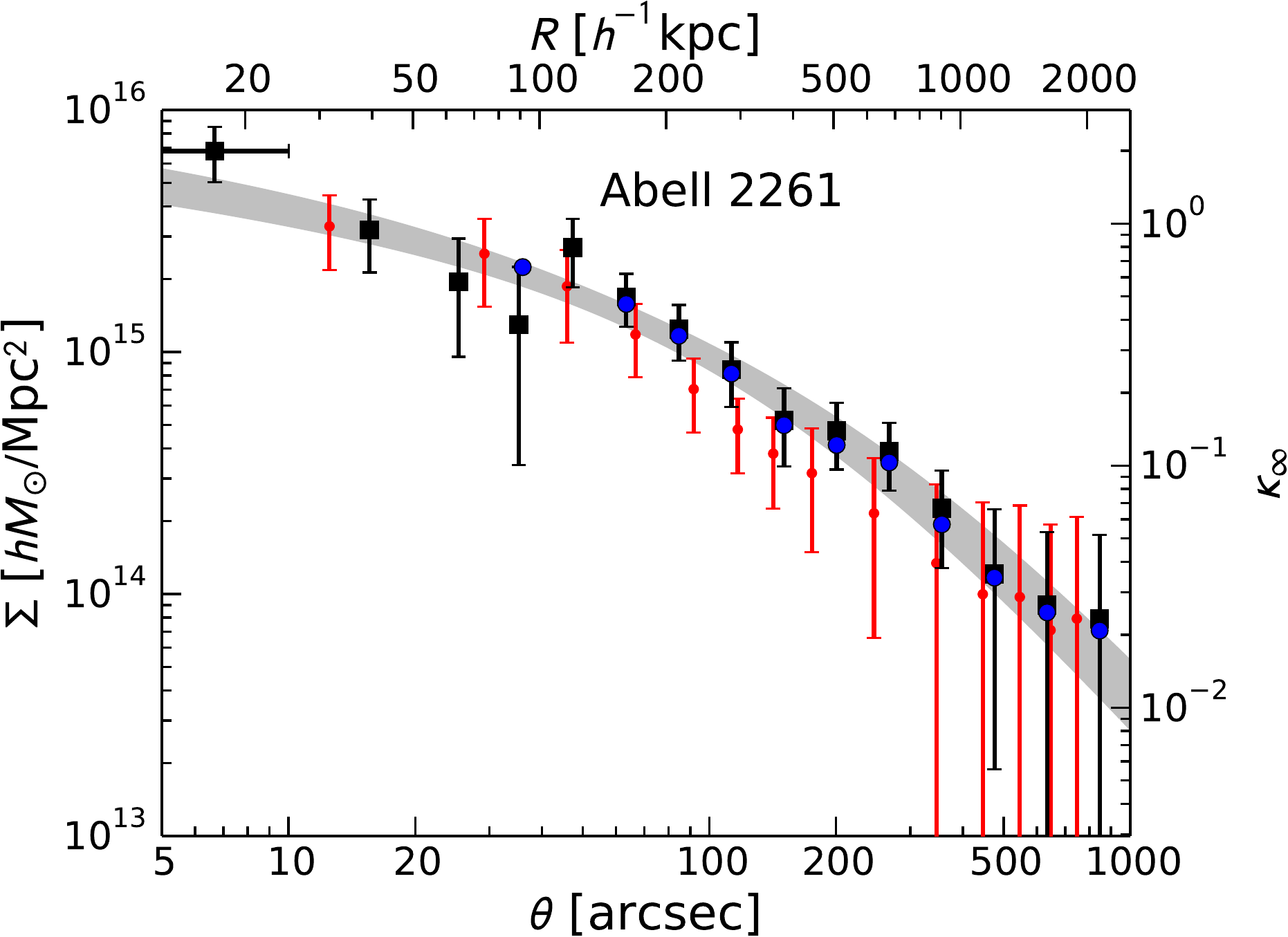}&
  \includegraphics[width=0.39\textwidth,angle=0,clip]{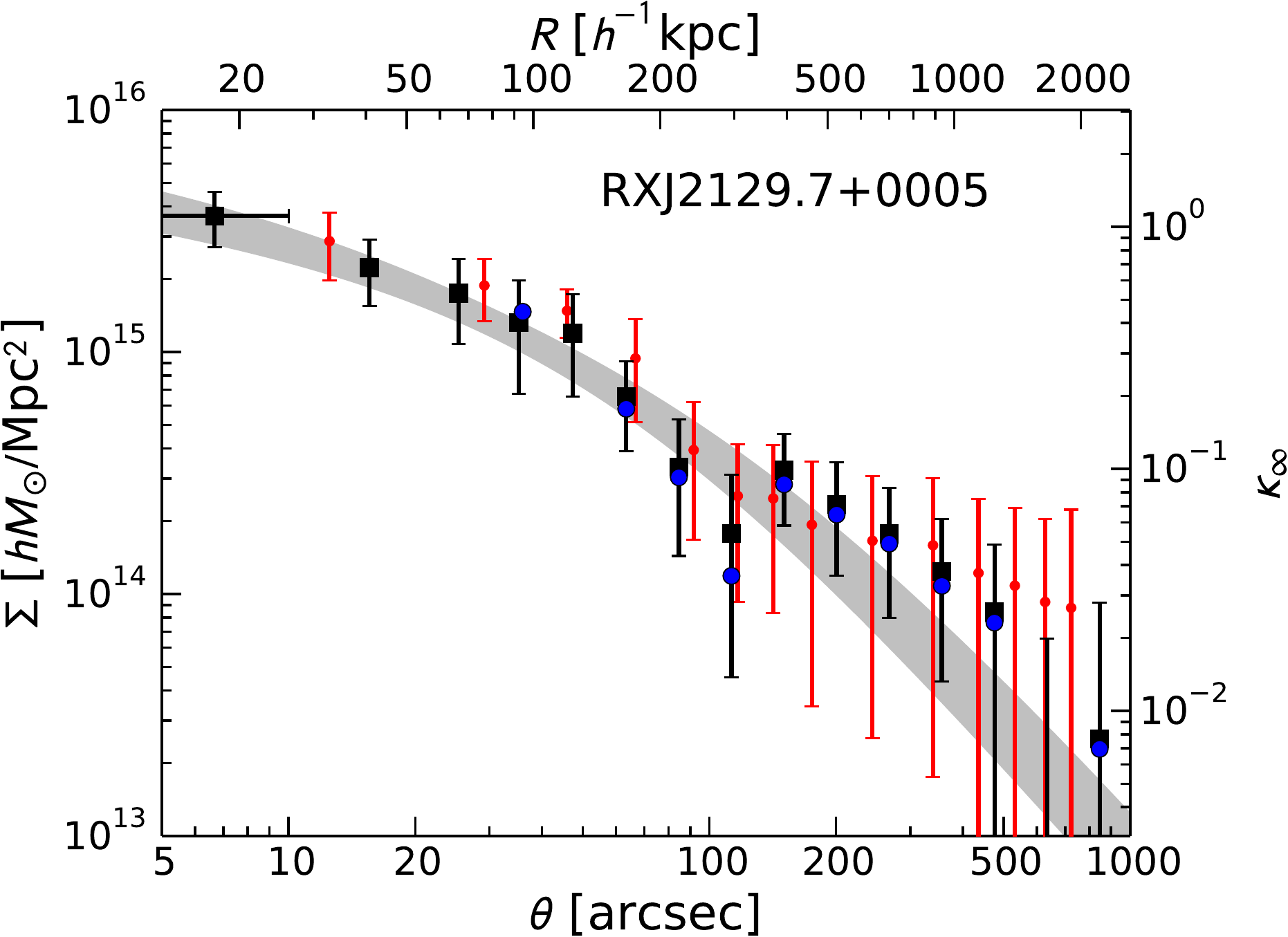} \\
 \end{array}
 $
 $
 \begin{array}
  {c@{\hspace{.1in}}c@{\hspace{.1in}}c}
  \includegraphics[width=0.39\textwidth,angle=0,clip]{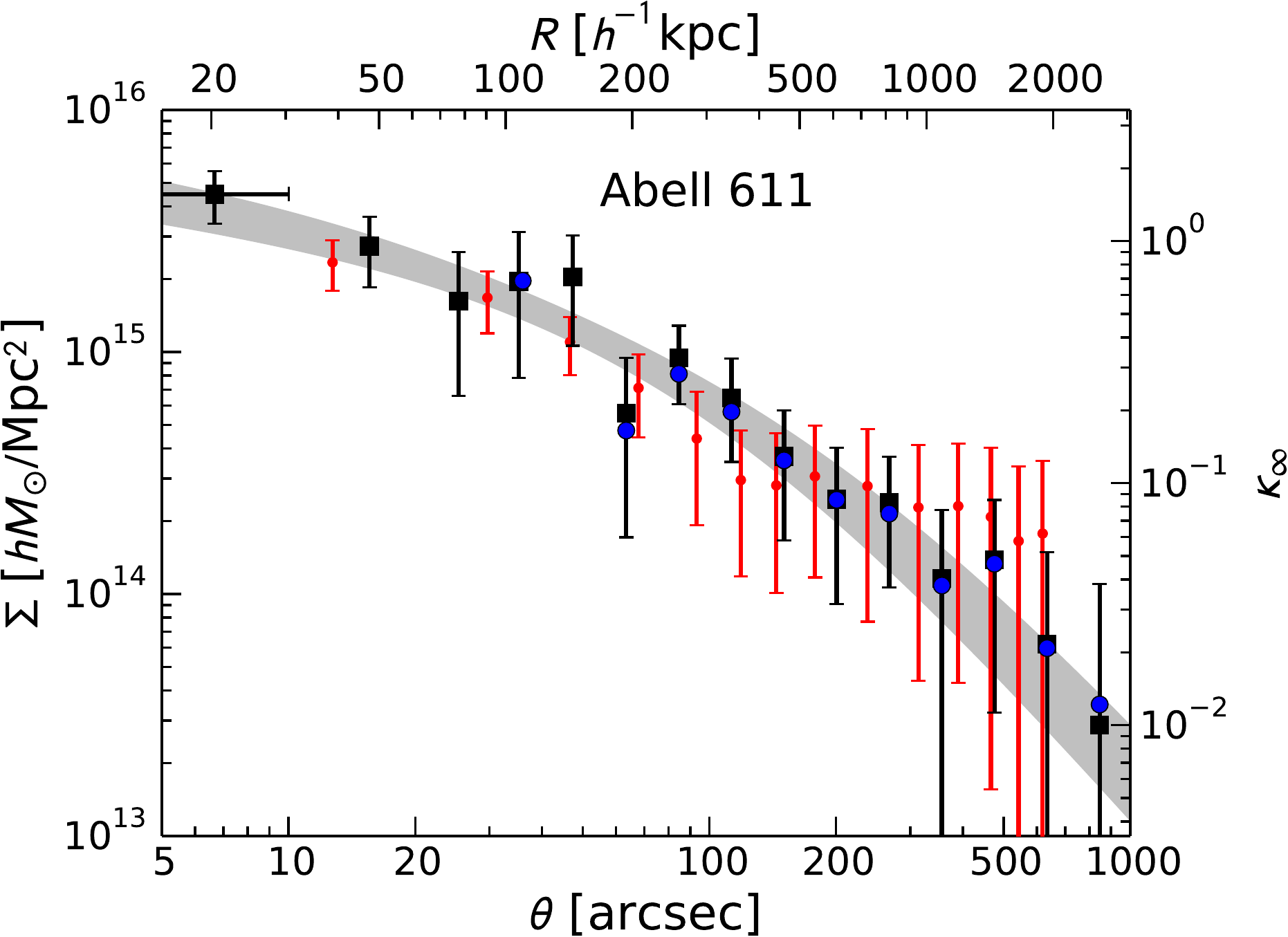} &
  \includegraphics[width=0.39\textwidth,angle=0,clip]{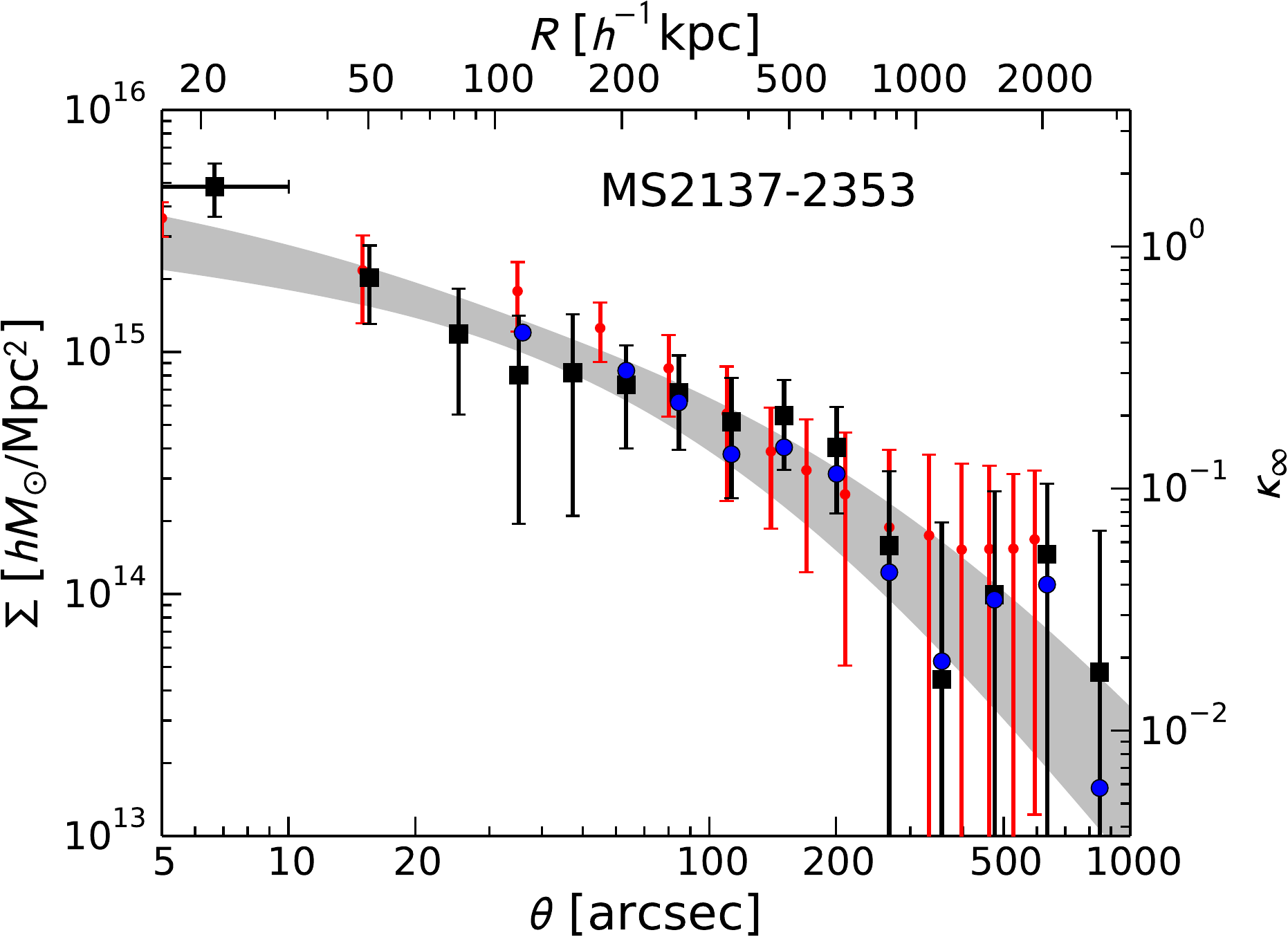} \\
 \end{array}
 $
 $
 \begin{array}
  {c@{\hspace{.1in}}c@{\hspace{.1in}}c}
  \includegraphics[width=0.39\textwidth,angle=0,clip]{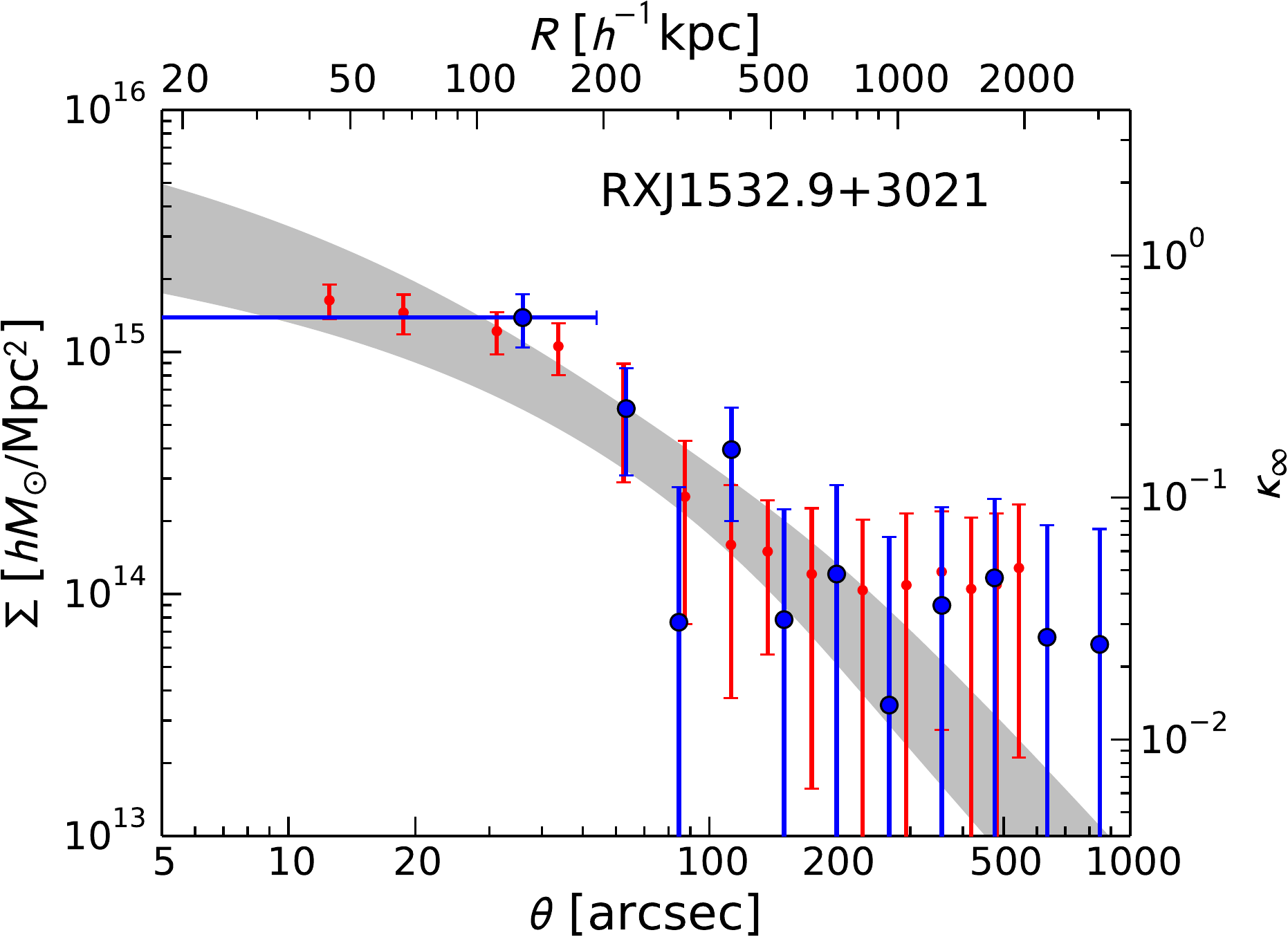}&
  \includegraphics[width=0.39\textwidth,angle=0,clip]{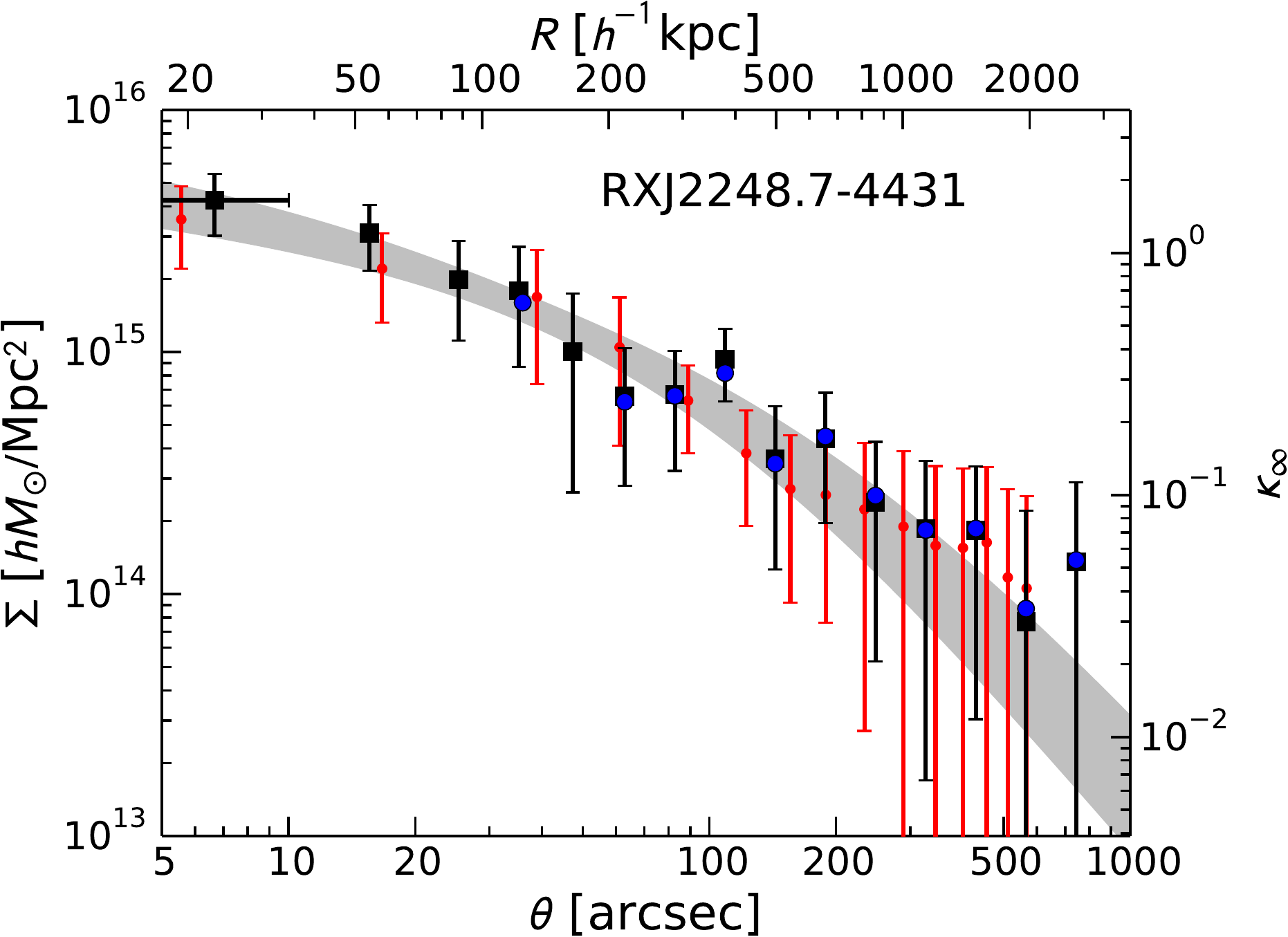}
 \end{array}
 $
 \end{center}
\caption{\label{fig:kappa}
Surface mass density profiles derived from a joint analysis of {\em HST}
 strong/weak-shear lensing and ground-based weak shear/magnification
 lensing data (black squares) we have obtained for a sample of 
 (a), (b) 16 X-ray-regular and (c) 4 high-magnification clusters selected
 from the CLASH survey. 
For each cluster, the central bin $\Sigma(<\theta_\mathrm{min})$ is
 marked with a horizontal bar. The location of each binned $\Sigma$
 point represents the area-weighted center of the radial band (Appendix
 \ref{appendix:estimators}).  
The error bars represent the $1\sigma$ uncertainty from the diagonal
 part of the total covariance matrix including statistical, systematic,
 projected uncorrelated LSS, and intrinsic-variance contributions, 
$C=C^\mathrm{stat}+C^\mathrm{sys}+C^\mathrm{lss}+C^\mathrm{int}$
 (Section \ref{subsubsec:cmat}).
The gray area in each plot shows the best-fit NFW profile (68\%
 CL) for the observed $\Sigma$ profile. 
The results are compared to the shear+magnification results (blue
 circles) of \citet{Umetsu2014clash} and those from a {\sc SaWLens} (red
 dots) analysis of \citet{Merten2015clash}. 
The scale on the right vertical axis shows the corresponding
lensing convergence $\kappa_\infty$ scaled to the reference
 far-background source plane.} 
\end{figure*}

\begin{figure*}[!htb] 
\addtocounter{figure}{-1}
 \begin{center}
 $
 \begin{array}
  {c@{\hspace{.1in}}c@{\hspace{.1in}}c}
   \includegraphics[width=0.39\textwidth,angle=0,clip]{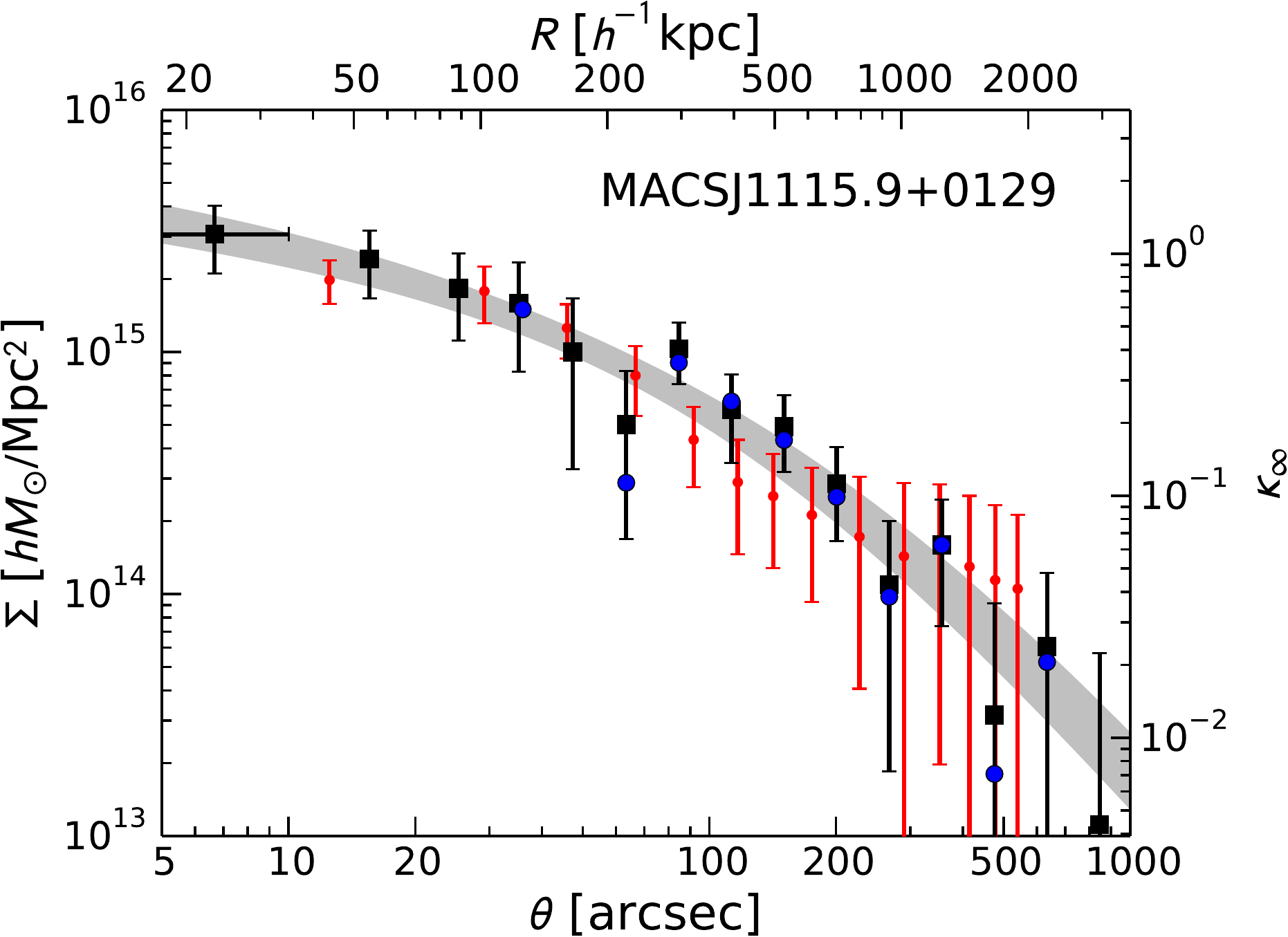}& 
   \includegraphics[width=0.39\textwidth,angle=0,clip]{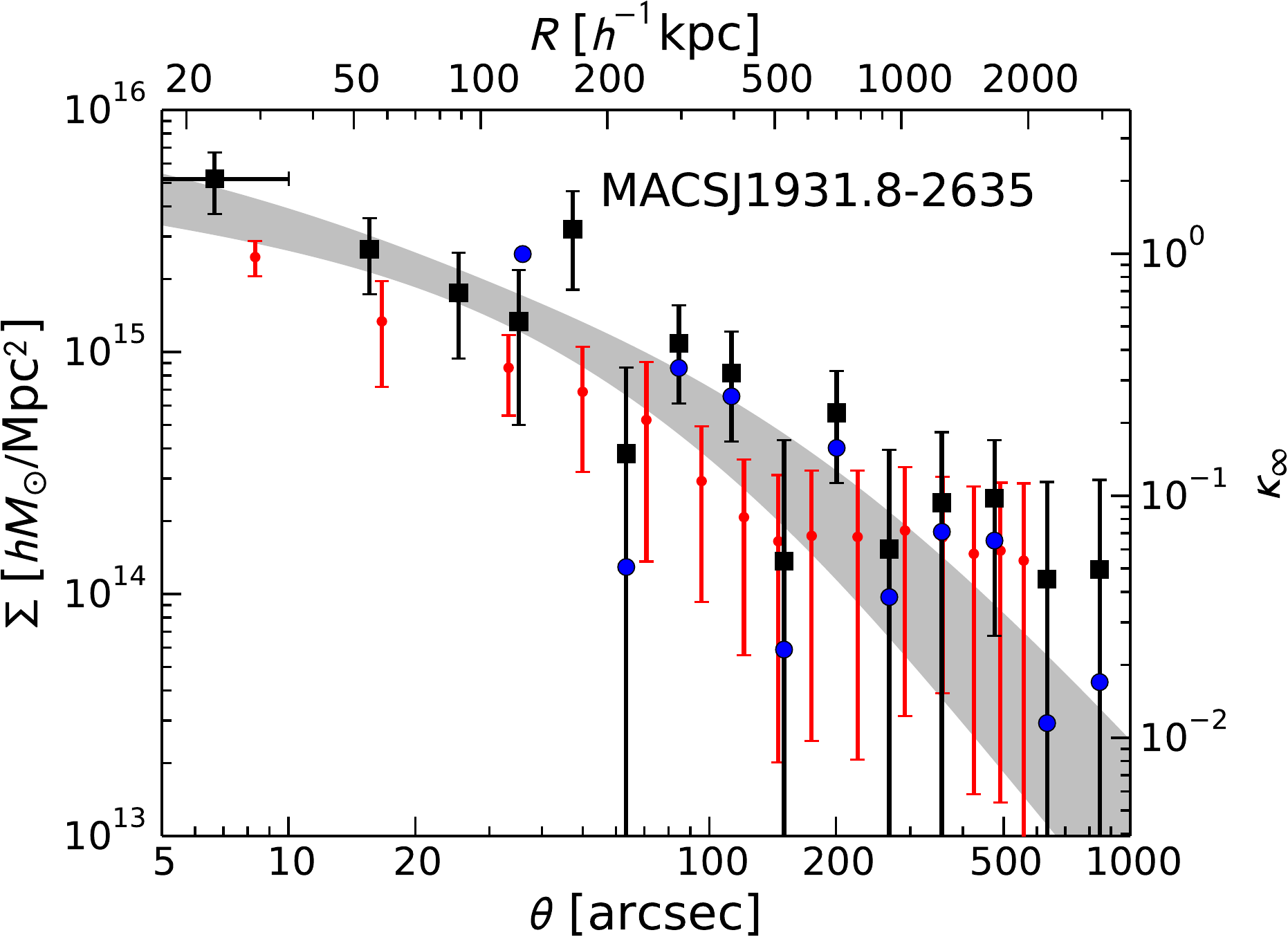}\\
  \end{array}
 $
 $
 \begin{array}
  {c@{\hspace{.1in}}c@{\hspace{.1in}}c}
   \includegraphics[width=0.39\textwidth,angle=0,clip]{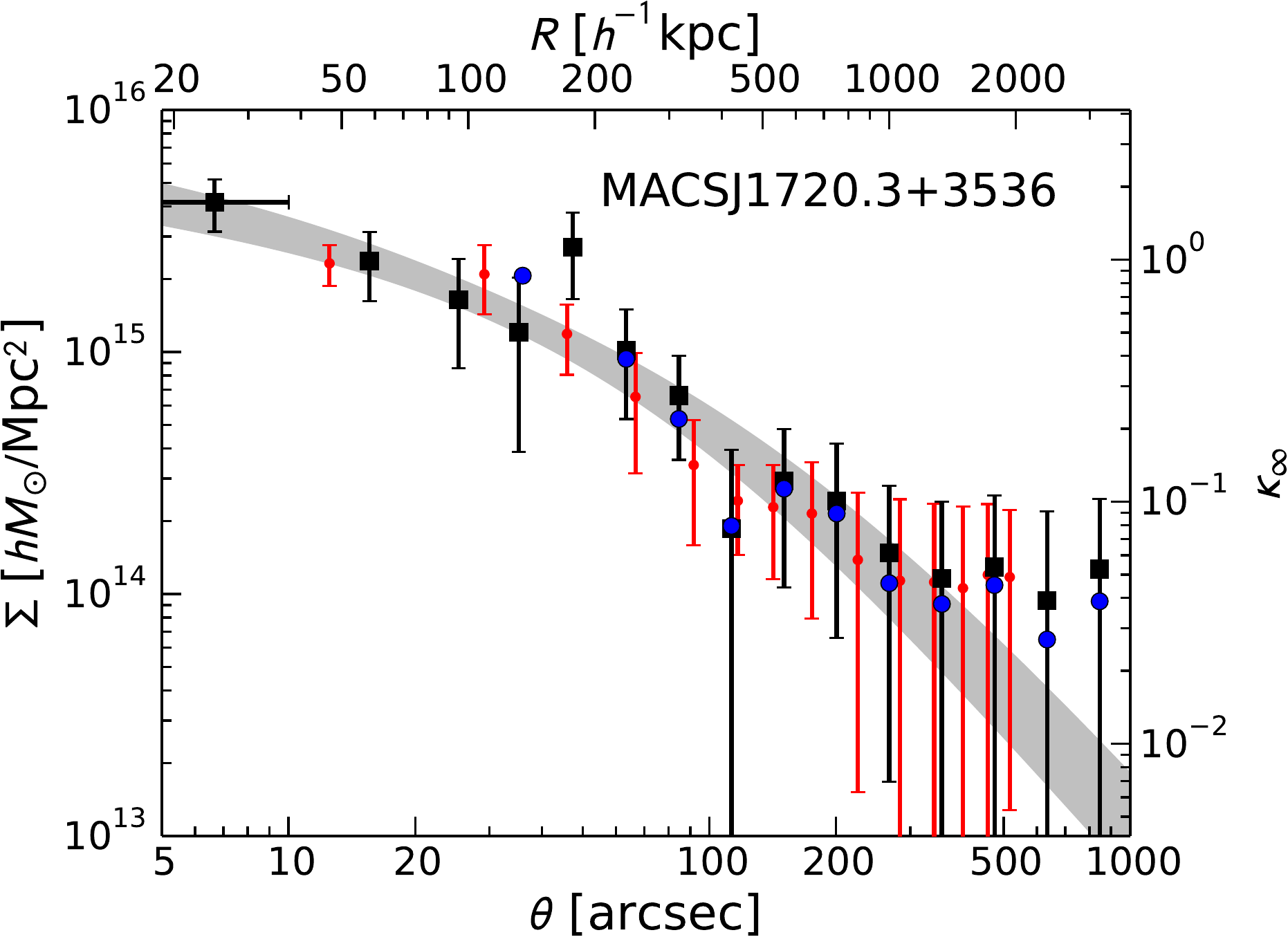}&  
   \includegraphics[width=0.39\textwidth,angle=0,clip]{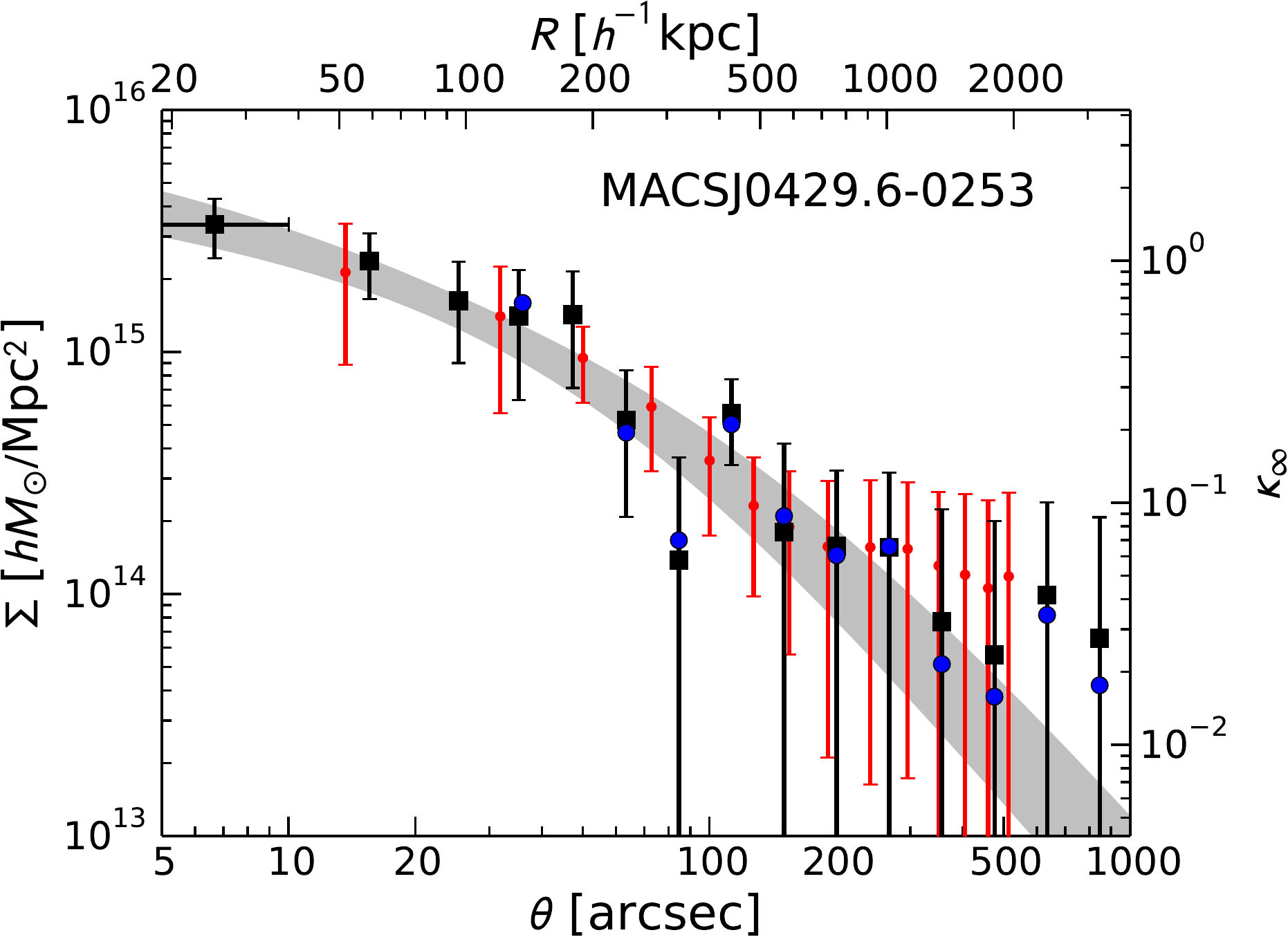}\\
  \end{array}
 $
 $
 \begin{array}
  {c@{\hspace{.1in}}c@{\hspace{.1in}}c}
   \includegraphics[width=0.39\textwidth,angle=0,clip]{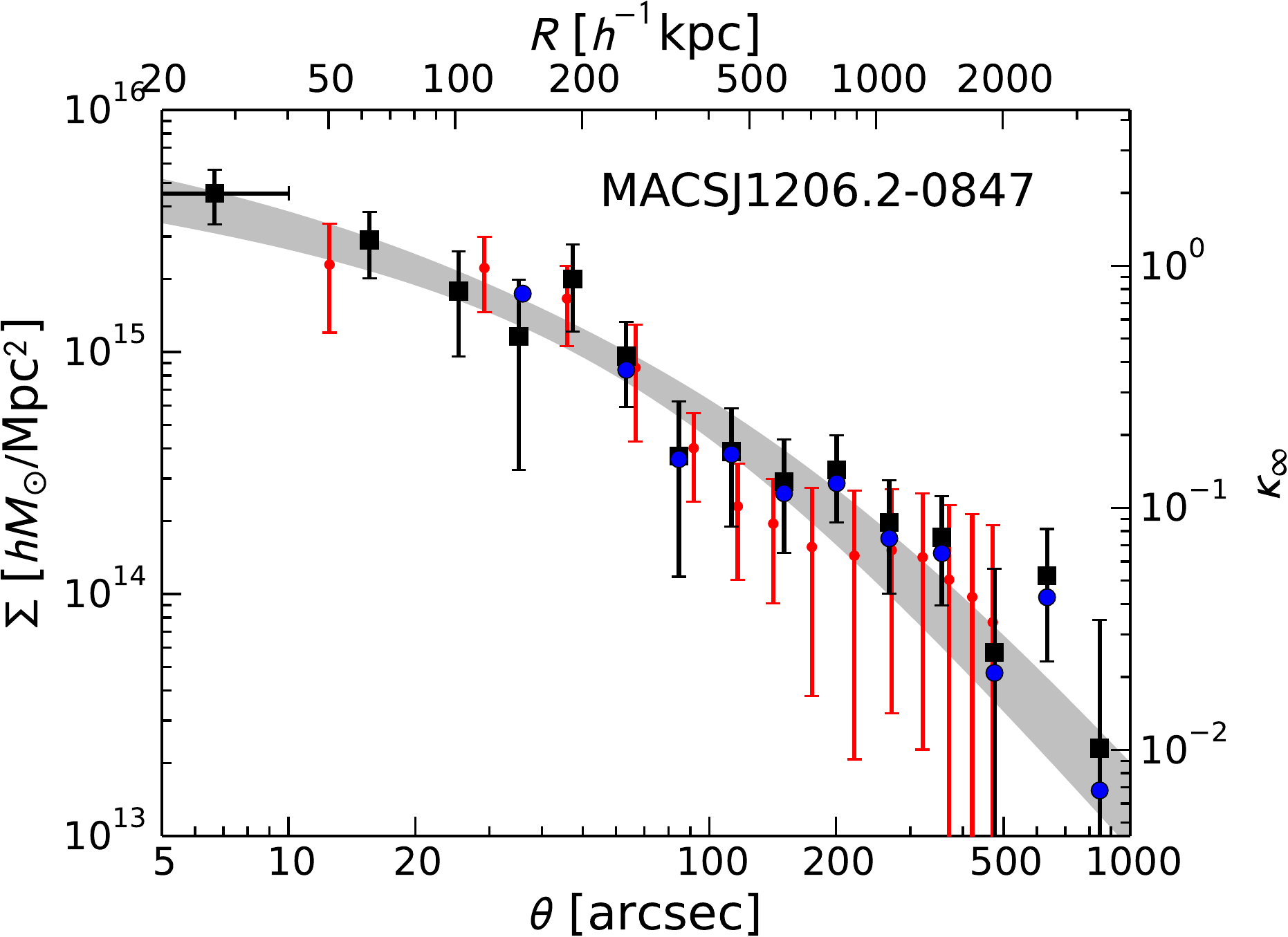}& 
   \includegraphics[width=0.39\textwidth,angle=0,clip]{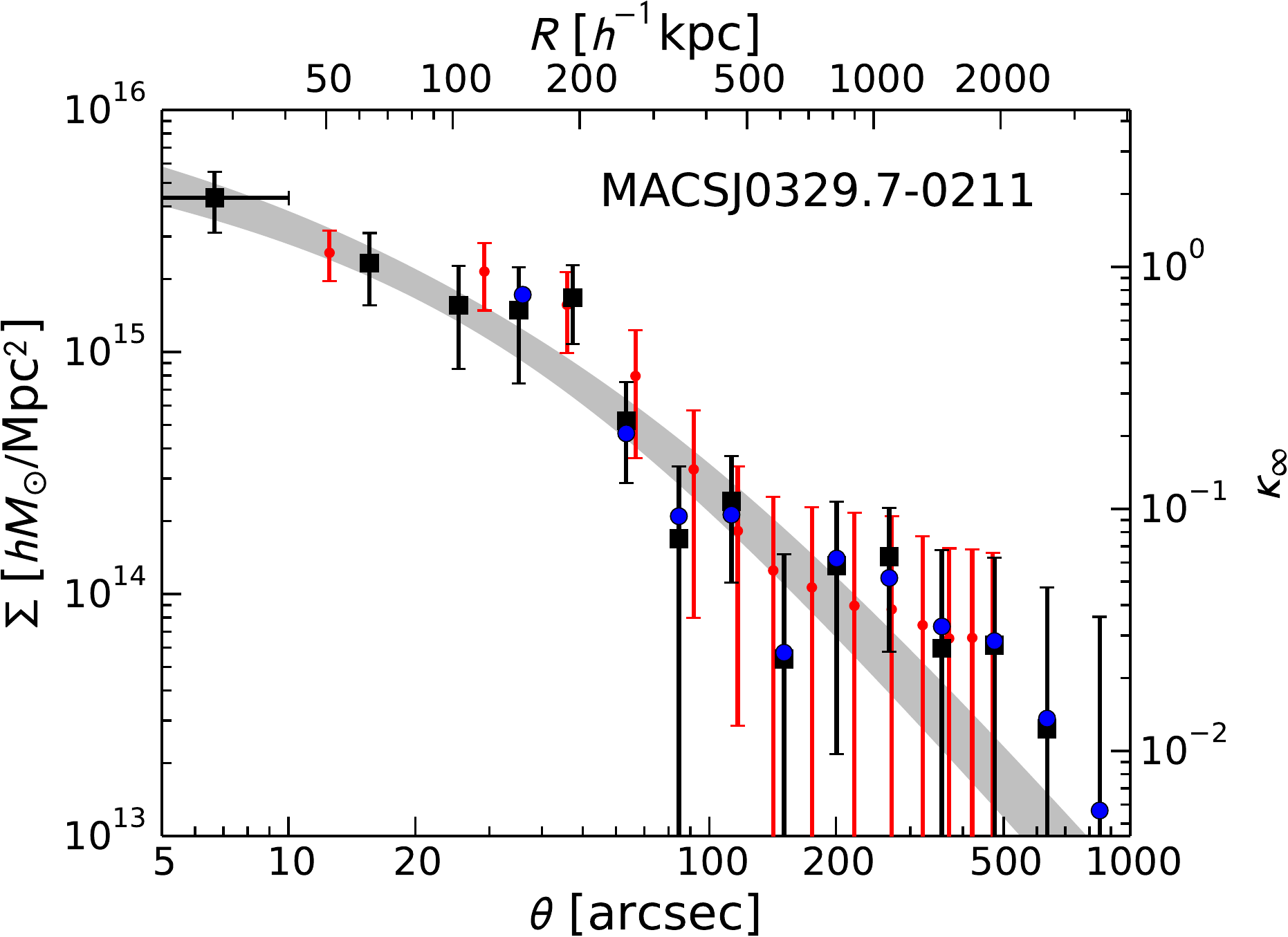}\\
 \end{array}
 $
 $
 \begin{array}
  {c@{\hspace{.1in}}c@{\hspace{.1in}}c}
   \includegraphics[width=0.39\textwidth,angle=0,clip]{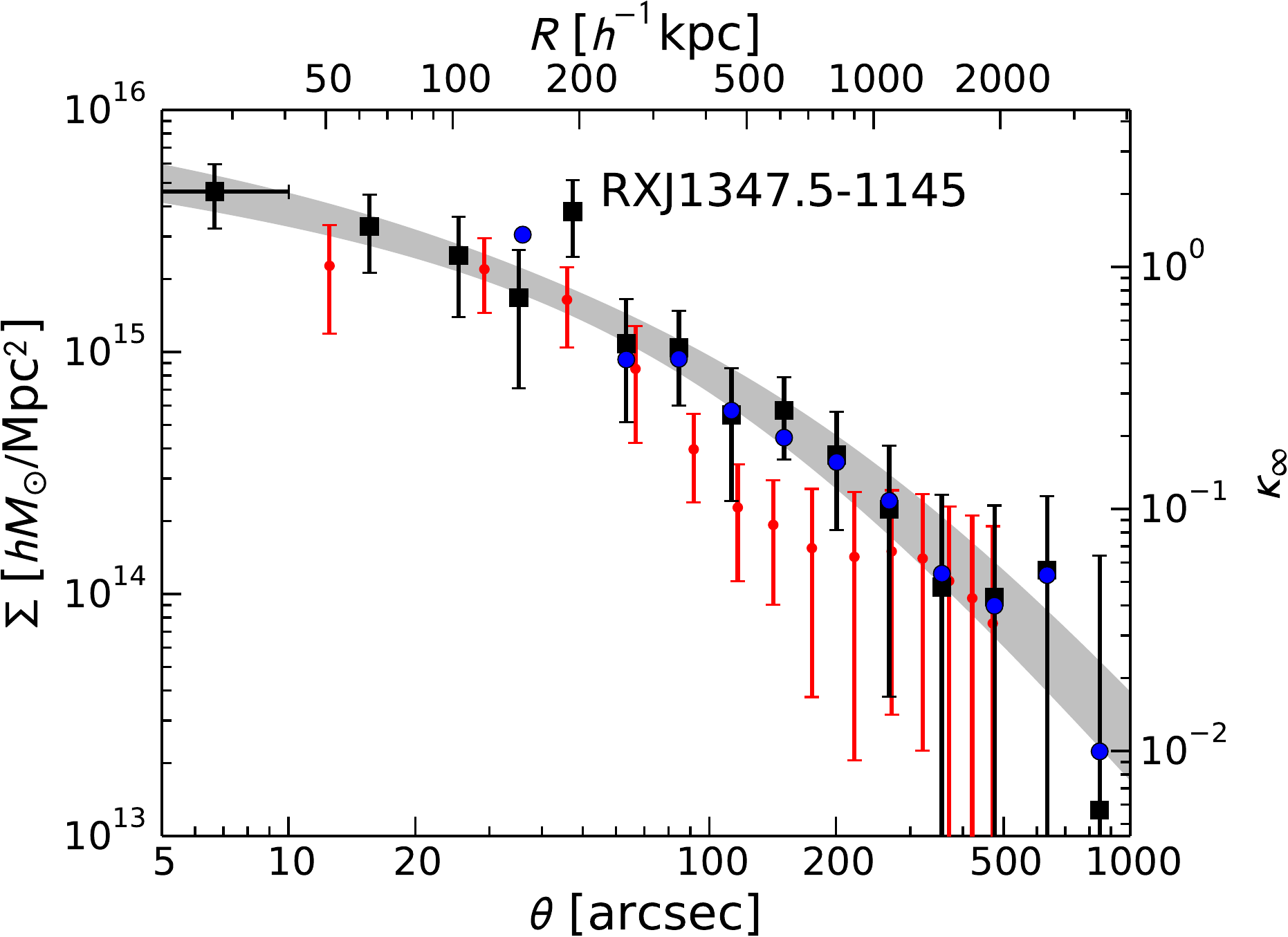} &
   \includegraphics[width=0.39\textwidth,angle=0,clip]{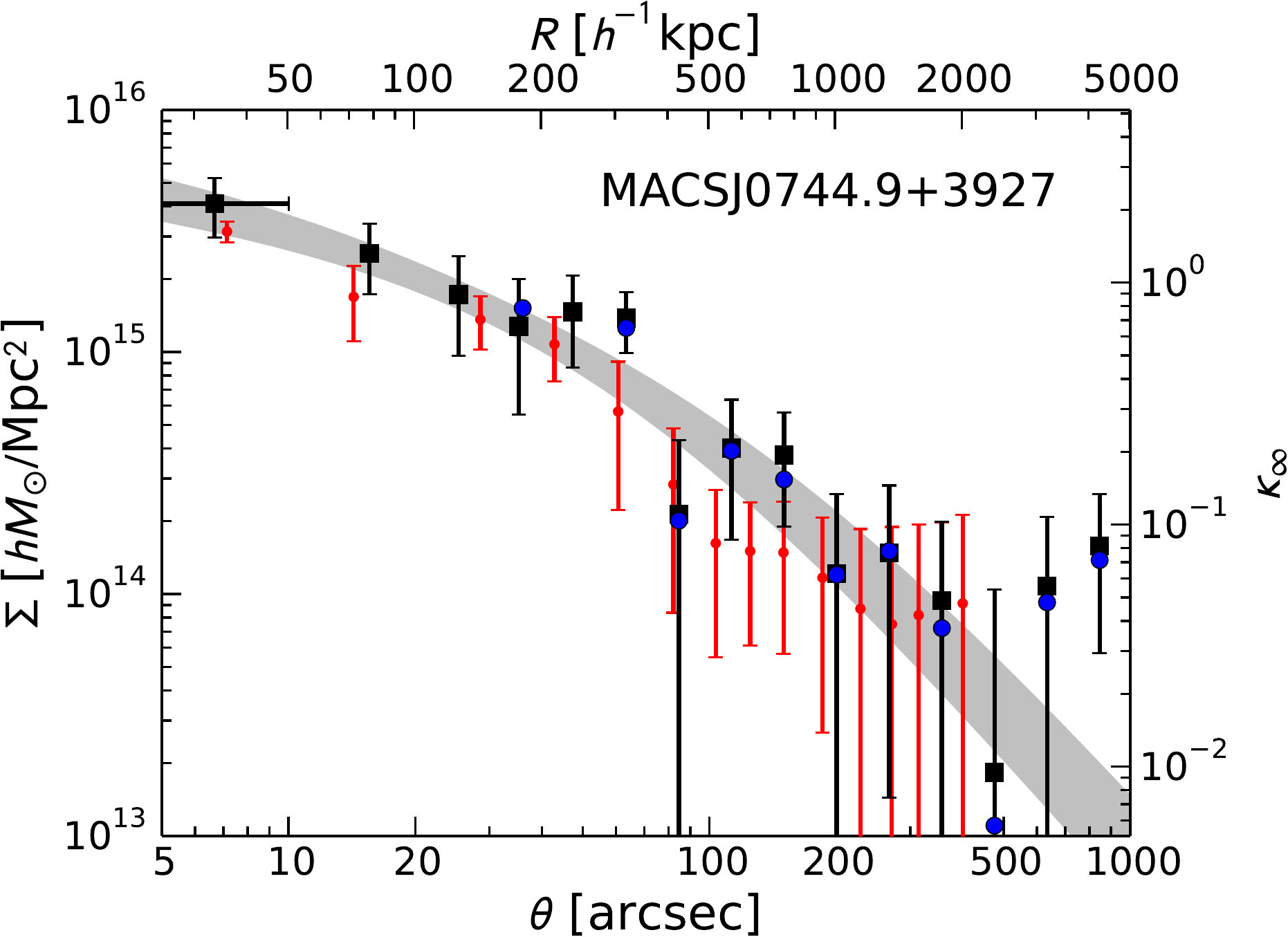} 
 \end{array}
 $
 \end{center}
  \caption{(Continued.)}
\end{figure*}


\begin{figure*}[!htb] 
\addtocounter{figure}{-1}
 \begin{center}
 $
 \begin{array}
  {c@{\hspace{.1in}}c@{\hspace{.1in}}c}
   \includegraphics[width=0.39\textwidth,angle=0,clip]{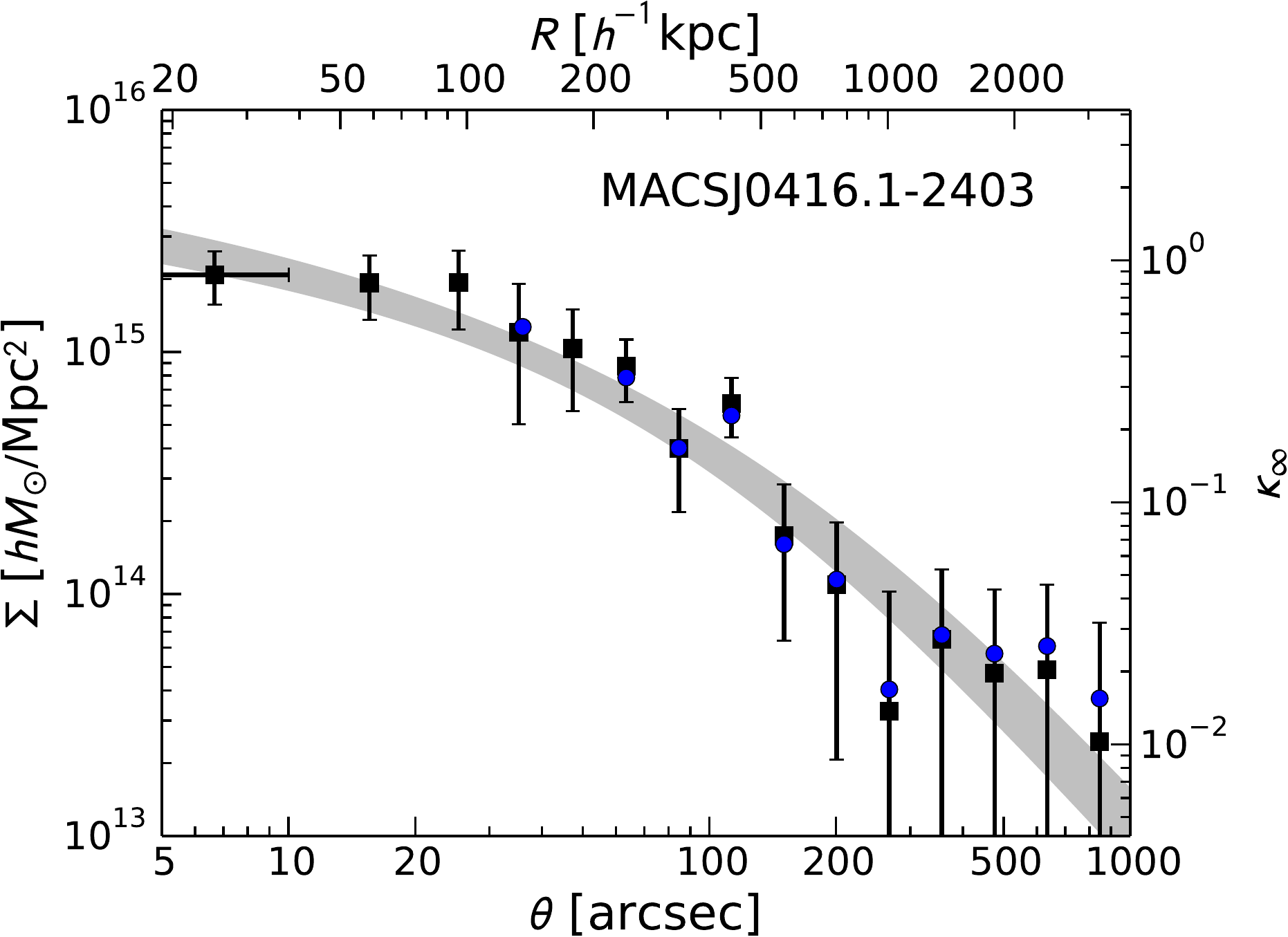}& 
   \includegraphics[width=0.39\textwidth,angle=0,clip]{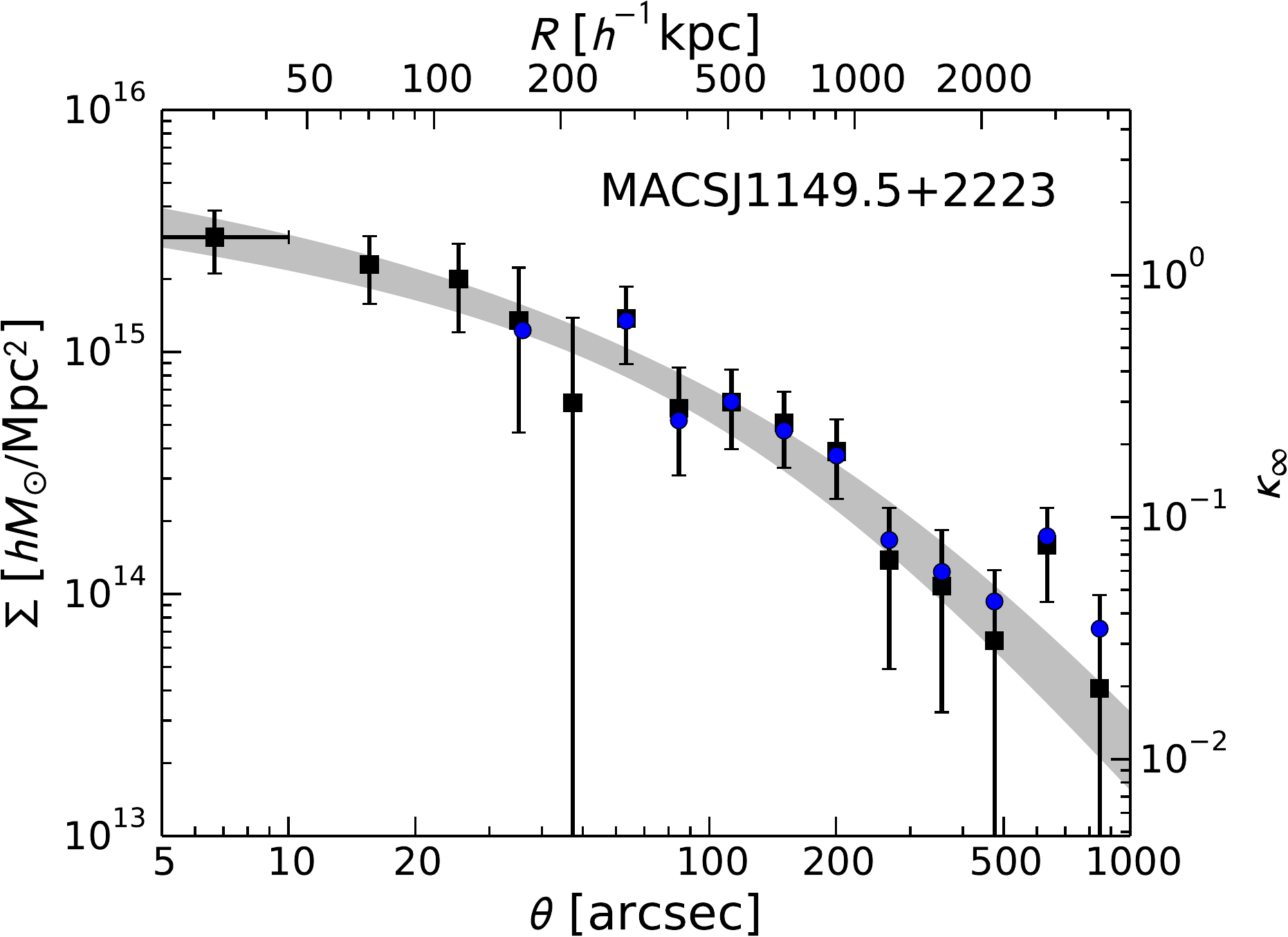}\\
   \includegraphics[width=0.39\textwidth,angle=0,clip]{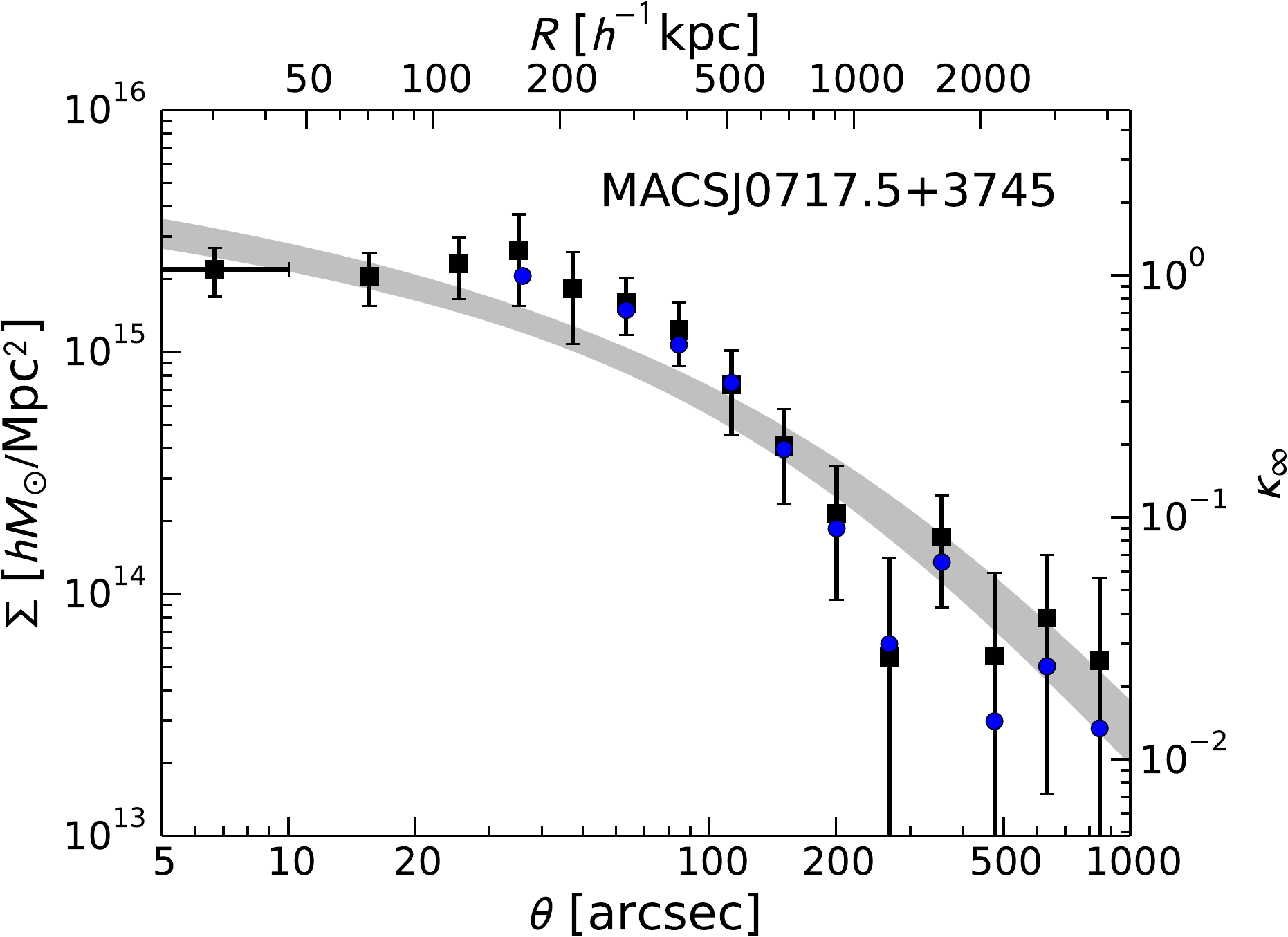}&
   \includegraphics[width=0.39\textwidth,angle=0,clip]{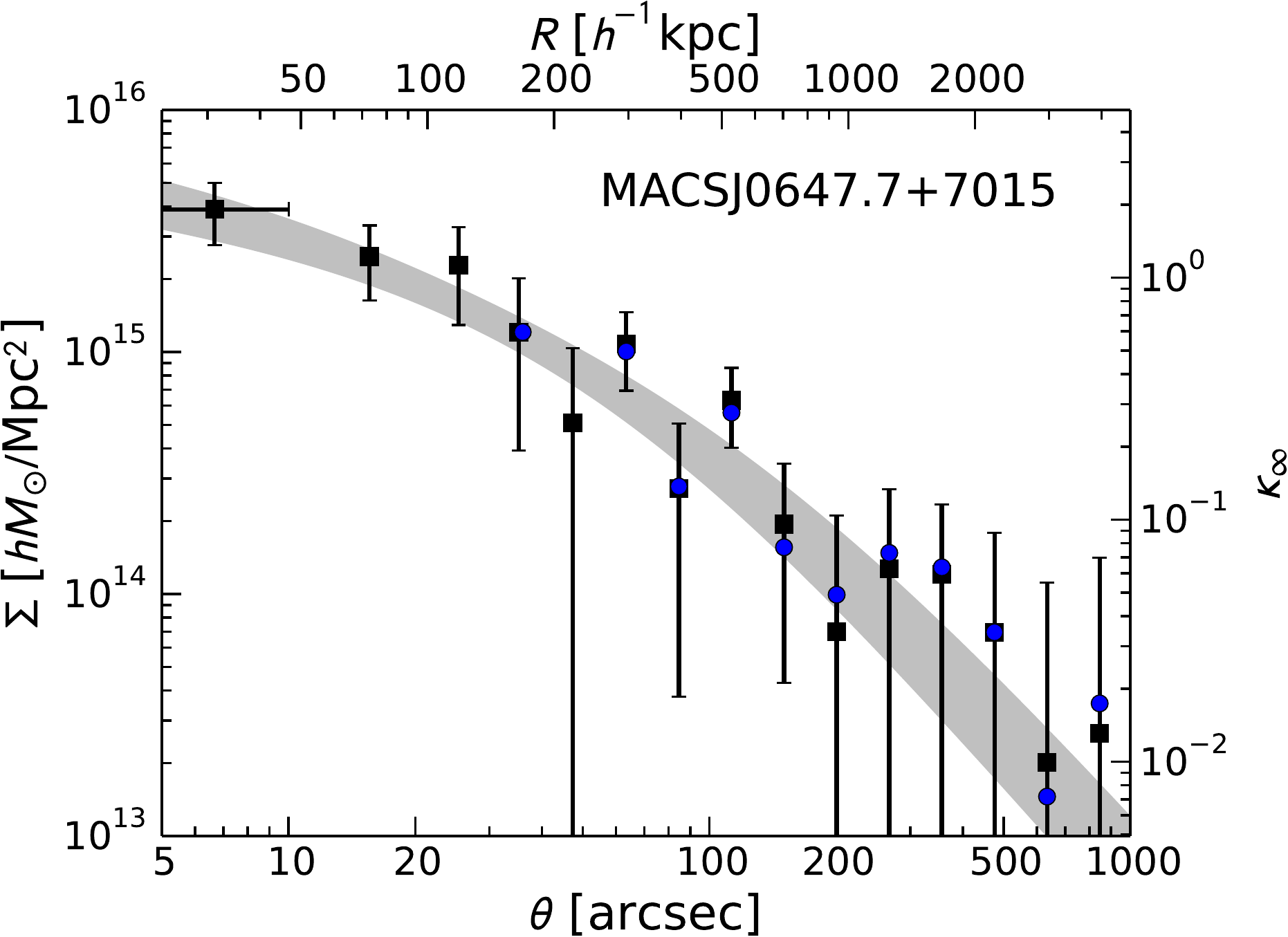} 
 \end{array}
 $
 \end{center}
  \caption{(Continued.)}
\end{figure*}



  
\end{document}